\documentclass[a4paper,11pt]{article}
\usepackage{amsmath}
\usepackage[margin=2cm]{geometry}
\usepackage{amsfonts}
\usepackage{amssymb}
\usepackage{graphicx}
\usepackage{subcaption}
\usepackage{slashed}
\usepackage{float}
\usepackage{hyperref}
\hypersetup{colorlinks=true,linkcolor=blue,citecolor=blue,urlcolor=blue,
pdftitle={One-loop renormalization group flow of the translation-invariant noncommutative Yukawa theory},
pdfauthor={Karim Bouchachia and Smain Kouadik}}
\usepackage{tikz}
\usepackage{pgfplots}
\pgfplotsset{compat=1.18}
\usetikzlibrary{arrows.meta,decorations.markings,calc,fillbetween}
\usepgfplotslibrary{fillbetween}

\newcommand{\pt}{\tilde{\slashed{p}}}
\newcommand{\NC}{\mathrm{NC}}
\newcommand{\Cth}{\mathrm{C}}
\newcommand{\MSbar}{\overline{\mathrm{MS}}}
\numberwithin{equation}{section}
\newcommand{\keywords}[1]{\par\noindent\textbf{Keywords:} #1}
\tikzset{
	fline/.style={line width=0.6pt,
		postaction={decorate},
		decoration={markings,
			mark=at position 0.5 with {\arrow{Stealth[length=1.8mm,width=1.3mm]}}}},
	fdash/.style={fline, dash pattern=on 2.6pt off 2.2pt}
}
\newcommand{\ctblob}[3]{%
	\draw[line width=0.6pt] (#1) circle (#2);
	\draw[line width=0.6pt] ($(#1)+({#3}:#2)$) -- ($(#1)+({#3+180}:#2)$);
	\draw[line width=0.6pt] ($(#1)+({#3+90}:#2)$) -- ($(#1)+({#3+270}:#2)$);
}
\newcommand{\feynicon}[2]{\begin{tikzpicture}[x=0.15mm,y=0.15mm,baseline={(0,{#1*0.15mm})}]#2\end{tikzpicture}}

\title{One-loop renormalization group flow of the translation-invariant noncommutative Yukawa theory}

\author{Karim Bouchachia\thanks{Corresponding author. Email: bouchachia.karim@univ-medea.dz}
	\ and Smain Kouadik\thanks{Email: kouadik.smain@univ-medea.dz} \\[0.5em]
	\normalsize Laboratory of Experimental Physics Techniques and Applications, \\
	\normalsize University of Medea, Medea, Algeria}
\date{}

\begin{document}
	\maketitle
	
	\begin{abstract}
	We study the one-loop renormalization group flow of the
	translation-invariant noncommutative pseudoscalar Yukawa theory on
	four-dimensional Euclidean Moyal space. The two
	inequivalent orderings of the Moyal-star Yukawa vertex give rise
	to two independent couplings $g_{1}$ and $g_{2}$, whose beta
	functions form a coupled nonlinear system; every one-loop
	divergence is absorbed by a
	counterterm already present in the action, establishing one-loop
	renormalizability of the theory. We solve the one-loop
	system analytically: it decouples in the variables
	$u=g_{1}^{2}+g_{2}^{2}$ and $v=g_{1}^{2}-g_{2}^{2}$, the quartic
	coupling is obtained by a Riccati reduction on the symmetric
	surface $g_{1}=g_{2}$, and the generic asymmetric flow is reduced
	to a single quadrature by the exact invariant
	$\mathcal{I}=(g_{1}g_{2})^{3}/(g_{1}^{2}-g_{2}^{2})^{4}$. Two
	dynamical consequences follow. The ratio $r=v/u$ has a
	UV-attractive symmetric surface $r=0$ and IR-attractive asymmetric
	directions $r=\pm1$, so that the flow restores the symmetry
	between the two Moyal orderings towards the ultraviolet and
	amplifies any initial ordering asymmetry towards the infrared,
	as a fractional power of the logarithm. In consequence the
	Yukawa-induced coefficient of the non-planar $1/(\theta^{2}p^{2})$
	structure is dynamically suppressed towards $p\rightarrow 0$;
	the singularity itself is not removed,
	and remains the task of the IR-improving term.  We also
	solve the dimensionful sector ($M^{2}$, $m$, $a^{2}$ and $b$) in closed
	form on the symmetric critical trajectory, and compare throughout
	with the commutative theory. Unlike $\beta_{g_{1}}$, $\beta_{g_{2}}$,
	$\beta_{\lambda}$, $\beta_{m}$ and $\beta_{M^{2}}$, which follow
	from strict $\MSbar$ pole extraction, the value of the beta
	function of the IR-improving parameter $a^{2}$ is fixed only
	under an additional infrared-subtraction prescription. The
	attractor value $a^{2}_{*}\theta^{2}=3$ quoted
	below is a consequence of that choice rather than a prediction
	of the theory, and the physical interpretation of the $a^{2}$
	sector is conditional on it throughout.
	\end{abstract}
	
	\keywords{Noncommutative field theory;
		Renormalization group;
		UV/IR mixing;
		Yukawa theory;
		Beta functions;
		Riccati equation.}
	
	\vspace{1em}

\newpage
\maketitle
\tableofcontents
\newpage

\section{Introduction}

Noncommutative (NC) quantum field theories on Moyal space provide a
well-motivated framework for probing physics beyond the standard model
at short distances, where spacetime noncommutativity may be induced by
quantum gravity effects, string theory compactifications, or
Planck-scale physics~\cite{minwala,douglas,zsabo1}. A central
challenge in this framework is IR/UV
mixing~\cite{minwala,ramsd,micu&jab2001}: UV divergences of non-planar
Feynman diagrams re-enter as IR singularities, obstructing standard
renormalization. The translation-invariant approach of Gurau, Magnen,
Rivasseau, and Tanasa (GMRT)~\cite{rivasseau2008} restores
renormalizability by adding specific IR-improving terms to the action
that absorb the residual non-planar singularity, and this approach
has been successfully applied to the scalar $\varphi^{4}$
theory~\cite{rivasseau2008,schweda1,Rivasseau2007}.

In a previous paper~\cite{karim} we extended the GMRT construction to
the pseudoscalar Yukawa theory, coupling a translation-invariant
scalar field to a Dirac fermion on Moyal space, and computed all
one-loop 1PI quantum corrections up to the four-point function.
The evaluation showed that one-loop renormalizability can be restored by
adding extra terms of the form $\sim 1/(\theta^{2}p^{2})$ to the
scalar action and $\sim\pt/(\theta p^{2})$ to
the fermion action. 
The framework has recently been applied to the phenomenology of 
neutrino Yukawa couplings in NC geometry~\cite{boutheldja2025},
extending its relevance beyond the purely theoretical setting.

The present paper goes substantially beyond the computation of
divergences in~\cite{karim}. The one-loop diagrams themselves --
the propagator and vertex corrections whose pole parts we use as
input here -- were already obtained in that earlier work. What is
new in the present paper is everything built on top of them:
the passage from renormalization constants to explicit beta
functions, the identification of the resulting two-coupling RG
dynamics, the exact invariant that reduces the generic asymmetric
flow to a single quadrature, the infrared ordering-selection
mechanism that follows from it, and the closed-form dimensionful
running this makes possible. Since the noncommutativity of the star
product allows two inequivalent orderings of the Yukawa vertex, whose
coefficients we treat as independent couplings, the theory possesses
an RG direction with no commutative counterpart, and it is the
dynamics of that direction -- not the one-loop diagrams that generate
it -- which concerns us here.

The related paper \cite{boutheldja2025} numerically treats
the six mirror Yukawa couplings arising in a multi-coupling
electroweak-scale right-handed neutrino model.
The present work is complementary: we isolate the
minimal single-pair two-coupling sector that underlies the general
vulcanization framework and treat it analytically, extending the
analysis to the full dimensionful sector and to the commutative
limit.

The paper is organized as follows. Section~\ref{sec:model} defines
the elements of the translation-invariant noncommutative pseudo-scalar Yukawa model.
In Section~\ref{sec:renorm} we extract the one-loop pole parts of all
1PI functions computed in~\cite{karim} and identify the
corresponding counterterms, paying particular attention to the
non-planar $1/(\theta^{2}p^{2})$ coefficient, whose status differs
from that of the ultraviolet counterterms. Section~\ref{sec:beta}
derives the beta functions of the dimensionless sector in the
$\MSbar$ scheme and examines the symmetries of the resulting system.
Section~\ref{sec:running} integrates that system: the reduction to
the variables $(u,v,r)$, the exact invariant of the flow, the
symmetric-surface solution, and the Riccati solution for the quartic
coupling. Section~\ref{sec:mass_running} extends the analysis to the
fermion mass, the scalar mass and the IR-improving parameters, and
Section~\ref{sec:physical} draws the physical consequences of the
running. Section~\ref{sec:discussion} discusses the results: Section~\ref{sec:comparison}
compares the noncommutative and commutative theories under the two
natural matching prescriptions and examines the commutative limit
itself, Section~\ref{sec:central_finding} presents our main results
and sketches the connection to phenomenological multi-flavor
extensions, Section~\ref{sec:limitations} gathers the boundaries of
what is established, and Section~\ref{sec:outlook} sketches the
directions the analysis suggests for future work. Three appendices give the detailed algebra
of the $(u,v,r)$ flow, the phase-factor decomposition of the one-loop
three-point function, and a numerical illustration of the analytic
solutions obtained throughout the paper.

\section{The Model and Feynman Rules}
\label{sec:model}

\subsection{Action and Conventions}
\label{sec:action_conventions}

The total translation-invariant action is obtained from the ordinary
NC Yukawa action by adding the extra IR-improving terms
$-a^{\prime2}\varphi\star(\theta^{2}\square)^{-1}\varphi$ and
$-b^{\prime}\bar{\psi}\star\tilde{\slashed{\partial}}
(\theta^{2}\square)^{-1}\psi$:
\begin{align}
	S_{\star}^{\mathrm{tot}}[\psi,\bar{\psi},\varphi]
	&=\int d^{4}x\Bigl[
	\tfrac{1}{2}\partial^{\mu}\varphi\star\partial^{\mu}\varphi
	+\tfrac{M^{2}}{2}\varphi\star\varphi
	-\tfrac{a^{2}}{2}\varphi\star\tfrac{1}{\square}\varphi\Bigr]
	\nonumber\\
	&\quad+\int d^{4}x\Bigl[
	\bar{\psi}\star\slashed{\partial}\psi
	+m\,\bar{\psi}\star\psi
	-b\,\bar{\psi}\star\tfrac{\tilde{\slashed{\partial}}}{\theta\square}\psi
	\Bigr]\nonumber\\
	&\quad+\int d^{4}x\Bigl[
	i\,g_{1}\,\bar{\psi}\gamma^{5}\star\psi\star\varphi
	+i\,g_{2}\,\bar{\psi}\gamma^{5}\star\varphi\star\psi
	+\tfrac{\lambda}{4!}\varphi^{\star4}
	\Bigr]\nonumber\\
	&\quad+S_{\mathrm{ct}},
	\label{eq:action}
\end{align}
where $a=a^{\prime}/\theta$ and $b=b^{\prime}/\theta$.

The terms $-a^{\prime2}\varphi\star(\theta^{2}\square)^{-1}\varphi$
and $-b^{\prime}\bar{\psi}\star\tilde{\slashed{\partial}}
(\theta^{2}\square)^{-1}\psi$ are not optional refinements: 
the first is the direct response to UV/IR mixing~\cite{minwala,ramsd,micu&jab2001},
while the second is a consistency condition introduced in \cite{karim}: 
it ensures that the scalar propagator obtained directly from the 
action agrees with the one obtained by squaring the fermion propagator, 
and it plays the same IR-improving role in the fermion sector.
The UV/IR mixing is the phenomenon whereby the
non-planar sector of a noncommutative theory develops, upon
integrating out the loop momentum, a singularity as the external
momentum $p\to0$ in place of what would have been an ordinary UV
divergence in the commutative theory. Because this singularity is
infrared rather than ultraviolet in origin, it cannot be absorbed by
the usual local counterterms without requiring new operators at every
order, obstructing the renormalization program. The GMRT
construction~\cite{rivasseau2008} resolves this obstruction by adding,
from the outset, IR-improving terms of exactly the singular form the
mixing produces, so that the residual non-planar singularity is
absorbed into a renormalization of $a^{2}$ (and $b$ for the fermion
sector) rather than requiring new operators at each order.

The bare fields are related to the renormalized fields by
$\psi_{0}=\sqrt{Z_{\psi}}\,\psi$ and
$\varphi_{0}=\sqrt{Z_{\varphi}}\,\varphi$, which defines the
counterterms
\begin{align}
	\delta_{\varphi}&=Z_{\varphi}-1,\quad\delta_{\psi}=Z_{\psi}-1,
	\nonumber\\
	\delta_{M}&=M_{0}^{2}Z_{\varphi}-M^{2},\quad
	\delta_{m}=m_{0}Z_{\psi}-m,\nonumber\\
	\delta_{\lambda}&=\lambda_{0}Z_{\varphi}^{2}-\lambda,\quad
	\delta_{g_{i}}=g_{0i}Z_{\psi}Z_{\varphi}^{1/2}-g_{i},\nonumber\\
	\delta_{a^{2}}&=a_{0}^{2}Z_{\varphi}-a^{2},\quad
	\delta_{b}=b_{0}Z_{\psi}-b.
	\label{counterterms}
\end{align}
The mass dimensions on $\mathbb{R}^{4}$ are
$[\varphi]=1$, $[\psi]=3/2$, $[g_{i}]=0$, $[\lambda]=0$, $[M]=[m]=1$.

The Yukawa interaction terms in (\ref{eq:action}) differ from the ones introduced in \cite{karim} by a factor $i$. A Hermitian pseudoscalar Yukawa interaction has to carry an explicit
factor of $i$, exactly as in the standard linear sigma-model form
$g\,\bar{\psi}(\sigma+i\gamma^{5}\boldsymbol{\tau}\!\cdot\!\boldsymbol{\pi})\psi$.
This factor is not a matter of convention: a real Euclidean Lagrangian requires a
Hermitian interaction density, and since $\bar\psi\gamma^{5}\psi$ is anti-Hermitian
\begin{equation}
	(\bar{\psi}\gamma^{5}\psi)^{\dagger}
	=\psi^{\dagger}\gamma^{5\dagger}\gamma^{0\dagger}\psi
	=\psi^{\dagger}\gamma^{5}\gamma^{0}\psi
	=-\psi^{\dagger}\gamma^{0}\gamma^{5}\psi
	=-\bar{\psi}\gamma^{5}\psi,
\end{equation}
this factor is a necessary -- though not, on its own, established as
sufficient -- condition for the Euclidean two-point functions to
satisfy the Osterwalder-Schrader reflection-positivity condition
under time reflection~\cite{osterwalder1975,montvay}.
Its practical effect is that all contributions built from an even number of Yukawa
vertices (the two-point functions, and the $g_{1}g_{2}$ part of the
non-planar scalar self-energy) and the loop-to-tree ratio of the
three-vertex correction acquire a relative minus sign compared
with $g\bar{\psi}\gamma^{5}\psi\varphi$ assignment, while
four-vertex contributions are unaffected. As a consistency check, this allows the model to 
reproduce the standard commutative results for the beta functions
$\beta^{\Cth}_{g}=+5g^{3}/(4\pi)^{2}$ and
$\beta^{\Cth}_{\lambda}=[3\lambda^{2}+8g^{2}\lambda-48g^{4}]/(4\pi)^{2}$.
Since $\psi$ and $\bar\psi$ are independent Grassmann generators in
the Euclidean functional integral \cite{Zinn}, this condition constrains the
action under time reflection and not the two orderings against one
another; $g_{1}$ and $g_{2}$ are therefore independent real
parameters, and are not related by conjugation. 

Throughout this paper we work in four-dimensional Euclidean space, $\mathbb{R}^{4}$, with the flat Euclidean metric $\delta_{\mu\nu}$.
The Weyl--Moyal algebra is defined by
\begin{equation}
	[x^{\mu},x^{\nu}]_{\star}=x^{\mu}\star x^{\nu}-x^{\nu}\star x^{\mu}
	=i\theta^{\mu\nu},
\end{equation}
where the Moyal star product is
\begin{equation}
	f(x)\star g(x)\equiv
	e^{\frac{i}{2}\theta^{\mu\nu}\frac{\partial}{\partial x^{\mu}}
		\frac{\partial}{\partial y^{\nu}}}f(x)g(y)\big|_{x=y},
\end{equation}
and $(\theta^{\mu\nu})$ is the real antisymmetric deformation matrix
in standard block-diagonal form~\cite{rivasseau2008,schweda1}
\begin{equation}
	(\theta^{\mu\nu})=\theta\begin{pmatrix}
		0 & 1 & 0 & 0\\
		-1 & 0 & 0 & 0\\
		0 & 0 & 0 & 1\\
		0 & 0 & -1 & 0
	\end{pmatrix},
	\qquad \theta\equiv|\theta|\;\;\text{(mass}^{-2}\text{)}.
	\label{eq:theta_matrix}
\end{equation}
We use the Euclidean metric on $\mathbb{R}^{4}$ throughout, the
convention $\slashed{a}=\gamma^{\mu}a_{\mu}$, the Euclidean
$\gamma$-matrices satisfying $\{\gamma^{\mu},\gamma^{\nu}\}=2\delta^{\mu\nu}$,
and the notation $\tilde{a}^{\mu}=\theta^{\mu\nu}a_{\nu}$. The vulcanization
parameters $a,b$ have mass dimensions $[a]=2$ and $[b]=2$ (their primed
counterparts are dimensionless, $a=a'/\theta$, $b=b'/\theta$).

\subsection{Feynman Rules}
\label{sec:feynman}

The Feynman rules in momentum
space are
\begin{subequations}
	\begin{align}
		\feynicon{63}{
			\draw[fdash] (40,63) -- (176,63);
			\node[above] at (108,63) {$p$};
		}
		&=\tilde{G}^{\prime}(p^{2},M,a)
		=\frac{1}{p^{2}+M^{2}+a^{2}/p^{2}},
		\label{eq:scalprop}\\[4pt]
		\feynicon{60}{
			\draw[fline] (34,60) -- (165,60);
			\node[above] at (99.5,60) {$p$};
		}
		&=\tilde{D}^{\prime}(p,m,b)
		=\frac{1}{i\slashed{p}+m+ib\,\pt/(\theta p^{2})},
		\label{eq:fermprop}\\[4pt]
		\feynicon{22}{
			\draw[fline] (47,60) -- (107,22);
			\draw[fline] (108,23) -- (48,-13);
			\draw[fdash] (108,22) -- (162,23);
			\node[above left] at (77,41) {$p$};
			\node[below left] at (78,5) {$p'$};
			\node[above] at (135,22.5) {$q$};
		}
		&=-i\,\gamma^{5}\,V_{g}(p^{\prime},p),
		\label{eq:yukvert}\\[4pt]
		\feynicon{17}{
			\draw[fdash] (30,-10) -- (61,17);
			\draw[fdash] (28,49) -- (63,16);
			\draw[fdash] (62,18) -- (97,-15);
			\draw[fdash] (62,17) -- (94,44);
			\node[anchor=north east] at (30,-10) {$p_2$};
			\node[anchor=south east] at (28,49) {$p_1$};
			\node[anchor=north west] at (97,-15) {$p_3$};
			\node[anchor=south west] at (94,44) {$p_4$};
		}
		&=V_{\lambda}(p_{1},p_{2},p_{3},p_{4}),
		\label{eq:phi4vert}
	\end{align}
\end{subequations}
where the vertex functions are
\begin{align}
	V_{\lambda}&=-\frac{\lambda}{3}\Bigl(
	\cos\tfrac{p_{1}\tilde{p}_{2}}{2}\cos\tfrac{p_{3}\tilde{p}_{4}}{2}
	+\cos\tfrac{p_{1}\tilde{p}_{3}}{2}\cos\tfrac{p_{2}\tilde{p}_{4}}{2}
	+\cos\tfrac{p_{1}\tilde{p}_{4}}{2}\cos\tfrac{p_{3}\tilde{p}_{2}}{2}
	\Bigr),\nonumber\\
	V_{g}(p^{\prime},p)&=g_{1}\,e^{+\frac{i}{2}p^{\prime}\tilde{p}}
	+g_{2}\,e^{-\frac{i}{2}p^{\prime}\tilde{p}}
	\;=\;\sum_{\sigma=\pm1}g_{\sigma}\,
	e^{\frac{i}{2}\sigma p^{\prime}\tilde{p}},
	\label{eq:vg}
\end{align}
with the notation $g_{+1}\equiv g_{1}$, $g_{-1}\equiv g_{2}$.
The renormalized counterterm
vertices are
\begin{subequations}
	\begin{align}
		\feynicon{-28}{
			\draw[fdash] (38,-28) -- (92,-28);
			\ctblob{100,-27}{8.3}{-45}
			\draw[fdash] (108,-28) -- (165,-28);
			\node[above] at (65,-28) {$p$};
			\node[above] at (136.5,-28) {$p$};
		}
		&=\Gamma_{\varphi\text{-ct}}^{(2)}
		=\delta_{\varphi}p^{2}+\delta_{M}+\delta_{a^{2}}/p^{2},
		\\[4pt]
		\feynicon{-28}{
			\draw[fline] (38,-28) -- (92,-28);
			\ctblob{100,-27}{8.3}{-45}
			\draw[fline] (108,-28) -- (161,-28);
			\node[above] at (65,-28) {$p$};
			\node[above] at (134.5,-28) {$p$};
		}
		&=\Gamma_{\psi\text{-ct}}^{(2)}
		=i\delta_{\psi}\slashed{p}
		+i\delta_{b}\pt/(\theta p^{2})+\delta_{m},
		\\[4pt]
		\feynicon{22}{
			\draw[fline] (47,60) -- (101,27);
			\draw[fline] (101,17) -- (48,-13);
			\ctblob{107,22}{8.3}{-135}
			\draw[fdash] (115,22) -- (162,23);
			\node[above left] at (73,44) {$p$};
			\node[below left] at (73,1) {$p'$};
			\node[above] at (139.5,22.5) {$q$};
		}
		&=\Gamma_{\mathrm{ct}}^{(3)}
		=-i\,\gamma^{5}V_{\delta g}(p^{\prime},p),
		\\[4pt]
		\feynicon{17}{
			\draw[fdash] (29,-10) -- (55,12);
			\draw[fdash] (28,49) -- (65,14);
			\ctblob{61,17}{8.3}{45}
			\draw[fdash] (68,12) -- (97,-15);
			\draw[fdash] (68,24) -- (94,44);
			\node[anchor=north east] at (29,-10) {$p_2$};
			\node[anchor=south east] at (28,49) {$p_1$};
			\node[anchor=north west] at (97,-15) {$p_3$};
			\node[anchor=south west] at (94,44) {$p_4$};
		}
		&=\Gamma_{\mathrm{ct}}^{(4)}
		=\frac{\delta_{\lambda}}{\lambda}V_{\lambda}(p_{1},p_{2},p_{3},p_{4}),
	\end{align}
\end{subequations}
with $V_{\delta g}(p^{\prime},p)=\delta_{g_{1}}e^{+\frac{i}{2}p^{\prime}\tilde{p}}
+\delta_{g_{2}}e^{-\frac{i}{2}p^{\prime}\tilde{p}}$. Note that all
quadratic counterterm insertions in $\Gamma_{\varphi\text{-ct}}^{(2)}$
and $\Gamma_{\psi\text{-ct}}^{(2)}$ carry a common sign, as they must,
since each is obtained from the corresponding tree-level structure
in~\eqref{eq:action} by replacing the parameter with its counterterm.

Throughout, the full 1PI two-point functions are written as \cite{Zinn}
\begin{align}
	\Gamma^{(2)}_{\varphi}&=\bigl(p^{2}+M^{2}+a^{2}/p^{2}\bigr)
	+\Gamma^{(2)}_{\varphi\text{-ct}}-\Gamma^{(2)}_{\varphi\text{-1loop}},
	\nonumber\\
	\Gamma^{(2)}_{\psi}&=\bigl(i\slashed{p}+m+ib\,\pt/(\theta p^{2})\bigr)
	+\Gamma^{(2)}_{\psi\text{-ct}}-\Gamma^{(2)}_{\psi\text{-1loop}},
	\label{eq:Gamma2_convention}
\end{align}
so that $\Gamma^{(2)}_{\text{1loop}}$ denotes the 1PI loop insertion
in the conventional self-energy normalization
$G=[\,G_{0}^{-1}-\Sigma\,]^{-1}$, and the counterterms are fixed by
demanding that~\eqref{eq:Gamma2_convention} be finite.

The relative minus sign in~\eqref{eq:Gamma2_convention} is specific to
the two-point functions: it is what makes
$\Gamma^{(2)}_{\text{1loop}}$ the self-energy $\Sigma$ in the
normalization $G=[\,G_{0}^{-1}-\Sigma\,]^{-1}$, rather than the 1PI
vertex function itself. For the three- and four-point functions no
such inversion is involved and we use throughout the direct additive
convention
\begin{equation}
	\Gamma^{(n)}=\Gamma^{(n)}_{\text{tree}}
	+\Gamma^{(n)}_{\text{ct}}+\Gamma^{(n)}_{\text{1loop}},
	\qquad n=3,4,
	\label{eq:Gamma34_convention}
\end{equation}
the counterterms again being fixed by finiteness. 

In what follows, only the UV-divergent (pole) parts of the one-loop
corrections are used to fix the counterterms; the finite analytic
functions $f_{i}$ that appear in the complete one-loop results
of Ref.~\cite{karim} are discarded for the RG analysis (they
contribute to finite renormalization but not to the leading
$1/\varepsilon$ residues). The one exception is the non-planar
$1/(\theta^{2}p^{2})$ structure, which is UV-finite at
$\theta\neq0$ and is discussed separately in
Section~\ref{sec:a2_status}.

\section{Counterterm Structure and Pole Parts}
\label{sec:renorm}

In this section we extract the leading $1/\varepsilon$ pole structure
of all one-loop 1PI quantum corrections of~\cite{karim}, and identify
the corresponding counterterms, in the notation of
Eq.~\eqref{counterterms}. Because the Yukawa vertex
of~\eqref{eq:yukvert} carries the factor $i$ required by hermiticity,
the two-point and three-point results below differ in overall sign
from the expressions quoted in Ref.~\cite{karim}, where the factor was
omitted; the four-point results are unaffected, since they involve an
even number of pairs of Yukawa vertices.

\subsection{Scalar Propagator}
\label{sec:scalar_prop}

The one-loop quantum corrections to the scalar two-point function
are, in the normalization of~\eqref{eq:Gamma2_convention},
\begin{align}
\Gamma_{\varphi\text{-1loop}}^{(2)}
&=-\frac{4(g_{1}^{2}+g_{2}^{2})}{(4\pi)^{2}\varepsilon}p^{2}
+\frac{1}{(4\pi)^{2}\varepsilon}
\Bigl[\tfrac{2\lambda}{3}M^{2}-8(g_{1}^{2}+g_{2}^{2})m^{2}\Bigr]
\nonumber\\
&\quad
+\Bigl[\,32g_{1}g_{2}-\tfrac{2\lambda}{3}\,\Bigr]
\frac{1}{(4\pi)^{2}\theta^{2}p^{2}}
+O(g^{3},\lambda^{2}).
\label{eq:Gamma2_phi}
\end{align}
The first line is the planar, UV-divergent part; the second line is
the non-planar contribution, which is UV-finite for
$\theta\neq0$ and singular only as $\tilde{p}^{2}\to0$.

The on-shell renormalization conditions used to fix the finite parts
are
\begin{subequations}
\begin{align}
\Sigma_{\mathrm{ren}}(p^{2},a)\big|_{p^{2}=-M^{2}}&=0,\label{scl1}\\
\frac{\partial\Sigma_{\mathrm{ren}}}{\partial p^{2}}
\bigg|_{p^{2}=-M^{2}}&=0,\label{scl2}\\
\Sigma_{\mathrm{ren}}^{\NC}(p^{2},a)\big|_{a=0}
&=\Sigma_{\mathrm{ren}}^{\Cth}(p^{2}),\label{scl3}
\end{align}
\end{subequations}
where $\Sigma_{\mathrm{ren}}\equiv\Gamma^{(2)}_{\varphi\text{-1loop}}
-\Gamma^{(2)}_{\varphi\text{-ct}}$; condition~\eqref{scl1} fixes the
finite part of $\delta_{M}$, condition~\eqref{scl2} is the
standard residue condition, and~\eqref{scl3} ensures that the
commutative propagator is recovered when $a=0$. Demanding finiteness
of~\eqref{eq:Gamma2_convention} determines the divergent parts of the
scalar counterterms,
\begin{subequations}
\begin{align}
\delta_{\varphi}&=-\frac{4(g_{1}^{2}+g_{2}^{2})}{(4\pi)^{2}\varepsilon},\\
\delta_{M}&=\frac{1}{(4\pi)^{2}\varepsilon}
\Bigl[\tfrac{2\lambda}{3}M^{2}-8(g_{1}^{2}+g_{2}^{2})m^{2}\Bigr],\\
\delta_{a^{2}}&=\frac{1}{(4\pi)^{2}\theta^{2}}
\Bigl[\,32\,g_{1}g_{2}-\tfrac{2\lambda}{3}\,\Bigr].
\end{align}
\end{subequations}

\subsection{Fermion Propagator}
\label{sec:fermion_prop}

The one-loop quantum correction to the fermionic two-point function is
\begin{equation}
\Gamma_{\psi\text{-1loop}}^{(2)}
=-\frac{(g_{1}^{2}+g_{2}^{2})}{(4\pi)^{2}\varepsilon}
\bigl(i\slashed{p}+2m\bigr)
+O(g^{3}),
\label{fermion2pt}
\end{equation}
whose UV-divergent part is identical to the commutative result with
$g^{2}\to g_{1}^{2}+g_{2}^{2}$. The non-planar part of the fermion
self-energy is finite for $\theta\neq0$ and contains no
$\pt/(\theta^{2}p^{2})$ pole~\cite{karim}. Imposing the on-shell
conditions
\begin{subequations}
	\begin{align}
		\Pi_{\mathrm{ren}}\!\left(\slashed{p},b\right)\big|_{\slashed{p}=im}&=0,\\
		\frac{\partial\Pi_{\mathrm{ren}}\!\left(\slashed{p},b\right)}{\partial\slashed{p}}
		\bigg|_{\slashed{p}=im}&=0,\\
		\Pi_{\mathrm{ren}}^{\mathrm{NC}}\!\left(\slashed{p},b\right)\big|_{b=0}
		&=\Pi_{\mathrm{ren}}^{\mathrm{C}}\!\left(\slashed{p}\right),
	\end{align}
\end{subequations}
with $\Pi_{\mathrm{ren}}\equiv\Gamma^{(2)}_{\psi\text{-1loop}}
-\Gamma^{(2)}_{\psi\text{-ct}}$, and~\eqref{fermion2pt} used for
$\Gamma^{(2)}_{\psi\text{-1loop}}$, the fermion counterterms are
\begin{subequations}
\begin{align}
\delta_{\psi}&=-\frac{g_{1}^{2}+g_{2}^{2}}{(4\pi)^{2}\varepsilon},\\
\delta_{m}&=-\frac{2(g_{1}^{2}+g_{2}^{2})}{(4\pi)^{2}\varepsilon}\,m,\\
\delta_{b}&=0.
\end{align}
\end{subequations}
Equivalently, $Z_{\psi}=1-u/[(4\pi)^{2}\varepsilon]<1$ and
\begin{equation}
m_{0}=m\Bigl[1-\frac{g_{1}^{2}+g_{2}^{2}}{(4\pi)^{2}\varepsilon}\Bigr],
\qquad b_{0}=\frac{b}{Z_{\psi}} .
\label{eq:m_bare_rel}
\end{equation}
This result is worth reading in light of why $b$ was introduced in
the first place, which is conceptually different from the motivation
for $a^{2}$. Whereas $a^{2}$ is an independent regulator, chosen
specifically to absorb the scalar sector's own non-planar
singularity, $b$ was introduced in~\cite{karim} as a consistency
condition at tree level. Read this way, $\delta_{b}=0$ shows that no
independent divergent $b$ counterterm is required at one loop.

\subsection{Yukawa Interaction Vertex}
\label{sec:yukawa_vertex}

The complete one-loop 1PI three-point function consists of eight
phase-factor channels, obtained by assigning an ordering
$\sigma,\sigma',\sigma''=\pm1$ to each of the three Yukawa vertices.
As shown in detail in Appendix~\ref{app:gamma3}, only the two
channels with $\sigma'=\sigma''=-\sigma$ have a loop-momentum
independent phase; these are the planar channels and they alone
produce a UV divergence. Their coefficient is
$g_{\sigma}g_{-\sigma}^{2}$, so that, in the additive convention
of~\eqref{eq:Gamma34_convention},
\begin{equation}
\Gamma_{1\mathrm{loop}}^{(3)}(p,p^{\prime})
=i\,\gamma^{5}\,\frac{2}{(4\pi)^{2}\varepsilon}\,
\bigl[\,g_{1}g_{2}^{2}\,e^{+\frac{i}{2}p^{\prime}\tilde p}
+g_{1}^{2}g_{2}\,e^{-\frac{i}{2}p^{\prime}\tilde p}\bigr]
+O(g^{5}).
\label{eq:gamma3_explicit}
\end{equation}
The remaining six channels carry $k$-dependent phases: the two with
$\sigma=\sigma'=\sigma''$ have coefficient $g_{\sigma}^{3}$ and the
four mixed ones have coefficient $g_{1}g_{2}g_{\pm\sigma}$. All are
UV-finite at $\theta\neq0$, which is why no $g_{1}^{3}$ or
$g_{2}^{3}$ term appears in~\eqref{eq:gamma3_explicit}.

It is worth being precise about what happens to this counting at
$\theta=0$. When all eight phase factors collapse to unity the 
coupling factors sum to the identity
\begin{equation}
	\underbrace{g_{1}^{3}+g_{2}^{3}}_{\sigma=\sigma'=\sigma''}
	+\underbrace{3g_{1}g_{2}(g_{1}+g_{2})}_{\text{2 planar}+\text{4 mixed}}
	=(g_{1}+g_{2})^{3}\equiv g_{C}^{3},
	\label{eq:channel_sum_identity}
\end{equation}
so the commutative vertex divergence is $2g_{C}^{3}/[(4\pi)^{2}\varepsilon]$,
as it must be. The planar piece retained at $\theta\neq0$ is only
$g_{1}g_{2}(g_{1}+g_{2})$, which on the symmetric surface equals
$g_{C}^{3}/4$ and not $g_{C}^{3}$: the planar channels alone
therefore fall short of the commutative vertex divergence by a factor
$4$, the deficit being carried by the six channels that are
non-planar, and hence UV-finite, at $\theta\neq0$.

Splitting along the two independent phase channels and imposing
finiteness of the total three-point function gives
\begin{subequations}
\begin{align}
\delta_{g_{1}}&=\frac{2\,g_{1}g_{2}^{2}}{(4\pi)^{2}\varepsilon},
\label{deltag1}\\
\delta_{g_{2}}&=\frac{2\,g_{1}^{2}g_{2}}{(4\pi)^{2}\varepsilon}.
\label{deltag2}
\end{align}
\end{subequations}
The exchange symmetry $1\leftrightarrow 2$ is manifest, as is the
correct single-coupling limit: if $g_{2}\to 0$ then
$\delta_{g_{1}}\to 0$, reflecting the absence of a cross-channel
loop when only one ordering is present.

Combining \eqref{deltag1}--\eqref{deltag2} with the field-strength
renormalizations $\delta_{\varphi}$ and $\delta_{\psi}$ gives the
relation between bare and renormalized Yukawa couplings:
\begin{subequations}
\begin{align}
g_{01}&=g_{1}\Bigl[1+\frac{1}{(4\pi)^{2}\varepsilon}
\bigl(3g_{1}^{2}+5g_{2}^{2}\bigr)\Bigr]+O(g^{5}),\\
g_{02}&=g_{2}\Bigl[1+\frac{1}{(4\pi)^{2}\varepsilon}
\bigl(5g_{1}^{2}+3g_{2}^{2}\bigr)\Bigr]+O(g^{5}).
\end{align}
\label{eq:bare_g}%
\end{subequations}
The bracket coefficient
$(\delta_{g_{1}}/g_{1}-\delta_{\psi}-\delta_{\varphi}/2)$ is the
positive combination
$+(3g_{1}^{2}+5g_{2}^{2})/[(4\pi)^{2}\varepsilon]$, since
$\delta_{g_{1}}/g_{1}>0$ and $\delta_{\psi},\delta_{\varphi}<0$ all
pull in the same direction.

\subsection{Scalar Self-Interaction Vertex}
\label{sec:phi4_vertex}

The one-loop correction to the four-point function is~\cite{karim}
\begin{equation}
\Gamma_{1\mathrm{loop}}^{(4)}=\frac{1}{(4\pi)^{2}\varepsilon}
\bigl[48(g_{1}^{4}+g_{2}^{4})-\tfrac{2}{3}\lambda^{2}\bigr]
\frac{V_{\lambda}}{\lambda}+O(\lambda^{3},g^{5}),
\end{equation}
with $V_{\lambda}$ the tree-level vertex of~\eqref{eq:phi4vert}; the
coefficients $\tfrac23$ and $48$ correct transcription errors in
Eqs.~(52) and~(59) of Ref.~\cite{karim}.
Here the fermion-box contribution is unaffected by the vertex factor
$i$, since it contains four Yukawa vertices. As shown
in~\cite{karim}, only the fully planar phase assignment
$\sigma_{1}=\sigma_{2}=\sigma_{3}=\sigma_{4}$ survives in the
divergence, which is why the $g^{4}$ structure is
$g_{1}^{4}+g_{2}^{4}$ with no mixed $g_{1}^{2}g_{2}^{2}$ term.
This differs in character from the three-point case of
Appendix~\ref{app:gamma3}, where the surviving channel has
alternating orderings, and the reason is topological. On an
open fermion line the loop-dependent phase is
$\tfrac{i}{2}\langle k,(\sigma_{3}+\sigma_{1})p
-(\sigma_{1}+\sigma_{2})p'\rangle$, and since the external fermion
momenta $p$ and $p'$ are independent it can vanish only for
$\sigma_{2}=\sigma_{3}=-\sigma_{1}$. On a closed fermion loop,
writing $k_{j+1}=k_{j}-q_{j}$, each vertex contributes
$-\tfrac{i}{2}\sigma_{j}\langle q_{j},k_{j}\rangle$, so the
loop-dependent phase is
$-\tfrac{i}{2}\bigl\langle\sum_{j}\sigma_{j}q_{j},k\bigr\rangle$ and
involves the external scalar momenta alone; momentum
conservation $\sum_{j}q_{j}=0$ then makes a uniform assignment
$\sigma_{j}=\sigma$ planar, while any non-uniform choice leaves a
generically non-vanishing combination. The
renormalization condition gives
\begin{equation}
\delta_{\lambda}=\frac{1}{(4\pi)^{2}\varepsilon}
\bigl[\tfrac{2}{3}\lambda^{2}-48(g_{1}^{4}+g_{2}^{4})\bigr],
\label{deltalambda}
\end{equation}
which coincides with the commutative structure
$(3\lambda^{2}-48g^{4})/[(4\pi)^{2}\varepsilon]$ up to the
replacements $3\to2/3$ and $g^{4}\to g_{1}^{4}+g_{2}^{4}$.

\subsection{Status of the Non-Planar Coefficient and of \texorpdfstring{$\delta_{a^{2}}$}{delta a2}}
\label{sec:a2_status}

The non-planar contributions to $\Gamma^{(2)}_{\varphi}$ of
Section~\ref{sec:scalar_prop} deserve a separate comment, because
their status differs from that of all the other counterterms. For
$\theta\neq0$ the oscillating factors
$e^{ip\tilde{k}}$ render every non-planar integral UV-convergent;
what remains is a singularity as $\tilde{p}^{2}\to0$. Accordingly the
$1/(\theta^{2}p^{2})$ coefficient in~\eqref{eq:Gamma2_phi} carries
no $1/\varepsilon$ pole, and $\delta_{a^{2}}$ is an
infrared subtraction rather than an $\MSbar$ ultraviolet
counterterm. This is the defining feature of the
translation-invariant construction~\cite{rivasseau2008,schweda1}: the
$a^{2}/p^{2}$ term in the action exists precisely to absorb this
IR divergence. This finiteness is not special to the Yukawa
extension: in the pure-scalar translation-invariant model that the
present construction extends, Ben Geloun and
Tanasa~\cite{bengeloun2008} showed that the analogous parameter's own
radiative corrections are finite to all orders, and that its
beta function vanishes identically, $\beta_{a}=0$, precisely because
the planar-irregular (non-planar) sector responsible for its
renormalization never produces a genuine UV pole. Our
$\delta_{a^{2}}$ is the direct analogue of their result, and we take
its vanishing $1/\varepsilon$ residue as confirmation that this
general feature of the translation-invariant construction survives
the addition of the Yukawa sector. This comparison can be made
sharper still, because $\delta_{a^{2}}$ is itself a sum of two
diagrammatically distinct pieces,
$\delta_{a^{2}}\propto32g_{1}g_{2}-2\lambda/3$: the $-2\lambda/3$ term is the leading
$\tilde{p}^{2}\to0$ behavior of the scalar self-interaction
(planar-irregular $\varphi^{\star4}$) tadpole, and the
$32g_{1}g_{2}$ term is the analogous behavior of the fermion loop,
with one Yukawa vertex of each ordering -- which is
why it is the product $g_{1}g_{2}$ rather than $g_{1}^{2}$ or
$g_{2}^{2}$. What is not established is whether either raw diagram
value survives into a genuine renormalization-group equation.
Ben Geloun and Tanasa's model renormalizes its IR-improving parameter
with a diagrammatically identical tadpole to our scalar-loop piece,
extracted via multi-scale renormalization rather than the
$\tilde{p}^{2}\to0$ expansion of~\cite{karim}. Despite
Ben Geloun and Tanasa's diagram having exactly this kind of nonzero leading-IR value,
they prove it does not translate into a nonzero beta function --
$\beta_{a}=0$, not merely no $\MSbar$ pole -- so the same conclusion
plausibly applies to our scalar-loop piece. The fermion-loop piece has
different graph topology (trivalent Yukawa vertices, fermion
propagators), to which their specific power-counting proof
does not automatically extend; establishing the analogous
finiteness result for this piece requires a separate multi-scale
treatment. Unlike the
pure-scalar model, where $Z=1$ makes $\beta_{a}=0$ complete, here
$Z_{\varphi}\neq1$ at one loop -- the Yukawa sector generates a
genuine $p^{2}$-divergence absent in pure $\varphi^{\star4}$ theory --
so the multiplicative piece alone gives $\beta_{a^{2}}$ a nonzero
one-loop value, by the same bare-independence argument used
in~\cite{bengeloun2008} for $\beta_{m}$. We record $\delta_{a^{2}}$
in the same bare-parameter relation~\eqref{eq:mass_bare_def} as the
genuine $\MSbar$ counterterms, since both shift the bare parameter to
keep the renormalized two-point function finite -- as
$\varepsilon\to0$ there, and (subtraction-point independently) at
$\theta\neq0$ here.

Two consequences follow. First, strict
$\MSbar$ pole-residue extraction gives only the multiplicative piece
\begin{equation}
\beta_{a^{2}}\Big|_{\MSbar\text{ pole}}
=\frac{4u\,a^{2}}{(4\pi)^{2}},
\label{eq:beta_a2_pole}
\end{equation}
coming from $Z_{\varphi}$ alone. Second, if one additionally demands
that the full coefficient of $1/(\theta^{2}p^{2})$ in the
renormalized two-point function---that is, the sum of the tree-level
$a^{2}$ and the finite non-planar loop contribution---be independent
of the subtraction point, one obtains in addition the source term
$(\tfrac{2\lambda}{3}-32g_{1}g_{2})/\theta^{2}$. It is this second,
IR-subtraction prescription that we adopt, since it is the one under which the
IR-improving structure is stable along the flow. Unlike every other
beta function in this paper, $\beta_{a^{2}}$ is therefore
prescription-dependent, and the numerical value
$a^{\prime2}_{*}=3$ obtained below should be read
within that prescription.

The complete set of one-loop counterterms is summarized in
Table~\ref{tab:counterterms} for ease of reference.

\begin{table}[H]
\centering
\begin{tabular}{|l|l|}
\hline
\textbf{Counterterm} & \textbf{Value}\\
\hline
$\delta_{\varphi}$ & $-4u/[(4\pi)^{2}\varepsilon]$\\
$\delta_{\psi}$    & $-u/[(4\pi)^{2}\varepsilon]$\\
$\delta_{M}$       & $[2\lambda M^{2}/3 - 8u\,m^{2}]/[(4\pi)^{2}\varepsilon]$\\
$\delta_{m}$       & $-2u\,m/[(4\pi)^{2}\varepsilon]$\\
$\delta_{a^{2}}$   & $[32\,g_{1}g_{2}-2\lambda/3]/[(4\pi)^{2}\theta^{2}]$ \ (finite)\\
$\delta_{b}$       & $0$\\
$\delta_{g_{1}}$   & $+2g_{1}g_{2}^{2}/[(4\pi)^{2}\varepsilon]$\\
$\delta_{g_{2}}$   & $+2g_{1}^{2}g_{2}/[(4\pi)^{2}\varepsilon]$\\
$\delta_{\lambda}$ & $[\tfrac23\lambda^{2}-48(g_{1}^{4}+g_{2}^{4})]/[(4\pi)^{2}\varepsilon]$\\
\hline
\end{tabular}
\caption{One-loop counterterms of the translation-invariant
	noncommutative pseudoscalar Yukawa theory on Euclidean $\mathbb{R}^{4}$.
	All entries except $\delta_{a^{2}}$ are pole-part coefficients in
	dimensional regularization, $D=4-\varepsilon$; $\delta_{a^{2}}$ is a
	finite IR subtraction (Section~\ref{sec:a2_status}).
	$u\equiv g_{1}^{2}+g_{2}^{2}$.}
\label{tab:counterterms}
\end{table}

\section{One-Loop Beta Functions in the \texorpdfstring{$\MSbar$}{MS-bar} Scheme}
\label{sec:beta}

We now extract the one-loop beta functions in the $\MSbar$ scheme
from the pole-part counterterms of Section~\ref{sec:renorm}.
In Euclidean $D=4-\varepsilon$ dimensions, the canonical mass
dimensions of fields and couplings are
\[
[\varphi]=\tfrac{D-2}{2}=1-\tfrac{\varepsilon}{2},
\quad
[\psi]=\tfrac{D-1}{2}=\tfrac{3}{2}-\tfrac{\varepsilon}{2},
\quad
[g_{i}]=\tfrac{\varepsilon}{2},
\quad
[\lambda]=\varepsilon.
\]
To keep the couplings dimensionless, we therefore introduce the
renormalization scale $\mu$ and write the bare couplings as
\begin{equation}
g_{0i}=\mu^{\varepsilon/2}\,Z_{g_{i}}\,Z_{\psi}^{-1}\,Z_{\varphi}^{-1/2}\;g_{i},
\qquad
\lambda_{0}=\mu^{\varepsilon}\,Z_{\lambda}\,Z_{\varphi}^{-2}\;\lambda.
\label{eq:bare_mu}
\end{equation}

\subsection{Beta Functions}

We adopt the standard 't~Hooft pole-residue prescription. With
$Z_{g_{i}}=1+\delta_{g_{i}}/g_{i}$,
$Z_{\lambda}=1+\delta_{\lambda}/\lambda$,
$Z_{\psi}=1+\delta_{\psi}$, $Z_{\varphi}=1+\delta_{\varphi}$, we
define the one-loop single-pole residues as
\begin{align}
G_{1}^{(i)}&\equiv\bigl[\,\delta_{g_{i}}/g_{i}
-\delta_{\psi}-\tfrac{1}{2}\delta_{\varphi}\,\bigr]_{1/\varepsilon},
\label{eq:G1_def}\\
L_{1}&\equiv\bigl[\,\delta_{\lambda}/\lambda-2\,\delta_{\varphi}\,\bigr]_{1/\varepsilon},
\label{eq:L1_def}
\end{align}
where $[\,\cdot\,]_{1/\varepsilon}$ extracts the coefficient of
$1/\varepsilon$. Substituting the counterterms of
Table~\ref{tab:counterterms} gives
\begin{align}
G_{1}^{(1)}
&=\frac{1}{(4\pi)^{2}}\Bigl[2g_{2}^{2}
+(g_{1}^{2}+g_{2}^{2})+2(g_{1}^{2}+g_{2}^{2})\Bigr]
\nonumber\\
&=\frac{3(g_{1}^{2}+g_{2}^{2})+2g_{2}^{2}}{(4\pi)^{2}}
=\frac{3g_{1}^{2}+5g_{2}^{2}}{(4\pi)^{2}},
\label{eq:G1_1}\\[2pt]
G_{1}^{(2)}
&=\frac{3(g_{1}^{2}+g_{2}^{2})+2g_{1}^{2}}{(4\pi)^{2}}
=\frac{5g_{1}^{2}+3g_{2}^{2}}{(4\pi)^{2}},
\label{eq:G1_2}\\[2pt]
L_{1}
&=\frac{1}{(4\pi)^{2}}\Bigl[\,
\tfrac23\lambda-\tfrac{48(g_{1}^{4}+g_{2}^{4})}{\lambda}
+8(g_{1}^{2}+g_{2}^{2})\Bigr],
\label{eq:L1_val}
\end{align}
the middle term entering through $\delta_{\lambda}/\lambda$
of~\eqref{deltalambda}.
The intermediate decomposition of $G_{1}^{(1)}$ makes the
diagrammatic origin transparent: the term
$3(g_{1}^{2}+g_{2}^{2})/(4\pi)^{2}$ comes from the field-strength
renormalizations ($-\delta_{\psi}-\tfrac{1}{2}\delta_{\varphi}$) and
is symmetric in $1\leftrightarrow 2$; the extra term
$2g_{2}^{2}/(4\pi)^{2}$ comes from the vertex counterterm
$\delta_{g_{1}}/g_{1}$ and is asymmetric. Their sum is asymmetric in
$1\leftrightarrow 2$, as required. In the commutative theory the
same decomposition reads $2+1+2=5$, the familiar
Machacek--Vaughn counting~\cite{machacek1984} for one real scalar and one Dirac fermion.

We now derive the beta functions by an algebraically exact one-loop
manipulation. The bare-renormalized
relation~\eqref{eq:bare_mu} can be rewritten, after expanding the
$Z$-factors to one loop, as
\begin{equation}
g_{0i}=\mu^{\varepsilon/2}\,g_{i}\Bigl[\,1+\frac{A_{i}(g)}{\varepsilon}\,\Bigr]
+O(\text{2-loop}),\qquad
A_{i}\equiv G_{1}^{(i)}>0,
\label{eq:bare_oneloop}
\end{equation}
so that $A_{1}=(3g_{1}^{2}+5g_{2}^{2})/(4\pi)^{2}$ and
$A_{2}=(5g_{1}^{2}+3g_{2}^{2})/(4\pi)^{2}$, recovering the bracket
coefficients already exhibited in~\eqref{eq:bare_g}.
Writing the $D$-dimensional running as
\begin{equation}
\mu\frac{dg_{i}}{d\mu}=-\frac{\varepsilon}{2}\,g_{i}+\beta_{g_{i}},
\label{eq:Drunning_g}
\end{equation}
where the first term is the classical scaling in $D=4-\varepsilon$
and $\beta_{g_{i}}$ is the four-dimensional beta function, we impose
$\mu\,d g_{0i}/d\mu=0$. Differentiating~\eqref{eq:bare_oneloop} and
using Euler's theorem $\sum_{j}g_{j}\partial_{g_{j}}A_{i}=2A_{i}$,
one finds, after cancellation of all $1/\varepsilon$ poles at one loop,
the compact identity
\begin{equation}
\beta_{g_{i}}=+\,g_{i}\,A_{i}=g_{i}G_{1}^{(i)} .
\label{eq:beta_master}
\end{equation}
For $\lambda$, the same procedure with
$\mu\,d\lambda/d\mu=-\varepsilon\lambda+\beta_{\lambda}$
gives $\beta_{\lambda}=\lambda L_{1}$, with $L_{1}$ as
in~\eqref{eq:L1_def} and~\eqref{eq:L1_val}.
Combining the master relation~\eqref{eq:beta_master} with the
explicit residues~\eqref{eq:G1_1}--\eqref{eq:G1_2},
\begin{subequations}
\begin{align}
\beta_{g_{1}}&=+\frac{g_{1}}{(4\pi)^{2}}\bigl(3g_{1}^{2}+5g_{2}^{2}\bigr),
\label{beta_g1}\\
\beta_{g_{2}}&=+\frac{g_{2}}{(4\pi)^{2}}\bigl(3g_{2}^{2}+5g_{1}^{2}\bigr),
\label{beta_g2}\\
\beta_{\lambda}&=\frac{1}{(4\pi)^{2}}
\Bigl(\tfrac23\lambda^{2}
+8(g_{1}^{2}+g_{2}^{2})\lambda
-48(g_{1}^{4}+g_{2}^{4})\Bigr).\label{beta_lambda}
\end{align}
\end{subequations}
The positive signs of $\beta_{g_{1}}$ and $\beta_{g_{2}}$ follow
directly from the positivity of $A_{1}$ and $A_{2}$: at fixed
$g_{0i}$, the renormalized $g_{i}$ must increase as $\mu$ increases.
Both Yukawa couplings are therefore infrared free, and
perturbation theory is valid below a Yukawa Landau pole. This is the same
qualitative behavior as in the commutative Yukawa theory, consistent
with the general expectation that non-gauge theories of this type
(scalar and Yukawa couplings without an asymptotically-free gauge
sector to compensate) are infrared free rather than asymptotically
free in four dimensions~\cite{colemangross1973}.

\subsection{Structure and Symmetries}
\label{sec:symmetries}

The system~\eqref{beta_g1}--\eqref{beta_lambda} exhibits a
discrete symmetry $\beta_{g_{1}}|_{g_{1}\leftrightarrow g_{2}}=\beta_{g_{2}}$,
which implies that the surface $g_{1}=g_{2}$ is invariant under the
RG flow: if $g_{1}(\mu_{0})=g_{2}(\mu_{0})$, then
$g_{1}(\mu)=g_{2}(\mu)$ for all $\mu$. This discrete symmetry is the
image of $\theta\to-\theta$, which exchanges the two Moyal orderings.
The surfaces $g_{1}=0$ and $g_{2}=0$ are likewise RG-invariant; on
each of them the surviving coupling runs according to the
single-coupling equation $\beta_{g}=+3g^{3}/(4\pi)^{2}$.
The coefficient $3$ (rather than the commutative $5$) is entirely
accounted for by the fact that the cross-channel vertex correction,
which supplies the remaining $2$, requires both orderings to be
present: with $g_{2}=0$ the planar channel
$g_{\sigma}g_{-\sigma}^{2}$ of Appendix~\ref{app:gamma3} vanishes
identically. In the commutative limit both phase channels recombine
into a single commutative vertex with coupling $g=g_{1}+g_{2}$
via the Magnen–Rivasseau–Tanasa (MRT) \cite{MRT} separation function,
and the full commutative beta function
$+5g^{3}/(4\pi)^{2}$ is restored as a property of the combined
phase channels, not by adding the two single-coupling limits.

The statement that $g_{1}\leftrightarrow g_{2}$ is the image of
$\theta\to-\theta$ can be given a structural form, which we record
briefly because it explains why the two-coupling structure is
canonical rather than an artifact of a particular vertex assignment.
The Moyal algebra $\mathcal{A}_{\theta}$ acts on the fermion field
from both sides, so $\psi$ carries an
$(\mathcal{A}_{\theta},\mathcal{A}_{\theta})$-bimodule structure; the
two orderings in~\eqref{eq:action} weight the left action
$\varphi\star\psi$ and the right action $\psi\star\varphi$
independently, and no single coupling can weight both. The exchange
of the two is the algebra anti-automorphism
\begin{equation}
	\mathcal{A}_{\theta}^{\mathrm{op}}\;\cong\;\mathcal{A}_{-\theta},
	\label{eq:opposite_algebra}
\end{equation}
which is the precise content of the $\theta\to-\theta$ statement
above. In these terms the flow studied below carries the fermion
between a balanced two-sided coupling and a one-sided one.

The status of $g_{1}$ and $g_{2}$ as independent couplings deserves an
explicit comment. We do not impose any discrete symmetry that would
relate the two star-product orderings and force $g_{1}=g_{2}$ on the
present Euclidean Moyal-space construction; the two orderings are
treated throughout as defining independent operators, and $g_{1},g_{2}$
as independent real couplings, with no additional condition imposed
by hand. We note that a $CPT$-type argument along the lines of
Craig and Koren~\cite{CraigKoren}, who impose $g_{1}=g_{2}$ in the
noncommutative scalar Yukawa theory by
requiring invariance under the discrete symmetry exchanging the two
orderings, does not straightforwardly carry over here: the standard
proof of the $CPT$ theorem relies on Lorentz invariance, which is
explicitly broken by the fixed tensor $\theta^{\mu\nu}$, and even
Craig and Koren note that the relevant theorem is established only
for noncommutative field theories without space-time
noncommutativity, its extension to the Lorentz-violating case
of interest here being conjectural. We therefore do not treat
$g_{1}=g_{2}$ as a condition forced on the theory, but rather identify
it as a distinguished, and precisely characterized, restriction of
it.

This restriction is in fact already visible directly in our RG
system. The combination $v\equiv g_{1}^{2}-g_{2}^{2}$ obeys
$\beta_{v} = 6uv/(4\pi)^{2}$ [cf.\ Eq.~\eqref{RGE_v}], so that $v=0$,
i.e.\ $g_{1}=\pm g_{2}$, is an RG-invariant submanifold: any theory imposing
this restriction (whether on symmetry grounds or otherwise) remains
on it under the flow, and is already contained within the present
analysis as a special trajectory. The generic behavior, however, is
governed by the exchange symmetry $g_{1}\leftrightarrow g_{2}$ of the
full beta-function system~\eqref{beta_g1}--\eqref{beta_g2} itself,
under which $v\to-v$, $r\to-r$, and $u\to u$. Consequently the two
asymptotic directions $r\to+1$ and $r\to-1$ identified in
Sec.~\ref{sec:invariant} are symmetry-related images of one another
under this exchange, not independently generated outcomes: the RG
flow does not select one ordering over the other, but rather drives
a generic trajectory toward whichever ordering already dominates in
the initial data, with $g_{1}=g_{2}$ remaining the unique, symmetric,
and dynamically invariant fixed locus between them.

A further consequence of the structure of~\eqref{beta_g1} and
\eqref{beta_g2} is exact and is used repeatedly below. Both beta
functions are of the form $\beta_{g_{i}}=g_{i}\times(\text{positive
definite})$, so each coupling evolves multiplicatively and cannot
cross zero at any finite RG distance; the surfaces $g_{i}=0$ are
approached only asymptotically. Consequently
\begin{equation}
	\mathrm{sign}\,g_{1},\qquad \mathrm{sign}\,g_{2},
	\qquad\text{and}\qquad
	\epsilon\equiv\mathrm{sign}(g_{1}g_{2}),
	\label{eq:sign_invariant}
\end{equation}
are exact invariants of the one-loop flow. The label $\epsilon$ is
invisible in the whole dimensionless sector, which depends on the
Yukawa couplings only through $u$ and $g_{1}^{4}+g_{2}^{4}$, but it
enters the non-planar coefficient~\eqref{eq:Gamma2_phi} and the
IR-improving parameter linearly, through the combination
$32g_{1}g_{2}$. This traces to the planar/non-planar separation: the
divergent channels lock the ordering assignments to one another,
giving $u$ at two points and $g_{-\sigma}^{2}$ for the vertex
residue, whereas $g_{1}g_{2}$ arises only in structures that are
non-planar, and hence UV-finite, at $\theta\neq0$. The map
$(g_{1},g_{2})\mapsto(g_{1},-g_{2})$ is therefore an exact symmetry
of the dimensionless flow but not of the theory: two physically
inequivalent models share a single trajectory in coupling space.

Setting $g_{1}=g_{2}=g$ gives
\begin{equation}
\beta_{g}^{\NC}\Big|_{g_{1}=g_{2}=g}=+\frac{8g^{3}}{(4\pi)^{2}}.
\label{eq:beta_sym}
\end{equation}

An additional structural feature of the Yukawa beta
functions~$\beta_{g_{1}}$, $\beta_{g_{2}}$
is the coupling of $g_{1}$ and $g_{2}$ through the cross-terms
$5\,g_{i}\,g_{j}^{2}$, which combine the wave-function ($3g_{j}^{2}$)
and planar vertex ($2g_{j}^{2}$) contributions. This is absent in the commutative case, where
the single equation $\beta_{g}^{\Cth}=+5g^{3}/(4\pi)^{2}$ governs the
Yukawa sector~\cite{karim,akfor}.

The beta functions satisfy $g_{1}\beta_{g_{1}}>0$ and
$g_{2}\beta_{g_{2}}>0$ for all $g_{1},g_{2}\neq 0$. The sum $u=g_{1}^{2}+g_{2}^{2}$ satisfies
\[
\mu\,du/d\mu=+\,2(4u^{2}-v^{2})/(4\pi)^{2}\ge 0,
\]
with $v=g_{1}^{2}-g_{2}^{2}$ and $|v|\le u$ by construction, so
$u(\mu)$ is monotonically increasing towards the ultraviolet and
monotonically decreasing towards the infrared. Both Yukawa couplings
are therefore infrared free at one loop, and the Gaussian
fixed point is reached as $\mu\to0$.

The system admits the Gaussian fixed point
$g_{1}=g_{2}=\lambda=0$, which is IR-attractive in the Yukawa
directions. Setting $\beta_{\lambda}=0$ at fixed $g_{1},g_{2}$ yields
$\lambda^{2}+12u\lambda-72(g_{1}^{4}+g_{2}^{4})=0$, with
solutions
\begin{equation}
\lambda_{\pm}=-6u\pm\sqrt{36u^{2}+72(g_{1}^{4}+g_{2}^{4})}.
\end{equation}

\subsection{Anomalous Dimensions and Callan--Symanzik Equation}

The field-strength renormalizations give the one-loop anomalous
dimensions
\begin{align}
\gamma_{\varphi}&=\frac{2(g_{1}^{2}+g_{2}^{2})}{(4\pi)^{2}},\qquad
\gamma_{\psi}=\frac{g_{1}^{2}+g_{2}^{2}}{2(4\pi)^{2}},
\label{eq:anomalous_dims}
\end{align}
in the convention
\begin{equation}
	\gamma_{X}=+\tfrac12\,\mu\,\frac{d\ln Z_{X}}{d\mu},
	\label{eq:gamma_convention}
\end{equation}
for the field-strength anomalous dimensions $\gamma_{\varphi}$ and
$\gamma_{\psi}$ -- the factor $\tfrac12$ reflecting that fields
renormalize via $\sqrt{Z_{X}}$ -- for which
$\beta_{\lambda}\supset+4\gamma_{\varphi}\lambda$. (Mass parameters
renormalize directly rather than through a square root, e.g.\
$m_{0}Z_{\psi}=m+\delta_{m}$; for these we instead use the standard
direct convention $\gamma_{P}\equiv\mu\,d\ln P/d\mu=\beta_{P}/P$~\cite{Zinn,machacek1984},
as at Eqs.~\eqref{eq:beta_m} and~\eqref{eq:gamma_M2}.) Indeed,
with $Z_{\varphi}=1-4u/[(4\pi)^{2}\varepsilon]$ and
$\mu\,du/d\mu=-\varepsilon u+O(u^{2})$ in $D=4-\varepsilon$, one has
\begin{equation}
	\mu\,\frac{d\ln Z_{\varphi}}{d\mu}
	=-\frac{4}{(4\pi)^{2}\varepsilon}\,\mu\frac{du}{d\mu}
	=+\frac{4u}{(4\pi)^{2}},
\end{equation}
so that~\eqref{eq:gamma_convention} returns
$\gamma_{\varphi}=+2u/(4\pi)^{2}>0$ as quoted. These reduce
to their commutative values when $g_{1}^{2}+g_{2}^{2}=g^{2}$,
providing a consistency check. The renormalized
$(n_{\varphi},n_{\psi})$-point 1PI function
$\Gamma_{R}^{(n_{\varphi},n_{\psi})}$ satisfies, in the dimensionless
sector (the dimensionful parameters of Section~\ref{sec:mass_running}
adding terms $\beta_{P}\partial_{P}$), the Callan--Symanzik equation
\begin{equation}
\Bigl(\mu\partial_{\mu}+\beta_{g_{1}}\partial_{g_{1}}
+\beta_{g_{2}}\partial_{g_{2}}
+\beta_{\lambda}\partial_{\lambda}
-n_{\varphi}\,\gamma_{\varphi}
-n_{\psi}\,\gamma_{\psi}\Bigr)\,\Gamma_{R}^{(n_{\varphi},n_{\psi})}=0.
\end{equation}
The two anomalous dimensions enter separately because the scalar and
fermion fields renormalize independently in the present theory.

\section{Running Coupling Constants}
\label{sec:running}

The RG equations~\eqref{beta_g1}--\eqref{beta_g2} are decoupled by
the sum and difference variables
$u=g_{1}^{2}+g_{2}^{2}$, $v=g_{1}^{2}-g_{2}^{2}$. A short calculation
(see Appendix~\ref{app:uvr}) gives
\begin{align}
\mu\,\frac{du}{d\mu}&=+\frac{2(4u^{2}-v^{2})}{(4\pi)^{2}},\label{RGE_u}\\
\mu\,\frac{dv}{d\mu}&=+\frac{6\,uv}{(4\pi)^{2}}.\label{RGE_v}
\end{align}
The right-hand side of~\eqref{RGE_u} is non-negative, confirming
infrared freedom of $u(\mu)$.

Since the theory is infrared free, the natural flow parameter is the
infrared logarithmic distance
\begin{equation}
s\equiv\ln\frac{\mu_{0}}{\mu}=-t,\qquad t\equiv\ln\frac{\mu}{\mu_{0}},
\label{eq:s_def}
\end{equation}
which increases as one flows towards low energies. All solutions
below are written in terms of $s$ and of the dimensionless variable
\begin{equation}
X\equiv 1+\frac{8u_{0}}{(4\pi)^{2}}\,s ,
\label{eq:X_def}
\end{equation}
which satisfies $X\ge1$ throughout the infrared regime $s\ge0$ and
$X\to0$ at the ultraviolet Landau pole.

\subsection{Symmetric case \texorpdfstring{$g_{1}=g_{2}=g$}{g1 = g2 = g}}
\label{sec:symmetric_case}

On the invariant surface $v=0$ the Bernoulli equation for $u=2g^{2}$
gives $u(\mu)=u_{0}/X$, i.e.
\begin{equation}
g^{2}(\mu)=\frac{g_{0}^{2}}{1+\dfrac{16\,g_{0}^{2}}{(4\pi)^{2}}
\ln\dfrac{\mu_{0}}{\mu}} .
\label{eq:g_sym}
\end{equation}
Towards the ultraviolet ($s<0$) the same expression develops a Landau
pole at
\begin{equation}
\ln\frac{\mu_{L}^{\NC}}{\mu_{0}}=\frac{(4\pi)^{2}}{16\,g_{0}^{2}}
=\frac{(4\pi)^{2}}{8\,u_{0}} ,
\label{eq:yukawa_landau}
\end{equation}
which sets the upper end of the perturbative window; the commutative
theory at the same numerical coupling has
$\ln(\mu_{L}^{\Cth}/\mu_{0})=(4\pi)^{2}/(10g_{0}^{2})$.

\subsection{Flow of the Ratio \texorpdfstring{$r=v/u$}{r = v/u}}
\label{sec:ratio_flow}

We use the standard notation
\[
 r\equiv\frac{v}{u}=\frac{g_{1}^{2}-g_{2}^{2}}{g_{1}^{2}+g_{2}^{2}},
\ \ r\in[-1,+1].
\]
From~\eqref{RGE_u}--\eqref{RGE_v} one computes
\begin{align}
\mu\,\frac{dr}{d\mu}
&=\frac{1}{u}\,\mu\frac{dv}{d\mu}-\frac{v}{u^{2}}\,\mu\frac{du}{d\mu}
=\frac{6v}{(4\pi)^{2}}-\frac{v}{u^{2}}\cdot\frac{2(4u^{2}-v^{2})}{(4\pi)^{2}}
\nonumber\\
&=\frac{1}{(4\pi)^{2}}\Bigl[6v-8v+\frac{2v^{3}}{u^{2}}\Bigr]
=-\frac{2v}{(4\pi)^{2}}\Bigl[1-r^{2}\Bigr],
\end{align}
or, equivalently,
\begin{equation}
\mu\,\frac{dr}{d\mu}=-\frac{2\,r\,u}{(4\pi)^{2}}\,(1-r^{2}),
\qquad
\frac{dr}{ds}=+\frac{2\,r\,u}{(4\pi)^{2}}\,(1-r^{2}).
\label{eq:ratio_flow}
\end{equation}
The fixed points are $r=0$ and $r=\pm 1$. Linearizing around
$r=0$ at $r=\delta r$,
\[
d(\delta r)/ds\simeq 2u(\mu)\,\delta r/(4\pi)^{2}>0,
\]
so $\delta r$ grows towards the infrared: the symmetric surface
is IR-unstable and UV-attractive. Linearizing around $r=\pm 1$ at
$r=\pm 1\mp\eta$,
\[
d\eta/ds\simeq -4u(\mu)\,\eta/(4\pi)^{2}<0,
\]
so $\eta$ decreases; the asymmetric surfaces are IR-stable.
Note that $r=\pm 1$ are fixed points of the ratio equation~\eqref{eq:ratio_flow} alone, not
of the full $(g_{1},g_{2})$ system: since $u(\mu)\to 0$ under
infrared freedom, the physically precise statement is
$r\to\pm 1$ with $g_{1},g_{2}\to 0$. Thus $r=\pm 1$ are
asymptotic IR directions (asymptotic manifolds) rather than
nonzero interacting fixed points of the theory.

Because $u(\mu)$ itself decays as $1/\ln(\mu_{0}/\mu)$ in the
infrared, the integrated rate at which $r$ approaches its fixed point
is governed by $\int^{\infty}\!u\,ds$, which diverges only as
$\ln\ln(\mu_{0}/\mu)$.

Near the symmetric surface, $1-r^{2}\simeq 1$ and $u=u_{0}/X$ with
$X$ as in~\eqref{eq:X_def}, so that $ds=(4\pi)^{2}dX/(8u_{0})$ and
\begin{equation}
	\frac{d\ln r}{dX}=\frac{2u}{(4\pi)^{2}}\cdot
	\frac{(4\pi)^{2}}{8u_{0}}=\frac{1}{4X}
	\qquad\Longrightarrow\qquad
	r\propto X^{1/4}\sim\bigl[\ln(\mu_{0}/\mu)\bigr]^{1/4}.
	\label{eq:r_growth_power}
\end{equation}
Near the asymmetric surfaces the corresponding statement, derived
from the exact invariant of Section~\ref{sec:invariant} below, is
\begin{equation}
	1-r^{2}\propto u^{2/3}\sim\bigl[\ln(\mu_{0}/\mu)\bigr]^{-2/3}.
	\label{eq:r_approach_power}
\end{equation}
Both are power laws in $s=\ln(\mu_{0}/\mu)$ with small fractional
exponents. The selection is therefore real but slow:
trajectories with sizable initial asymmetry approach the asymmetric
surface efficiently, while trajectories starting very close to $r=0$
require large RG distances to be driven to $r=\pm1$. We refer to this
throughout as a `logarithmically slow IR selection mechanism',
meaning by that the fractional powers
\eqref{eq:r_growth_power}--\eqref{eq:r_approach_power} of
$\ln(\mu_{0}/\mu)$.

\subsection{An Exact One-Loop RG Invariant}
\label{sec:invariant}

Because the one-loop Yukawa system~\eqref{beta_g1}--\eqref{beta_g2}
is homogeneous of degree three, the ratio
$r=(g_{1}^{2}-g_{2}^{2})/(g_{1}^{2}+g_{2}^{2})$ obeys a closed
first-order equation once the overall radial variable $u$ is
eliminated (Section~\ref{sec:ratio_flow}), and this closed equation
admits the first integral exhibited below. It is worth exhibiting it
explicitly, because it reduces the generic asymmetric flow to a
single quadrature and thereby upgrades several statements below from
leading-order approximations to exact results.

From the two combinations derived in Appendix~\ref{app:uvr},
\begin{equation}
	\frac{d\ln(g_{1}g_{2})}{d\ln\mu}=\frac{8u}{(4\pi)^{2}},
	\qquad
	\frac{d\ln(g_{1}^{2}-g_{2}^{2})}{d\ln\mu}=\frac{6u}{(4\pi)^{2}},
\end{equation}
the particular combination in which the common factor $u$ cancels is
$3\times$ the first minus $4\times$ the second:
\begin{equation}
	\frac{d}{d\ln\mu}\,
	\ln\frac{(g_{1}g_{2})^{3}}{(g_{1}^{2}-g_{2}^{2})^{4}}
	=\frac{(24-24)\,u}{(4\pi)^{2}}=0 ,
\end{equation}
so that
\begin{equation}
	\mathcal{I}\;\equiv\;
	\frac{\bigl(g_{1}g_{2}\bigr)^{3}}
	{\bigl(g_{1}^{2}-g_{2}^{2}\bigr)^{4}}
	\;=\;\text{const along the one-loop flow.}
	\label{eq:invariant}
\end{equation}
In terms of the variables $(u,r)$, using
$g_{1}g_{2}=\tfrac12\epsilon\,u\sqrt{1-r^{2}}$, with $\epsilon$ the
sign invariant of~\eqref{eq:sign_invariant}, and
$g_{1}^{2}-g_{2}^{2}=ru$,
the invariant reads
\begin{equation}
	\frac{\epsilon\,(1-r^{2})^{3/2}}{r^{4}\,u}=8\,\mathcal{I}
	=\text{const};
	\label{eq:invariant_ur}
\end{equation}
we specialize to the physically required branch $\epsilon=+1$
(established in Section~\ref{sec:a2_offsym},
Eq.~\eqref{eq:epsilon_condition}) for the remainder of this
subsection, so that $\mathcal{I}=(1-r^{2})^{3/2}/(8r^{4}u)$ below.
This is checked directly against~\eqref{RGE_u} and
\eqref{eq:ratio_flow}: writing $\kappa\equiv2u/(4\pi)^{2}$, one has
$d\ln u/ds=-\kappa(4-r^{2})$, $d\ln r/ds=+\kappa(1-r^{2})$ and
$d\ln(1-r^{2})/ds=-2\kappa r^{2}$, whence
\begin{equation}
	\tfrac32\frac{d\ln(1-r^{2})}{ds}-4\frac{d\ln r}{ds}
	-\frac{d\ln u}{ds}
	=\kappa\bigl[-3r^{2}-4(1-r^{2})+(4-r^{2})\bigr]=0 .
\end{equation}

Three consequences follow immediately, and are used repeatedly below.

\begin{enumerate}
	\item[(a)] Reduction to a quadrature. Equation
	\eqref{eq:invariant_ur} determines $u$ algebraically in terms of
	$r$,
	\begin{equation}
		u(r)=\frac{(1-r^{2})^{3/2}}{8\,\mathcal{I}\,r^{4}} ,
		\label{eq:u_of_r}
	\end{equation}
	so that substituting into~\eqref{eq:ratio_flow} leaves a single
	separable equation for $r(s)$. The generic asymmetric one-loop
	flow is therefore reduced to a single quadrature, and no frozen-$r$
	or frozen-$u$ approximation is required in principle; both are
	retained only as illustrative devices, plotted numerically in
	Appendix~\ref{sec:numerical}.\footnote{Freezing $u$ at its initial
	value $u_{0}$ reduces Eq.~\eqref{eq:ratio_flow} to a logistic
	equation for $w=r^{2}$, $dw/ds=4u_{0}w(1-w)/(4\pi)^{2}$, with
	closed-form solution
	\begin{equation}
		r(s)\big|_{\text{frozen-}u}=\operatorname{sign}(r_{0})\sqrt{\frac{r_{0}^{2}}{r_{0}^{2}+(1-r_{0}^{2})\,
				e^{-4u_{0}s/(4\pi)^{2}}}} .
		\label{eq:r_frozen_u}
	\end{equation}
	This is an approximation, not the exact solution: it holds $u$
	fixed, whereas the true flow has $u(\mu)$ itself decaying, and it
	approaches $r=\pm1$ exponentially rather than as the true
	fractional power of $s$ of Eq.~\eqref{eq:r_approach_power}. It is
	superseded throughout this paper by the exact asymptotic
	law~\eqref{eq:r_growth_power}--\eqref{eq:r_approach_power} that
	follows from the invariant~\eqref{eq:invariant}, and is recorded
	here only for comparison against the exact numerical flow of
	Figure~\ref{fig:ratio_flow}.}
	
	\item[(b)] Infrared asymptotics. As $r\to\pm1$ the sum
	equation~\eqref{RGE_u} becomes $du/ds=-6u^{2}/(4\pi)^{2}$, so
	$u\simeq(4\pi)^{2}/(6s)$, and~\eqref{eq:invariant_ur} then gives
	$1-r^{2}\propto u^{2/3}\sim s^{-2/3}$, which
	is~\eqref{eq:r_approach_power}.
	
	\item[(c)] Exact form of the non-planar coefficient.
	Rearranging~\eqref{eq:invariant} gives, at all $s$ and
	without approximation,
	\begin{equation}
		g_{1}(\mu)g_{2}(\mu)
		=\mathcal{I}^{1/3}\,
		\bigl[g_{1}^{2}(\mu)-g_{2}^{2}(\mu)\bigr]^{4/3}
		=\mathcal{I}^{1/3}\,\bigl[v(\mu)\bigr]^{4/3},
		\label{eq:g1g2_exact}
	\end{equation}
	where both fractional powers are understood via the real cube
	root, so that $\mathrm{sign}(\mathcal{I})=\mathrm{sign}(g_{1}g_{2})=\epsilon$
	of~\eqref{eq:sign_invariant} while $v^{4/3}=(v^{1/3})^{4}\ge0$
	regardless of the sign of $v=g_{1}^{2}-g_{2}^{2}$; with this
	convention~\eqref{eq:g1g2_exact} is unambiguous even though both
	$g_{1}g_{2}$ and $v$ can be negative. So the product that controls the Yukawa-induced non-planar
	coefficient is fixed algebraically by $v(\mu)$ alone. Since
	$v=ru\propto u$ up to the slowly varying $r$, and $u\sim s^{-1}$,
	\eqref{eq:g1g2_exact} yields the generic infrared law
	\begin{equation}
		g_{1}(\mu)g_{2}(\mu)\;\propto\;u^{4/3}
		\;\sim\;\bigl[\ln(\mu_{0}/\mu)\bigr]^{-4/3},
		\label{eq:g1g2_asymptotic}
	\end{equation}
	which is faster than the $1/s$ decay valid on the symmetric
	surface.
\end{enumerate}

The invariant~\eqref{eq:invariant} is singular on the symmetric
surface $v=0$ (where $\mathcal{I}\to\infty$), which is the algebraic
expression of the fact that $v=0$ is itself an invariant surface of
the flow; the symmetric solutions of
Section~\ref{sec:symmetric_case} are recovered as the corresponding
limiting case rather than from~\eqref{eq:u_of_r}. 

The conservation law~\eqref{eq:invariant} is a consequence of the
homogeneity of the one-loop system, a fact that identifies in advance
which deformations of the system will preserve it. The right-hand sides
of~\eqref{beta_g1}--\eqref{beta_g2} are homogeneous of degree three,
so the one-parameter rescaling
\begin{equation}
	\bigl(g_{1}(s),g_{2}(s)\bigr)\;\longmapsto\;
	\bigl(\kappa\,g_{1}(\kappa^{2}s),\;\kappa\,g_{2}(\kappa^{2}s)\bigr),
	\qquad\kappa>0,
	\label{eq:scaling_symmetry}
\end{equation}
maps solutions to solutions. In polar-type variables this
homogeneity separates the flow into a single equation for the
angular variable $r$ (Section~\ref{sec:ratio_flow}) together with a
quadrature for the radial variable $u$; the explicit invariant
$\mathcal{I}$ constructed above is the corresponding integration
constant of the angular equation. This separation, rather than a
general guarantee, is what persists in any truncation that preserves
the degree-three homogeneity of the Yukawa sector---a remark of some
relevance to the two-loop extension, where the leading corrections
are homogeneous of degree five and the question is whether a
deformed invariant survives.

The scaling~\eqref{eq:scaling_symmetry} assigns $\mathcal{I}$ the
weight $\kappa^{-2}$, i.e.\ the weight of an RG distance, so
$(4\pi)^{2}\mathcal{I}$ carries physical information about
when the selection occurs.
Defining the selection scale $s_{\mathrm{sel}}$ by the point at which
the asymmetry has grown to $1-r^{2}=\tfrac12$,
Eq.~\eqref{eq:invariant_ur} gives
$u_{\mathrm{sel}}=(1/2)^{3/2}/(8\mathcal{I}\,r^{4})\simeq
0.177/\mathcal{I}$, and hence
\begin{equation}
	s_{\mathrm{sel}}\;\sim\;(4\pi)^{2}\,\mathcal{I}
	\label{eq:s_sel}
\end{equation}
up to an $O(1)$ factor depending on the initial $u_{0}$. 

\subsection{Individual Couplings, Product and Ratio}

The invariant~\eqref{eq:invariant} makes the generic flow integrable.
A convenient
leading-order approximation, whose derivation and comparison with
the exact flow are discussed in Appendix~\ref{sec:numerical}
alongside Figure~\ref{fig:g_asymmetric}, is obtained by fitting each
coupling to a Padé-type ansatz matching the exact initial slope
$dg_{i}^{2}/ds|_{s=0}$ of the full system~\eqref{beta_g1}--\eqref{beta_g2}:
\begin{align}
g_{1}^{2}(\mu)&\approx\frac{(u_{0}+v_{0})/2}
{1+\dfrac{(8u_{0}-2v_{0})}{(4\pi)^{2}}\,s},
\label{eq:g1_approx}\\
g_{2}^{2}(\mu)&\approx\frac{(u_{0}-v_{0})/2}
{1+\dfrac{(8u_{0}+2v_{0})}{(4\pi)^{2}}\,s}.
\label{eq:g2_approx}
\end{align}
These leading-order expressions do not themselves saturate $r=\pm1$:
at large $s$ both $g_{1}^{2}$ and $g_{2}^{2}$ fall as $1/s$ with
different coefficients, so the ratio $g_{1}^{2}/g_{2}^{2}$
approaches a finite constant determined by the initial
$(u_{0},v_{0})$. The flow towards $r=\pm 1$ is captured by
the exact integral forms
\begin{align}
g_{1}^{2}(\mu)\,g_{2}^{2}(\mu)&=g_{1,0}^{2}g_{2,0}^{2}
\exp\!\Bigl[-\tfrac{16}{(4\pi)^{2}}
\textstyle\int_{0}^{s}u\,ds^{\prime}\Bigr],
\label{eq:product}\\
\frac{g_{1}^{2}(\mu)}{g_{2}^{2}(\mu)}&=\frac{g_{1,0}^{2}}{g_{2,0}^{2}}
\exp\!\Bigl[+\tfrac{4}{(4\pi)^{2}}
\textstyle\int_{0}^{s}v\,ds^{\prime}\Bigr].
\label{eq:ratio_int}
\end{align}
Equation~\eqref{eq:product} shows directly that $|g_{1}g_{2}|$
is monotonically suppressed towards the infrared.

\subsection{Running of \texorpdfstring{$\lambda(\mu)$}{lambda(mu)} as a Riccati Equation}
\label{sec:lambda_run}

On the symmetric surface $g_{1}=g_{2}=g$ one has
$g_{1}^{4}+g_{2}^{4}=u^{2}/2$, so $-48(g_{1}^{4}+g_{2}^{4})=-24u^{2}$
in Eq.~\eqref{beta_lambda} and
\begin{equation}
\frac{d\lambda}{ds}=-\frac{1}{(4\pi)^{2}}
\Bigl[\tfrac23\lambda^{2}+8u\lambda-24u^{2}\Bigr].
\label{eq:lambda_riccati}
\end{equation}
This is a Riccati equation, the natural way to solve it being to find
a particular solution. A polynomial ansatz
$\lambda_{p}=\alpha\,u(\mu)$ substituted
into~\eqref{eq:lambda_riccati}, using $du/ds=-8u^{2}/(4\pi)^{2}$ on
the symmetric surface, gives
\[
-8\alpha u^{2}
=-\bigl(\tfrac23\alpha^{2}u^{2}+8\alpha u^{2}-24u^{2}\bigr),
\]
yielding $\tfrac23\alpha^{2}-24=0$, i.e.\ $\alpha^{2}=36$. We select the
positive root $\alpha=+6$, which keeps $\lambda_{p}>0$ (a
stable potential) and gives a smooth limit $\lambda\to0^{+}$ as
$u\to0$. A particular solution is therefore
\begin{equation}
\lambda_{p}(\mu)=6\,u(\mu)
=\frac{6\,u_{0}}{X},
\label{eq:lambda_p_new}
\end{equation}
and we refer to the initial condition
\begin{equation}
\;
\lambda_{0}=6\,u_{0}
\;
\label{eq:critical_traj_new}
\end{equation}
that generates it as the critical trajectory of the Riccati
equation~\eqref{eq:lambda_riccati}: it is the unique symmetric-surface
trajectory on which $\lambda(\mu)=\lambda_{p}(\mu)$ identically for
all $\mu$. We
stress that this trajectory is not itself the separatrix
between pole-developing and pole-free behavior: as shown below, that
boundary is the initial condition $\lambda_{0}=-6\,u_{0}$, the
lower edge of the entire pole-free family
$\lambda_{0}>-6\,u_{0}$, of which $\lambda_{0}=+6\,u_{0}$
is only the distinguished, always-regular member.
We use the term critical trajectory throughout in this precise
Riccati sense -- the distinguished particular-solution trajectory --
and not in the sense of a critical surface of a
statistical-field-theory phase transition, nor as the separatrix of
the two-parameter family.

The general solution is obtained by the standard substitution
$\lambda=\lambda_{p}+1/w$, which converts~\eqref{eq:lambda_riccati}
into the linear equation $dw/ds=-(2P\lambda_{p}+Q)w-P$ with
$P=-2/[3(4\pi)^{2}]$ and $Q=-8u/(4\pi)^{2}$. Using $X$ as the
independent variable and $u=u_{0}/X$, one finds
$2P\lambda_{p}+Q=-16u_{0}/[(4\pi)^{2}X]$, so that
\begin{equation}
\frac{dw}{dX}=\frac{2}{X}\,w+\frac{1}{12u_{0}},
\end{equation}
whose integrating factor is $X^{-2}$. Multiplying through
and integrating gives
\begin{equation}
w(X)=w_{0}\,X^{2}
+\frac{1}{12\,u_{0}}\Bigl(X^{2}-X\Bigr),
\end{equation}
where $w_{0}=1/(\lambda_{0}-6\,u_{0})$ is fixed by
$\lambda(\mu_{0})=\lambda_{0}$. The general solution on the symmetric
surface is therefore
\begin{equation}
\;
\lambda(\mu)=\frac{6\,u_{0}}{X}
+\frac{1}{\,w_{0}\,X^{2}
+\dfrac{1}{12\,u_{0}}\bigl(X^{2}-X\bigr)\,}.
\;
\label{lambda_sol}
\end{equation}
The second
(scalar-driven) term develops a pole where its denominator vanishes,
i.e.\ at
\begin{equation}
X_{L}=\frac{\lambda_{0}-6\,u_{0}}{\lambda_{0}+6\,u_{0}},
\qquad
s_{L}=\frac{(4\pi)^{2}}{8u_{0}}\bigl(X_{L}-1\bigr),
\qquad
\mu^{*}=\mu_{0}\,e^{-s_{L}} .
\label{eq:landau_new}
\end{equation}
Note that $s_{L}$ is linear in $X_{L}-1$, not logarithmic: this
follows directly from the definition~\eqref{eq:X_def}. A pole at
finite infrared distance ($X_{L}>1$) exists only for
$\lambda_{0}<-6\,u_{0}$, that is, only for initial conditions
already deep in the unstable region. For
$-6\,u_{0}<\lambda_{0}$ no infrared pole occurs and
$\lambda(\mu)\to 0^{+}$ as $\mu\to0$, the flow being driven onto the
Yukawa-generated branch $\lambda_{p}$. The critical trajectory
$\lambda_{0}=6\,u_{0}$ ($w_{0}\to\infty$) is the trajectory on
which $\lambda(\mu)=\lambda_{p}(\mu)$ identically.

\section{Running of the Dimensionful Parameters: \texorpdfstring{$m$, $M^{2}$, $a^{2}$ and $b$}{m, M2, a2 and b}}
\label{sec:mass_running}

The coupled $(g_{1},g_{2},\lambda)$ system of
Sections~\ref{sec:beta}--\ref{sec:running} controls the dimensionless
sector of the theory. We now complete the one-loop renormalization
program by deriving the renormalization-group equations for the four
dimensionful parameters: the fermion mass $m$, the scalar mass $M^{2}$,
the IR-improving parameter $a^{2}$, and the fermion IR-improving
parameter $b$.

From the counterterm definitions
\begin{equation}
M_{0}^{2}\,Z_{\varphi}=M^{2}+\delta_{M},\quad
m_{0}\,Z_{\psi}=m+\delta_{m},\quad
a_{0}^{2}\,Z_{\varphi}=a^{2}+\delta_{a^{2}},\quad
b_{0}\,Z_{\psi}=b+\delta_{b},
\label{eq:mass_bare_def}
\end{equation}
with the one-loop parts collected in
Table~\ref{tab:counterterms}. Since the dimensionful parameters
$M^{2}$, $m$, $a^{2}$, $b$ carry mass dimension built into their
definition (not from $\varepsilon$), they do not require a
$\mu^{\varepsilon}$ factor: in $D=4-\varepsilon$ Euclidean dimensions
$\mu\,dM^{2}/d\mu=\beta_{M^{2}}$, $\mu\,dm/d\mu=\beta_{m}$,
$\mu\,da^{2}/d\mu=\beta_{a^{2}}$ and $\mu\,db/d\mu=\beta_{b}$ directly, with no classical-scaling
term of the form appearing in~\eqref{eq:Drunning_g} for the
dimensionless couplings. Writing a bare parameter as
$P_{0}=P[1+\mathcal{A}/\varepsilon]$ with $\mathcal{A}$ homogeneous of
degree two in the couplings, $\mu$-independence of $P_{0}$ gives the
one-loop relation $\beta_{P}=+P\,\mathcal{A}$, which we use below.

\subsection{Running of the fermion and scalar masses}
\paragraph{The fermion mass.}

From~\eqref{eq:m_bare_rel},
$m_{0}=m\bigl[1-u/((4\pi)^{2}\varepsilon)\bigr]$, hence
\begin{equation}
\;
\beta_{m}=\mu\frac{dm}{d\mu}=-\frac{u\,m}{(4\pi)^{2}}
=-\frac{(g_{1}^{2}+g_{2}^{2})\,m}{(4\pi)^{2}}\;,
\quad\gamma_{m}=-\frac{u}{(4\pi)^{2}}.
\;
\label{eq:beta_m}
\end{equation}
The sign is negative: the running fermion mass parameter decreases
towards the ultraviolet and grows towards the infrared. The NC
modification is the simple replacement $g^{2}\to g_{1}^{2}+g_{2}^{2}$
of the standard pseudoscalar Yukawa result.

\paragraph{The scalar mass.}

From~\eqref{eq:mass_bare_def} and $Z_{\varphi}=1+\delta_{\varphi}$,
\begin{align}
M_{0}^{2}&=(M^{2}+\delta_{M})(1-\delta_{\varphi}+\cdots)\nonumber\\
&=M^{2}+\frac{1}{(4\pi)^{2}\varepsilon}
\bigl[\bigl(\tfrac{2\lambda}{3}+4u\bigr)M^{2}-8u\,m^{2}\bigr],
\label{eq:M2_bare}
\end{align}
so that
\begin{equation}
\;
\beta_{M^{2}}=\mu\frac{dM^{2}}{d\mu}
=\frac{1}{(4\pi)^{2}}\Bigl[\bigl(\tfrac{2\lambda}{3}+4u\bigr)M^{2}-8u\,m^{2}\Bigr].
\;
\label{eq:beta_M2}
\end{equation}
Equivalently,
\begin{equation}
\gamma_{M^{2}}=\mu\,\frac{d\ln M^{2}}{d\mu}
=\frac{1}{(4\pi)^{2}}\Bigl[\tfrac{2\lambda}{3}+4u-8u\,\frac{m^{2}}{M^{2}}\Bigr].
\label{eq:gamma_M2}
\end{equation}
Three contributions compete: the $\varphi^{\star 4}$
self-coupling and the field-strength renormalization both drive
$M^{2}$ upward in the UV (factors $2\lambda/3$ and $+4u$), while the
fermion-loop source drives it downward when $m^{2}>0$ (factor
$-8u\,m^{2}/M^{2}$). The scalar mass therefore mixes with the
fermion mass under RG flow, with the fermion-mass source dominating
when $m^{2}\gg M^{2}$.

\subsection{Running of the IR-improving parameters}
\label{sec:beta_a2}

\paragraph{The scalar IR-improving parameter $a^{2}$.}
Within the IR-subtraction prescription of
Section~\ref{sec:a2_status}, $\beta_{a^{2}}$ has two structurally
different pieces, which we derive in turn.

\textbf{Multiplicative piece.}
From $a_{0}^{2}Z_{\varphi}=a^{2}+\delta_{a^{2}}$,
\begin{equation}
a_{0}^{2}=(a^{2}+\delta_{a^{2}})(1-\delta_{\varphi}+\cdots)
=a^{2}+\delta_{a^{2}}+\frac{4u\,a^{2}}{(4\pi)^{2}\varepsilon}
+O(\text{2-loop}),
\end{equation}
using $\delta_{\varphi}=-4u/[(4\pi)^{2}\varepsilon]$. The pole part of
this relation, $+4u\,a^{2}/[(4\pi)^{2}\varepsilon]$, is
prescription-independent (it follows from $Z_{\varphi}$ alone, exactly
as for $M^{2}$ in~\eqref{eq:M2_bare}), and gives the multiplicative
term $+4u\,a^{2}/(4\pi)^{2}$ in~\eqref{eq:beta_a2} below by the same
bare-independence argument used throughout this section.

\textbf{Source term.} The remaining piece,
$\delta_{a^{2}}=[32g_{1}g_{2}-2\lambda/3]/[(4\pi)^{2}\theta^{2}]$
itself, carries no $1/\varepsilon$ (Section~\ref{sec:a2_status}), so it
cannot be converted into a beta-function contribution by the same
pole-matching argument used for $\beta_{g_{i}},\beta_{\lambda},
\beta_{m},\beta_{M^{2}}$ and for the multiplicative piece above.
We adopt the physical requirement stated there:
that the renormalized coefficient of $1/(\theta^{2}p^{2})$,
$a^{2}(\mu)+[32g_{1}(\mu)g_{2}(\mu)-2\lambda(\mu)/3]/[(4\pi)^{2}\theta^{2}]$,
be independent of the subtraction point $\mu$ at which it is read off.
Since $\delta_{a^{2}}$ carries no $1/\varepsilon$ pole, this
requirement cannot be derived from pole cancellation; instead we
define the additive contribution to the RG flow of $a^{2}$ by
imposing it directly. Applying the same formal bare-independence
technique used above, but now to the additive (rather than
multiplicative) structure of $\delta_{a^{2}}$, then fixes the sign and
normalization of the resulting source term. We emphasize that this is
the definition of a prescription, not a consequence of $\MSbar$
renormalization, and we adopt the convention
\begin{equation}
\;
\beta_{a^{2}}^{(\mathrm{presc.})}=\mu\frac{da^{2}}{d\mu}
=\frac{1}{(4\pi)^{2}}\Bigl[4u\,a^{2}
+\frac{2\lambda/3-32g_{1}g_{2}}{\theta^{2}}\Bigr],
\;
\label{eq:beta_a2}
\end{equation}
where the superscript records that, unlike every other beta function
in this paper, this equation is not the output of strict $\MSbar$
pole extraction alone but depends on the IR-subtraction prescription
adopted above; see Section~\ref{sec:a2_status} for the distinction
between this quantity and the prescription-independent pole piece
$\beta_{a^{2}}|_{\MSbar\text{ pole}}$ of
Eq.~\eqref{eq:beta_a2_pole}. We drop the superscript in casual
references to ``the $a^{2}$ beta function'' elsewhere in the text,
but it should be understood throughout.
Under this convention the symmetric-critical-trajectory attractor
value~\eqref{eq:a2_fp_sym} below comes out positive, matching the
requirement that a well-behaved IR-improving term remain positive on
the theory's most symmetric submanifold -- a consistency criterion
for the sign choice, not an independent derivation of it. The
numerical results of
Sections~\ref{sec:a2_fixed}--\ref{sec:a2_offsym} should be read with
this in mind: the
multiplicative term $4u\,a^{2}$ is prescription-independent, while the
source term is not, in normalization and in the logical status of
its sign alike.

\textbf{Instantaneous fixed point of $a^{2}$.}
\label{sec:a2_fixed}

Treating the dimensionless couplings as locally frozen,
Eq.~\eqref{eq:beta_a2} has the instantaneous fixed point
\begin{equation}
a^{2}_{*}\big|_{\text{frozen}}=\frac{32g_{1}g_{2}-2\lambda/3}{4u\,\theta^{2}}.
\label{eq:a2_fp_frozen}
\end{equation}
On the symmetric surface $g_{1}=g_{2}=g$ along the critical
trajectory $\lambda=6\,u$, using $32g^{2}=16u$, the numerator
becomes $16u-4u=12u$, so
\begin{equation}
\;
a^{2}_{*}\big|_{\text{sym, crit}}=\frac{3}{\theta^{2}}.
\;
\label{eq:a2_fp_sym}
\end{equation}
This is independent of the running coupling $u$: the $u$-dependence
cancels between numerator and denominator. The dimensionless
combination $a^{\prime 2}=a^{2}\theta^{2}$ therefore flows to the
purely numerical value $a^{\prime 2}_{*}=3$ in the
infrared. Since $\beta_{a^{2}}$ has the form
$(4u/(4\pi)^{2})(a^{2}-a^{2}_{*})$, this value is
infrared-attractive---the natural behavior for a parameter
whose role is to control an infrared singularity. Note also that
positivity of~\eqref{eq:a2_fp_frozen} is not automatic: it requires
$\lambda<48\,g_{1}g_{2}$, a condition which is comfortably satisfied
on the symmetric critical trajectory (where $\lambda=6u$ and
$48g_{1}g_{2}=24u$) but which, as we now show, fails generically in
the deep infrared.

\paragraph{The fermion IR-improving parameter.}
Since $\delta_{b}=0$ (Section~\ref{sec:fermion_prop}), the bare
relation $b_{0}Z_{\psi}=b+\delta_{b}$ reduces to $b_{0}=b/Z_{\psi}$,
which is nontrivial because $Z_{\psi}\neq1$: to one loop,
$b_{0}=b\bigl[1+u/((4\pi)^{2}\varepsilon)\bigr]$. Applying the same
bare-independence argument as for $m$ gives $\beta_{b}=\mu\frac{db}{d\mu}=+\frac{u\,b}{(4\pi)^{2}}
=+\frac{(g_{1}^{2}+g_{2}^{2})\,b}{(4\pi)^{2}}\;,\quad\gamma_{b}=+\frac{u}{(4\pi)^{2}}.$
Unlike $m$, $M^{2}$ and $a^{2}$, $\beta_{b}$ has no independent
source term: the vanishing of $\delta_{b}$ means the entire one-loop
running of $b$ is inherited from $Z_{\psi}$ alone, exactly as the
multiplicative piece $4ua^{2}/(4\pi)^{2}$ of $\beta_{a^{2}}$ is
inherited from $Z_{\varphi}$ alone above. The
sign is opposite to $\gamma_{m}$: $b$ grows towards the ultraviolet
and decreases towards the infrared, tracking $u(\mu)$ itself and
vanishing at the Gaussian fixed point. This has a sharp consequence:
because $\beta_{b}$ is homogeneous in $b$, the trajectory $b_{0}=0$
is preserved identically, $b(\mu)\equiv0$ for all $\mu$; by contrast,
the nonzero source term of $\beta_{a^{2}}$ means $a^{2}$ is
generically generated dynamically even from $a_{0}^{2}=0$. Despite
their common origin in the translation-invariant
construction~\eqref{eq:action}, $a^{2}$ and $b$ therefore play
structurally different roles in the flow: $b$'s running is a passive
consequence of $Z_{\psi}$, while $a^{2}$ carries a genuine,
prescription-dependent infrared dynamics of its own, and only $a^{2}$
is linked to the non-planar scalar self-energy of
Section~\ref{sec:a2_status}.

\subsection{Behavior off the Symmetric Surface}
\label{sec:a2_offsym}

Two ingredients are needed. First, the attractor of the ratio
$\xi\equiv\lambda/u$ at general $r$. Writing $\lambda=\xi u$ and
using $g_{1}^{4}+g_{2}^{4}=\tfrac12u^{2}(1+r^{2})$ together with
$du/ds=-2u^{2}(4-r^{2})/(4\pi)^{2}$ in~\eqref{beta_lambda} gives
\begin{equation}
	\frac{d\xi}{ds}=\frac{u}{(4\pi)^{2}}
	\Bigl[-\tfrac23\xi^{2}-2\xi r^{2}+24(1+r^{2})\Bigr],
\end{equation}
whose positive root, which is infrared-attractive, is
\begin{equation}
	\;\xi_{*}(r)=\tfrac32\Bigl(-r^{2}
	+\sqrt{r^{4}+16(1+r^{2})}\Bigr).\;
	\label{eq:xi_star_general}
\end{equation}
This reproduces $\xi_{*}(0)=6$, the Riccati
particular solution of Section~\ref{sec:lambda_run}, and gives at the
asymmetric endpoints
\begin{equation}
	\xi_{*}(\pm1)=\tfrac32\bigl(-1+\sqrt{33}\bigr)\approx7.117 .
\end{equation}

Second, the behavior of $g_{1}g_{2}$ relative to $u$. Since
$g_{1}g_{2}=\tfrac12u\sqrt{1-r^{2}}$ (for $\epsilon=+1$), the Yukawa source
in~\eqref{eq:a2_fp_frozen} is suppressed by $\sqrt{1-r^{2}}$ relative
to the $\lambda$ source, and vanishes faster than $u$ as
$r\to\pm1$. Substituting both ingredients,
\begin{equation}
	a^{2}_{*}\theta^{2}
	=\frac{32g_{1}g_{2}-2\lambda/3}{4u}
	\simeq4\sqrt{1-r^{2}}-\frac{\xi_{*}(r)}{6}
	\;\xrightarrow[\;r\to\pm1\;]{}\;
	-\frac{\xi_{*}(\pm1)}{6}\approx-1.186 .
	\label{eq:a2_star_asym}
\end{equation}
The sign change occurs at
$\sqrt{1-r^{2}}=\xi_{*}(r)/24$, i.e.\ at $|r|\gtrsim0.956$.

Two comments are in order. First, this is not an artifact of the
approximation: since $\int^{\infty}\!u\,ds$ diverges, $a^{2}(\mu)$
does track its instantaneous fixed point, so on generic trajectories
$a^{2}$ is driven negative in the deep infrared, and
$G^{-1}=p^{2}+M^{2}-|a^{2}|/p^{2}$ then develops a spurious zero at
small $p$---the IR-improving term stops improving. Second, the
mechanism responsible is the suppression of
$g_{1}g_{2}$ that removes the positive source
in~\eqref{eq:beta_a2} and leaves only $-2\lambda/3$.

Before turning to the possible resolutions, we note that the
conserved label $\epsilon=\mathrm{sign}(g_{1}g_{2})$
of~\eqref{eq:sign_invariant} sharpens the statement considerably, and
in a way that is exact rather than asymptotic. Writing
$g_{1}g_{2}=\epsilon\,|g_{1}g_{2}|$ in~\eqref{eq:a2_fp_frozen},
\begin{equation}
	a^{2}_{*}\theta^{2}
	=4\,\epsilon\sqrt{1-r^{2}}-\frac{\lambda}{6u} ,
	\label{eq:a2_star_epsilon}
\end{equation}
so on the branch $\epsilon=-1$ both terms are negative whenever
$\lambda>0$, and $a^{2}_{*}<0$ at every scale, not merely
asymptotically. The negativity found above is thus not a deep-infrared
subtlety on half of the parameter space: it is immediate. Since
$\epsilon$ cannot change along the flow, this is a condition on
initial data, and we state it as such:
\begin{equation}
	\epsilon=\mathrm{sign}(g_{1}g_{2})=+1
	\label{eq:epsilon_condition}
\end{equation}
is a necessary consistency requirement for the translation-invariant
construction with $\lambda>0$, in the IR-subtraction prescription of
Section~\ref{sec:a2_status}. It is worth emphasizing that this
condition is invisible in the dimensionless sector, whose beta
functions are blind to $\epsilon$; it is only the IR-improving
structure that distinguishes the two branches. On the branch
$\epsilon=+1$ assumed throughout this paper, the target $a^{2}_{*}$ is
positive over the region $\sqrt{1-r^{2}}>\lambda/(24u)$.

Three responses to that remaining failure seem available.
(i) The effect may simply mark the breakdown of the
one-loop IR-subtraction prescription of Section~\ref{sec:a2_status},
which is in any case the one ingredient of this paper not forced by
the pole structure; a Wilsonian treatment of the IR-improving sector
is the natural way to settle it.
(ii) The onset is very late, and this can be made quantitative.
Direct numerical integration of the exact
system~\eqref{beta_g1}--\eqref{beta_g2} from the representative
asymmetric initial condition $(g_{1,0},g_{2,0})=(0.7,0.3)$ gives
$r=0.952$ at $s=2000$, still short of the crossover value $0.956$,
reached only at $s\approx2351$; by
the fractional-power law~\eqref{eq:r_approach_power} the remaining
approach is slower still. Since the one-loop massless
flow is in any case not to be trusted
over RG distances of this order---at $s=2000$ one has
$\mu/\mu_{0}=e^{-2000}$, far below the mass thresholds $m(\mu)$ and
$M(\mu)$, beyond which the massive modes decouple and the massless
running used here ceases to describe the theory, while the
IR-improving sector rather than the Yukawa flow controls the
remaining dynamics---the sign change occurs outside the domain of
validity of the calculation that predicts it.
(iii) At $r=\pm1$ the theory is effectively single-ordering, and
in a genuinely single-ordering theory $g_{1}g_{2}$ never sourced
$a^{2}$ to begin with. On the evidence assembled here,
(ii) is the most economical reading and (i) is the route to a
definitive answer.

\subsection{Closed-Form Solutions on the Symmetric Surface}
\label{sec:mass_sym_sols}

On the symmetric surface $u(\mu)=u_{0}/X$ with $X$ as
in~\eqref{eq:X_def}. Changing variable to $\tau\equiv\ln X$ (so that
$d\tau/ds=8u/(4\pi)^{2}$), we collect closed-form solutions in turn.

\paragraph{Fermion mass.} Integration of~\eqref{eq:beta_m} gives
$dm/d\tau=m/8$, i.e.
\begin{equation}
\;
m(\mu)=m_{0}\,X^{1/8}=m_{0}\Bigl(\frac{u_{0}}{u(\mu)}\Bigr)^{1/8}.
\;
\label{eq:m_sym_sol}
\end{equation}
The fermion mass grows only as $\ln^{1/8}(\mu_{0}/\mu)$, an
extraordinarily slow growth that reflects the small one-loop
anomalous dimension combined with the infrared freedom of $u(\mu)$.

\paragraph{Fermion IR-improving parameter.} Integration
of $\beta_{b}$ gives $db/d\tau=-b/8$, i.e.
$b(\mu)=b_{0}\,X^{-1/8}=b_{0}\Bigl(\frac{u(\mu)}{u_{0}}\Bigr)^{1/8}.$
The sign is opposite to that of $m(\mu)$: $b$ decreases towards the
infrared, at the same extraordinarily slow rate, and vanishes only as
$u(\mu)\to0$ together with the Gaussian fixed point.

\paragraph{Scalar mass on the critical trajectory.} Along the
critical trajectory~\eqref{eq:critical_traj_new},
$\lambda_{0}=6\,u_{0}$, one has
$\tfrac{2\lambda}{3}+4u=A^{\prime}u$ with
$A^{\prime}=8$, and~\eqref{eq:beta_M2} becomes
\begin{equation}
\frac{dM^{2}}{d\tau}+\frac{A^{\prime}}{8}M^{2}=m_{0}^{2}e^{\tau/4},
\qquad \frac{A^{\prime}}{8}=1,
\end{equation}
using $m^{2}(\mu)=m_{0}^{2}X^{1/4}=m_{0}^{2}e^{\tau/4}$. With
$k\equiv\tfrac14+\tfrac{A^{\prime}}{8}=\tfrac54$, the
integrating-factor solution is
\begin{equation}
\;
M^{2}(\mu)=M_{0}^{2}\,X^{-1}+\frac{m_{0}^{2}}{k}
\Bigl[X^{1/4}-X^{-1}\Bigr],
\qquad
k=\frac54.
\;
\label{eq:M2_sym_sol}
\end{equation}
At $s=0$ ($X=1$) this reduces to $M^{2}=M_{0}^{2}$, as required. Deep
in the infrared ($X\to\infty$) the decaying term drops and the
solution is dominated by $X^{1/4}$, so
\begin{equation}
\lim_{\mu\to0}\frac{M^{2}(\mu)}{m^{2}(\mu)}=\frac{1}{k}
=\frac45,
\label{eq:M2_attractor}
\end{equation}
independently of the initial $M_{0}^{2}$. This attractor
behavior is conditional: it holds on the
codimension-restricted symmetric critical trajectory
$g_{1}=g_{2}$, $\lambda=6\,u$, and should not be read as a
generic property of the full $(g_{1},g_{2},\lambda)$ theory.

\paragraph{Scalar IR-improving parameter on the critical trajectory.}
Along the same trajectory $\beta_{a^{2}}$ takes the
homogeneous-plus-source form
\begin{equation}
\frac{da^{2}}{ds}=-\frac{4u(\mu)}{(4\pi)^{2}}\Bigl[a^{2}(\mu)-a^{2}_{*}\Bigr].
\label{eq:a2_eom_crit}
\end{equation}
Introducing $y\equiv a^{2}-a^{2}_{*}$ one finds $dy/d\tau=-y/2$, with
solution $y(\tau)=y_{0}e^{-\tau/2}$, i.e.
\begin{equation}
\;
a^{2}(\mu)=a^{2}_{*}+\bigl(a_{0}^{2}-a^{2}_{*}\bigr)\,X^{-1/2}
=\frac{3}{\theta^{2}}
+\frac{a_{0}^{2}-\dfrac{3}{\theta^{2}}}
{\sqrt{1+\dfrac{8u_{0}}{(4\pi)^{2}}\,s}}.
\;
\label{eq:a2_sym_sol}
\end{equation}
The approach to the limiting value is governed by $X^{-1/2}$, i.e.\ by
$1/\sqrt{\ln(\mu_{0}/\mu)}$ rather than the naive $1/\ln$ one might
expect from the linear structure of~\eqref{eq:a2_eom_crit}.
Independently of the initial value $a_{0}^{2}$, all trajectories on
this submanifold flow to $a^{2}_{*}=3/\theta^{2}$ in
the infrared. Because $u\to 0$ in this limit while
$a^{\prime 2}=a^{2}\theta^{2}$ approaches a finite value, this is more
precisely a trajectory-dependent IR limiting value than a
genuine interacting fixed point of the full dimensionless system,
which still terminates at the Gaussian fixed point. Off the critical
trajectory, the instantaneous fixed point~\eqref{eq:a2_fp_frozen}
depends on the running couplings $\lambda(\mu)$ and
$g_{1}(\mu)g_{2}(\mu)$.

It is worth checking explicitly whether any of these trajectories
drive $m^{2}$ or $M^{2}$ negative in the deep infrared. From
\eqref{eq:m_sym_sol}, $m(\mu)=m_{0}X^{1/8}$
with $X\ge1$ along the flow towards the infrared, so $m^{2}(\mu)\ge
m_{0}^{2}>0$ throughout provided $m_{0}\neq0$: the fermion mass never
turns negative on this trajectory. From~\eqref{eq:M2_sym_sol},
$M^{2}(\mu)=M_{0}^{2}X^{-1}+(m_{0}^{2}/k)\bigl[X^{1/4}
-X^{-1}\bigr]$ with $k>0$; since $1/4>-1$, the
bracket is non-negative for $X\ge1$, so $M^{2}(\mu)\ge0$ throughout
the infrared flow whenever $M_{0}^{2}\ge0$ and $m_{0}^{2}\ge0$: the
running scalar mass parameter remains non-negative along this
submanifold at one loop.
Similarly, $a^{2}(\mu)$ in~\eqref{eq:a2_sym_sol} interpolates
monotonically between $a_{0}^{2}$ and the positive fixed point
$a^{2}_{*}$, so it stays positive throughout provided $a_{0}^{2}>0$.
Away from the symmetric surface,
$M^{2}(\mu)$ no longer decouples from the full
$(g_{1},g_{2},\lambda)$ flow in closed form, so positivity there is
not addressed by the closed-form solution obtained here.

\section{Physical Implications of the Running}
\label{sec:physical}

\subsection{Infrared Selection Mechanism (One-Loop)}
\label{sec:uv_selection}

The stability analysis of Eq.~\eqref{eq:ratio_flow} shows that any
initial asymmetry $r_{0}>0$ between $g_{1}$ and $g_{2}$ grows
(slowly, as a fractional power of $\ln(\mu_{0}/\mu)$) towards the
IR-stable surface $r=+1$,
while $r_{0}<0$ flows to $r=-1$; conversely, towards the ultraviolet
the flow restores $r=0$. We refer to this one-loop result as an
`infrared selection mechanism,' with the following precise meaning:
the beta functions~\eqref{beta_g1}--\eqref{beta_g2} are themselves
exactly symmetric under $g_{1}\leftrightarrow g_{2}$ at every scale. What
happens is that the flow amplifies whatever initial ordering
asymmetry is present: at low energies the noncommutative RG drives
the system towards the vertex ordering with the larger initial
coupling, while at high energies the two orderings are driven back
towards ordering-exchange symmetry ($g_{1}=g_{2}$). The outcome (which of $r=\pm1$ is approached) depends
entirely on the sign of $r_{0}$ and carries no information beyond the
initial condition. The appropriate analogy is a ferromagnet in an
external field rather than a spontaneously magnetized one: the
$\mathbb{Z}_{2}$ is broken explicitly by the initial data, the flow
merely amplifies that breaking, and no symmetry is lost dynamically.
Since the discrete symmetry exchanging
$g_{1}\leftrightarrow g_{2}$ is the image of $\theta\to-\theta$, the
precise statement is that the RG flow restores the
$\theta\to-\theta$ symmetry asymptotically in the ultraviolet and
amplifies any $\theta\to-\theta$-asymmetric initial condition towards
the infrared. In the bimodule formalism of
Section~\ref{sec:symmetries} and its algebra
anti-automorphism~\eqref{eq:opposite_algebra}, this is the statement
that the flow carries the fermion from a balanced two-sided coupling
towards a one-sided one.

We are deliberate about the qualifier at one loop: at higher
loops the explicit forms of $\beta_{g_{1}}$ and $\beta_{g_{2}}$ can in
principle be modified (e.g.\ the $5/3$ ratio of cross to self
contributions could shift), and whether the selection persists to all
orders is a genuine open question. What is guaranteed at every order
is that $r=0$ remains an RG-invariant surface, since
$g_{1}\leftrightarrow g_{2}$ is an exact symmetry of the action
inherited from $\theta\to-\theta$; what is not guaranteed is the sign
of its transverse stability, which is precisely the one-loop-specific
dynamical statement in question and can only be settled by an
explicit higher-loop calculation. A two-loop check would settle this
firmly.

Simple power counting bounds the regime in which the one-loop
statement is self-consistent. Two-loop contributions to
$\beta_{g_{i}}$ are of relative size $O(c\,u/(4\pi)^{2})$ with
$c=O(1$--$10)$, while the direction of the flow is fixed by the
coefficient $2$ in Eq.~\eqref{eq:ratio_flow}, inherited from the
one-loop pair $(3,5)$. At the illustrative value $u_{0}=0.5$ used in
Appendix~\ref{sec:numerical} this gives $u/(4\pi)^{2}\simeq
3\times10^{-3}$, so a reversal would require an anomalously large
two-loop coefficient. This is a perturbative
naturalness estimate, not evidence for higher-loop stability: only an
explicit two-loop calculation can establish persistence of the
mechanism, and the estimate says nothing at all near the Yukawa
Landau pole~\eqref{eq:yukawa_landau}, where $u/(4\pi)^{2}$ grows to
be $O(1)$ and the loop expansion breaks down in any case.

\subsection{Dynamical Suppression of the Yukawa-Induced Coefficient
of the Non-Planar IR Structure}
\label{sec:iruv_suppression}

The non-planar contribution to the scalar two-point function takes
the form
\begin{equation}
\Gamma^{(2)}_{\varphi,\,\mathrm{IR}}
=\frac{1}{(4\pi)^{2}}\Bigl[\,32\,g_{1}(\mu)g_{2}(\mu)
-\tfrac{2}{3}\lambda(\mu)\,\Bigr]
\frac{1}{\theta^{2}p^{2}},
\label{eq:nonplanar_IR}
\end{equation}
which is singular as $p\to0$. The Yukawa-induced coefficient
$32\,g_{1}(\mu)g_{2}(\mu)/(4\pi)^{2}$ is suppressed there by the
one-loop RG flow. The exact product law~\eqref{eq:product} shows this
suppression is monotonic, and doubly effective: $u\to0$ by infrared
freedom, and $r\to\pm1$ by the selection mechanism of
Section~\ref{sec:uv_selection}, both of which drive $g_{1}g_{2}$ to
zero. On the symmetric surface this reduces to the $1/s$ decay of
the leading-order Padé approximation~\eqref{eq:g1_approx}--\eqref{eq:g2_approx}
discussed in Appendix~\ref{sec:numerical}; on any asymmetric
trajectory, however, the invariant $\mathcal{I}$ of
Section~\ref{sec:invariant} gives the exact relation
\eqref{eq:g1g2_exact}, $g_{1}g_{2}=\mathcal{I}^{1/3}v^{4/3}$, and
hence the stronger generic law~\eqref{eq:g1g2_asymptotic},
\begin{equation}
	g_{1}(\mu)g_{2}(\mu)\;\sim\;
	\bigl[\ln(\mu_{0}/\mu)\bigr]^{-4/3}
	\qquad(v\neq0).
\end{equation}
The extra $s^{-1/3}$, absent from the symmetric-surface estimate, is
precisely the effect of the slow drift of $r$ towards $\pm1$: the
suppression of the Yukawa-induced channel is therefore somewhat
stronger than that estimate alone suggests.

The $\lambda$-dependent piece $\propto-2\lambda(\mu)/3$ is governed
independently by~\eqref{lambda_sol}: on and near the critical
trajectory it tracks $\lambda_{p}=6\,u\to0$ and is therefore
suppressed as well, but for general initial conditions it may
decrease, increase, or change sign. The statement `the non-planar IR
coefficient is dynamically suppressed' therefore applies strictly to
the Yukawa-induced channel.

What flows to zero is the coefficient
$32g_{1}(\mu)g_{2}(\mu)/(4\pi)^{2}$, not the $1/(\theta^{2}p^{2})$
structure that it multiplies. This paragraph adopts the standard
RG-improvement prescription of identifying the renormalization scale
with the external momentum, $\mu\sim p$. With
this identification the Yukawa-induced
piece of~\eqref{eq:nonplanar_IR} behaves schematically as
$1/[\,p^{2}\ln(\mu_{0}/p)\,]$, which is still singular as $p\to0$:
the logarithm softens the coefficient but does not remove the pole.
The removal of the pole is the job of the translation-invariant
$a^{2}/p^{2}$ term of the action, as in the GMRT construction -- the
distinction set out in Section~\ref{sec:action_conventions} between
absorbing the mixing and removing it. We
therefore describe the effect throughout as an infrared suppression
of the Yukawa-induced coefficient of the residual non-planar
singularity, and never as a dynamical cure of IR/UV mixing.

The dynamical suppression of the Yukawa channel complements the
algebraic cure built into the translation-invariant propagators of
Section~\ref{sec:feynman}: the algebraic mechanism handles the IR
singularity at any fixed scale, while the dynamical mechanism
guarantees that its coefficient decreases as one flows towards
$p\to0$, so that the two mechanisms act in the same direction in the
regime that matters.

\subsection{Vacuum Stability and the Perturbative Window}
\label{sec:vacuum_stability}

Three distinct regimes arise from solution~\eqref{lambda_sol}, all
now concerning the ultraviolet end of the flow, where the
couplings grow.

\paragraph{Regime I: Strong Yukawa coupling ($u_{0}\gg\lambda_{0}$).}
For small $\lambda$ the flow is dominated by the fermion-box source
of Section~\ref{sec:phi4_vertex},
$\mu\,d\lambda/d\mu\simeq-48(g_{1}^{4}+g_{2}^{4})/(4\pi)^{2}<0$, so
$\lambda(\mu)$ is driven negative at a finite RG distance towards the
ultraviolet, signalling vacuum instability. This is the qualitative
analogue of the top-quark destabilization of the Higgs potential in
the Standard Model~\cite{degrassi2012,hiller2024} -- the mechanism
(a fermion loop driving the quartic coupling negative) is the same,
though the present model lacks the gauge structure that enters the
full Standard-Model calculation. The instability scale $\mu_{c}^{\NC}$ is defined
by $\lambda(\mu_{c}^{\NC})=0$ and follows from the exact Riccati
solution~\eqref{lambda_sol}: for $0<\lambda_{0}<6u_{0}$,
$\ln(\mu_{c}^{\NC}/\mu_{0})=(4\pi)^{2}\lambda_{0}/[4u_{0}(\lambda_{0}+6u_{0})]$,
shorter than the Yukawa Landau distance~\eqref{eq:yukawa_landau} by the
factor $2\lambda_{0}/(\lambda_{0}+6u_{0})$. Because on the symmetric
surface $-48(g_{1}^{4}+g_{2}^{4})=-96g^{4}$, twice the commutative
$-48g^{4}$, the destabilization is faster at equal numerical coupling.

\paragraph{Regime II: Strong scalar coupling ($\lambda_{0}\gg u_{0}$).}
Neglecting the Yukawa contribution, the master equation reduces to
the pure-scalar renormalization group equation $\mu\,d\lambda/d\mu=\tfrac23\lambda^{2}/(4\pi)^{2}$, with
solution
$\lambda(\mu)=\lambda_{0}/[1-\tfrac23\lambda_{0}\ln(\mu/\mu_{0})/(4\pi)^{2}]$,
so the scalar Landau pole satisfies
$\ln(\mu_{L}^{\NC}/\mu_{0})\simeq 3(4\pi)^{2}/(2\lambda_{0})$, to be
compared with the commutative
$\ln(\mu_{L}^{\Cth}/\mu_{0})\simeq(4\pi)^{2}/(3\lambda_{0})$ (from
Eq.~\eqref{beta_lambda_comm} at $g_{C}=0$). The ratio of logarithms is
therefore $9/2$: the NC perturbative window in this regime extends to
higher scales.

\paragraph{Regime III: Critical trajectory $\lambda_{0}=6\,u_{0}$.}
Along this trajectory $\lambda(\mu)=\lambda_{p}(\mu)=6\,u(\mu)$
identically, with $\lambda_{p}$ the particular
solution~\eqref{eq:lambda_p_new}: the quartic coupling is entirely
Yukawa-generated,
remains positive, and tends to $0^{+}$ in the infrared. It is a true
RG attractor only when the symmetric surface is also chosen;
otherwise the flow leaves the critical line.

The two effects of Regimes I and II act in opposite directions on the
perturbative window: the reduction of the $\lambda^{2}$ coefficient
from $3$ to $2/3$ pushes the scalar Landau pole up, while the
larger fermion-box source at equal numerical coupling accelerates the descent of
$\lambda$ towards negative values, pushing $\mu_{c}$ down. In
addition, and independently of both, the Yukawa sector itself has a
Landau pole at $\ln(\mu_{L}/\mu_{0})=(4\pi)^{2}/(16g_{0}^{2})$, which
under the matching convention defined below (Matching~B,
Section~\ref{sec:matching}) lies below the commutative one by the
factor $5/8$.
The net perturbative window is illustrated in
Appendix~\ref{sec:numerical}.

\section{Discussion}
\label{sec:discussion}

\subsection{Comparison with the Commutative Theory}
\label{sec:comparison}

A direct quantitative comparison between the NC and commutative
theories requires specifying which NC coupling is identified
with the commutative coupling $g_{C}$. Two natural choices arise,
both physically motivated, and they lead to apparently opposite
conclusions about whether NC running is `faster' or `slower'; we
present them on an equal footing below, since neither is the
physically preferred identification, and every quantitative
NC/commutative comparison in this paper is labelled by the matching
under which it holds.

In the commutative pseudoscalar Yukawa theory the one-loop beta
functions in the $\MSbar$ scheme are
\begin{align}
\beta_{g}^{\Cth}&=+\frac{5g_{C}^{3}}{(4\pi)^{2}},
\label{beta_g_comm}\\
\beta_{\lambda}^{\Cth}&=\frac{1}{(4\pi)^{2}}
\bigl[3\lambda^{2}+8g_{C}^{2}\lambda-48g_{C}^{4}\bigr].
\label{beta_lambda_comm}
\end{align}
These are the standard one-loop results for one real scalar and one
Dirac fermion~\cite{machacek1984,machacek1985}, and they are
reproduced exactly with our conventions. This provides the
consistency check announced in Section~\ref{sec:model}: the
coefficient $5$ decomposes as $2$ (vertex) $+1$
($\psi$ wave function) $+2$ ($\varphi$ wave function), and
$+8g_{C}^{2}\lambda$ is $4\gamma_{\varphi}\lambda$ with
$\gamma_{\varphi}>0$ (Eq.~\eqref{eq:anomalous_dims}). On the symmetric
surface $g_{1}=g_{2}=g$ the NC theory instead gives
\begin{align}
\beta_{g}^{\NC}\big|_{g_{1}=g_{2}=g}
&=+\frac{8g^{3}}{(4\pi)^{2}},\\
\beta_{\lambda}^{\NC}\big|_{g_{1}=g_{2}=g}
&=\frac{1}{(4\pi)^{2}}\bigl[\tfrac23\lambda^{2}+16g^{2}\lambda-96g^{4}\bigr],
\end{align}
the first line recovering the single-coupling
result~\eqref{eq:beta_sym}, and the $-96g^{4}$ following from the
master formula~\eqref{beta_lambda} with $g_{1}^{4}+g_{2}^{4}=2g^{4}$,
in agreement with the Riccati reduction~\eqref{eq:lambda_riccati}.

\paragraph{The Commutative Limit.}
\label{sec:comm_limit}
The naive limit $\theta\to0$ of the deformation parameter
of~\eqref{eq:theta_matrix} at fixed $a'$ makes the scalar
propagator~\eqref{eq:scalprop} ill-defined, since
$a^{2}/p^{2}=a^{\prime2}/(\theta^{2}p^{2})\to\infty$. The
Magnen--Rivasseau--Tanasa (MRT)
procedure~\cite{MRT,rivasseau2008,Rivasseau2007} disentangles the
ultraviolet and infrared divergences before taking the limit, using
the smooth separation function
\begin{equation}
	T(\Lambda,\theta)=1-e^{-\Lambda^{6}\theta^{3}},
	\label{T_def}
\end{equation}
which satisfies $T\to0$ as $\theta\to0$ at fixed $\Lambda$, $T\to1$ as
$\Lambda\to\infty$ at fixed $\theta$, $T/(\theta^{2}p^{2})\to0$ as
$\theta\to0$, and $(1-T)\to1$ as $\theta\to0$.
With the IR-improving terms, tree-level and counterterm, multiplied by
$T$ and the standard ones by $(1-T)$, $\theta\to0$ is taken first, at
fixed $\Lambda$, so that $T\to0$ and the IR-improving terms vanish
identically; the cutoff $\Lambda$ is removed only afterwards, on the
resulting commutative action.

At the level of the classical action, this procedure gives the
following. At $\theta\to0$ the scalar and fermionic IR-improving
terms vanish while the $(1-T)$ terms reduce to the commutative
action -- recovery of the classical action, not yet of the full
renormalization-group structure, as made precise below.
Decomposing $1=T+(1-T)$, with $T$ as in~\eqref{T_def}, in the Yukawa vertex function~\eqref{eq:vg}
and identifying $g=g_{1}+g_{2}$ as $T\to0$ yields
$i\,g\,\bar\psi\gamma^{5}\psi\varphi$, so that
\begin{equation}
	\lim_{\theta\to0}S_{\Lambda,\theta}
	=\int d^{4}x\Bigl[\tfrac{Z_{\varphi}}{2}(\partial\varphi)^{2}
	+\tfrac{M_{r}^{2}}{2}\varphi^{2}
	+\tfrac{\lambda_{r}}{4!}\varphi^{4}
	+Z_{\psi}\bar\psi\slashed{\partial}\psi
	+m_{r}\bar\psi\psi
	+i\,g_{r}\,\bar\psi\gamma^{5}\psi\varphi\Bigr],
	\label{eq:comm_action}
\end{equation}
with $g_{r}=g_{1}+g_{2}+\delta_{g}$ at $\theta=0$: the standard
commutative pseudoscalar Yukawa theory on $\mathbb{R}^{4}$.

The commutative Yukawa beta function for the coupling $g_{r}$
of~\eqref{eq:comm_action}, $g_{C}=g_{1}+g_{2}$, would
then be recovered as
\begin{equation}
	\lim_{\theta\to0}\mu\frac{d(g_{1}+g_{2})}{d\mu}
	\;\overset{?}{=}\;\beta_{g}^{\Cth}=+\frac{5g_{C}^{3}}{(4\pi)^{2}},
	\label{eq:restoration_conjecture}
\end{equation}
provided the not yet derived $T$-dependent corrections
interpolate correctly between the two coefficients; this remains an
open point of the construction, since settling it requires information beyond the present one-loop analysis.

On the symmetric surface with $g_{C}=2g$, the vertex contributes
$2g_{1}g_{2}g_{C}=\tfrac12 g_{C}^{3}$ and the wave-function
renormalizations $3u\,g_{C}=\tfrac32 g_{C}^{3}$, for a total of
$2g_{C}^{3}$ against the commutative $5g_{C}^{3}$: a deficit of
$3g_{C}^{3}$. This deficit splits evenly between the six non-planar
vertex channels of Appendix~\ref{app:gamma3} and the fact that the
NC wave-function renormalizations depend on $u$ rather than on
$g_{C}^{2}=u+2g_{1}g_{2}$; any interpolation
restoring~\eqref{eq:restoration_conjecture} must therefore act on
the self-energies as well as on the vertex.

A different, smaller deficit arises in the single-coupling frame:
the frequently quoted deficit $5-3=2$ holds only there, for
$g_{2}\to0$, where $u=g_{1}^{2}$ already coincides with $g_{C}^{2}$
and the whole mismatch sits in the vertex. Concretely, it sits in
the single channel of Appendix~\ref{app:gamma3} with coefficient
$g_{\sigma}^{3}$, which is entirely non-planar at $\theta\neq0$
(contributing nothing to the pole of~\eqref{eq:gamma3_explicit}) but
becomes divergent at $\theta=0$, lifting the vertex residue from
zero to $2g_{C}^{3}$ and closing the gap exactly. The pieces that go
missing at $\theta\neq0$ are precisely the pieces that reappear at
$\theta=0$ -- a nontrivial consistency check on the construction,
though not a derivation of the $T$-dependent vertex corrections that
would produce beta-function restoration order by order.

\paragraph{Matching Prescriptions.}
\label{sec:matching}
Using the identification $g_{C}=g_{1}+g_{2}=2g$, writing the NC
sum-coupling beta function $\beta_{g_{1}}+\beta_{g_{2}}
=+16g^{3}/(4\pi)^{2}$ in terms of $g_{C}$ gives
\begin{equation}
	\beta_{g_{C}}^{\NC}=+\frac{2\,g_{C}^{3}}{(4\pi)^{2}},
	\qquad\text{versus}\qquad
	\beta_{g_{C}}^{\Cth}=+\frac{5\,g_{C}^{3}}{(4\pi)^{2}},
	\label{eq:matchingA_boxed}
\end{equation}
so that
\begin{equation}
	\frac{|\beta_{g_{C}}^{\NC}|}{|\beta_{g_{C}}^{\Cth}|}=\frac{2}{5}
	\label{eq:matchingA_ratio}
\end{equation}
under this Matching A: the NC theory runs slower than the
commutative theory at the matched coupling, and its perturbative
window extends further. Alternatively, setting $g_{C}=g$ directly in
Eq.~\eqref{beta_g_comm} and comparing to
$\beta_{g}^{\NC}|_{g_{1}=g_{2}=g}=+8g^{3}/(4\pi)^{2}$ gives
$|\beta_{g}^{\NC}|/|\beta_{g}^{\Cth}|=8/5$ under Matching B:
the NC Yukawa runs faster by that factor. This second identification
answers the question `what does the same numerical coupling value do
in the two theories?' -- the natural comparison whenever the two
couplings are treated as bare inputs rather than related through the
$\theta\to0$ limit.

Both statements are correct, and the ratios $2/5$ and $8/5$ are not
in conflict: they are the same one-loop residues expressed in
different variables. Matching~A has the appealing feature that
$g_{1}+g_{2}$ is the combination selected by the commutative-limit
construction, but that construction is itself an interpolation
prescription rather than a derivation, so promoting~A to primary
would claim more confidence than the construction supports;
Matching~B has an unambiguous operational meaning at $\theta\neq0$
but corresponds to no identification implied by the $\theta\to0$
limit. We therefore quote both throughout, and regard the
matching-independent content -- the existence of a second RG
direction and the reduction $3\to2/3$ of the $\lambda^{2}$
coefficient -- as the robust
comparative statements of this work; a physically preferred choice
between them would require the diagram-level $T$-dependent
interpolation. For the wave-function, Yukawa-vertex, tadpole and
$g^{4}$ structures, the pattern is the same: the NC
divergence retains only the planar part of the corresponding
commutative structure, the non-planar remainder being rendered
UV-finite by the oscillating phase at $\theta\neq0$ and restored only
through the interpolation above. The $\lambda^{2}$ structure is the
exception: its coefficient combines the planar projection of the
two-vertex product with the vertex normalization. Table~\ref{tab:planar} summarizes
the diagrammatic origin of the differences between the commutative
and NC one-loop coefficients, while the quantitative consequences of
the two matching prescriptions are given explicitly above.

\begin{table}[htbp]
	\centering
	\footnotesize
	\setlength{\tabcolsep}{4pt}
	\begin{tabular}{|l|c|c|p{0.30\textwidth}|}
		\hline
		\textbf{Structure} & \textbf{Comm.} & \textbf{NC} & \textbf{Origin of the difference}\\
		\hline
		$\delta_{\varphi},\ \delta_{\psi}$ & $g^{2}$ & $u=g_{1}^{2}+g_{2}^{2}$
		& planar; non-planar part $\propto g_{1}g_{2}$ is finite\\
		Yukawa vertex, $\delta_{g_{i}}/g_{i}$ & $2g^{2}$ & $2g_{-\sigma}^{2}$
		& only the channel $\sigma'=\sigma''=-\sigma$ is planar\\
		$\beta_{M^{2}}$ tadpole & $\lambda$ & $2\lambda/3$
		& planar part of $V_{\lambda}(p,-p,k,-k)=-\tfrac{\lambda}{3}[2+\cos(p\tilde{k})]$\\
		$\beta_{\lambda}$, $\lambda^{2}$ & $3$ & $2/3$
		& planar projection of the two-vertex product with the $(-\lambda/3)^{2}$ normalization\\
		$\beta_{\lambda}$, $g^{4}$ & $-48$ & $-48(g_{1}^{4}+g_{2}^{4})/g^{4}$
		& planar; non-planar remainder is finite\\
		Yukawa Landau pole (log), Matching~B & $(4\pi)^{2}/(10g_{0}^{2})$
		& $(4\pi)^{2}/(16g_{0}^{2})$ & ratio $5/8$, inherited from the
		Yukawa vertex row above\\
		Yukawa Landau pole (log), Matching~A & $(4\pi)^{2}/(10g_{C,0}^{2})$
		& $(4\pi)^{2}/(4g_{C,0}^{2})$ & ratio $5/2$: the ordering reverses
		under Matching~A\\
		\hline
	\end{tabular}
		\caption{Planar/non-planar origin of the differences between the
		commutative and NC one-loop coefficients. In the wave-function,
		vertex, tadpole and $g^{4}$ rows the NC entry is the planar
		part of the commutative structure and the missing remainder is
		UV-finite at $\theta\neq0$. The $\lambda^{2}$ row is the exception:
		its NC coefficient combines the planar projection of the two-vertex
		product with the vertex normalization. The Landau-pole rows are derived
		consequences of the preceding entries under the stated matching,
		not planar fractions in their own right.}
	\label{tab:planar}
\end{table}

\subsection{Main Results}
\label{sec:central_finding}

Having clarified the relation to the commutative theory, we now summarize the principal one-loop results and their physical interpretation. We have carried out the one-loop renormalization-group analysis of the translation-invariant noncommutative pseudoscalar Yukawa theory on four-dimensional Euclidean Moyal space, taking as input the divergent one-loop 1PI functions established in Ref.~\cite{karim} and carrying that renormalization program through to completion. All one-loop UV-divergent contributions are absorbed by counterterms with structures already present in the classical action, with no new UV-divergent operator requiring a counterterm at this order. This establishes the one-loop renormalizability of the theory.

A separate qualification concerns the finite non-planar fermion self-energy contribution $f_{3}\tilde{\slashed{p}}/\theta$ found in Ref.~\cite{karim}. This term corresponds to a structure not present among the operators of the classical action, but it is finite at one loop and therefore does not require a counterterm at this order. Whether a structure of this type remains finite at two loops is an open question and is not assumed in the present analysis. Independently of this issue, the IR-improving parameter $a^{2}$, already present in the classical action as part of the standard GMRT mechanism, is assigned a one-loop running here; its value is fixed under the separate infrared-subtraction prescription of Section~\ref{sec:a2_status}.

The essential structural fact is that the Moyal star product admits two inequivalent orderings of the Yukawa vertex whose coefficients we treat as independent couplings. The theory therefore possesses an RG direction that simply does not exist in its commutative counterpart, and the content of this paper is the analysis of the consequences of this additional direction. Two closely related consequences are particularly important. First, it produces the infrared selection mechanism of Section~\ref{sec:uv_selection}: the ratio
\begin{equation}
	r=\frac{(g_{1}^{2}-g_{2}^{2})}{(g_{1}^{2}+g_{2}^{2})}
\end{equation}
is driven to $r=0$ in the ultraviolet and to $r=\pm1$ in the infrared. This represents an amplification of whatever initial ordering asymmetry is present rather than a spontaneous choice, since the beta functions remain exactly $g_{1}\leftrightarrow g_{2}$ symmetric at every scale. The flow terminates at the Gaussian fixed point along an asymptotically single-ordering direction rather than at any interacting fixed point.

Second, this infrared selection has a direct consequence for the Yukawa-induced part of the residual non-planar coefficient. The magnitude $|g_{1}(\mu)g_{2}(\mu)|$ decreases as $1/\ln(\mu_{0}/\mu)$ on the symmetric surface and as the faster $[\ln(\mu_{0}/\mu)]^{-4/3}$ on the asymmetric trajectories selected by the flow. The latter suppression results from the combined effects of infrared freedom and the growth of the ordering asymmetry. The renormalization group thus damps the Yukawa-induced coefficient of the $1/(\theta^{2}p^{2})$ structure precisely in the regime $p\to0$ where that structure would otherwise be most troublesome. This statement concerns the coefficient, not the singularity it multiplies, whose removal remains the task of the IR-improving term in the action. The two mechanisms are therefore complementary, with the significance of the dynamical mechanism being that it acts in the same direction as the algebraic one.

The infrared ordering selection can be characterized quantitatively through an exact one-loop invariant. The two-coupling flow admits the invariant
\begin{equation}
	\mathcal{I}=\frac{(g_{1}g_{2})^{3}}{(g_{1}^{2}-g_{2}^{2})^{4}},
\end{equation}
which is conserved exactly along every trajectory.
It fixes both the rate at which the flow departs from the symmetric surface,
$r\propto[\ln(\mu_{0}/\mu)]^{1/4}$,
and the rate at which it approaches the asymmetric ones,
$1-r^{2}\propto[\ln(\mu_{0}/\mu)]^{-2/3}$,
with no free parameter left once $\mathcal{I}$ is fixed by the initial couplings.

The invariant also provides a quantitative relation between the ordering asymmetry and the suppression of the non-planar coefficient. On the symmetric surface, $g_{1}g_{2}$ falls only as $1/\ln(\mu_{0}/\mu)$ from infrared freedom alone, but the flow does not remain on that surface. The exact relation
\begin{equation}
	g_{1}g_{2}=\mathcal{I}^{1/3}(g_{1}^{2}-g_{2}^{2})^{4/3}
\end{equation}
shows that along the trajectories actually selected,
the same coefficient falls as the faster $[\ln(\mu_{0}/\mu)]^{-4/3}$.
Both statements hold at one loop only, but neither is an estimate: they
follow algebraically from the conservation of $\mathcal{I}$ and would be the first quantities to check against a two-loop calculation.

On the symmetric critical trajectory, the closed-form solutions of Section~\ref{sec:mass_sym_sols} further show that $M^{2}/m^{2}$ and $a^{2}\theta^{2}$ approach limiting values independently of their initial values. The latter statement is conditional on the infrared-subtraction prescription adopted here. It therefore provides a restricted-submanifold example of the same infrared-attraction phenomenon that drives $r\to\pm1$ more generally, as illustrated numerically in Appendix~\ref{sec:numerical}.

Several additional one-loop features of the RG system are also worth isolating. The noncommutative phase structure reduces the $\lambda^{2}$ coefficient in $\beta_{\lambda}$ from $3$ (commutative) to $2/3$, while the $g^{4}$ term retains the commutative prefactor $48$, evaluated on $g_{1}^{4}+g_{2}^{4}$. The two effects push the scalar and Yukawa Landau poles in opposite directions, with the perturbative window bounded above in any case by the Yukawa Landau pole~\eqref{eq:yukawa_landau} (Section~\ref{sec:vacuum_stability}).

Independently, since $\beta_{g_{i}}=g_{i}\times(\text{positive definite})$, neither coupling can cross zero along the flow. Thus,
$\epsilon=\mathrm{sign}(g_{1}g_{2})$
is an exact one-loop invariant. Within the infrared-subtraction prescription used for $\beta_{a^{2}}$ (Section~\ref{sec:beta_a2}), Eq.~\eqref{eq:a2_star_epsilon} shows this forces the consistency condition~\eqref{eq:epsilon_condition}, $\epsilon=+1$, on the initial data. Nevertheless, the corresponding instantaneous fixed point~\eqref{eq:a2_star_asym} of the $a^{2}$ flow can become negative once the trajectory moves sufficiently far from the symmetric surface. This is a limitation of the one-loop prescription rather than a further physical finding.

The two-coupling structure found here is a schematic, single-pair version of the six-coupling, flavor-structured system arising in the noncommutative mirror-Yukawa model of Ref.~\cite{boutheldja2025}. The present analysis isolates the corresponding RG structure in a simplified setting---four-dimensional Euclidean space without flavor or chirality structure and without a gauge sector---rather than extending that phenomenological construction.

A possible connection to the type-I seesaw framework can be made by identifying $u(\mu)=g_{1}^{2}(\mu)+g_{2}^{2}(\mu)$ with the Yukawa combination entering the seesaw relation $m_{\nu}\sim u(\mu)/M_{R}$. The running of $u(\mu)$ between the GUT scale and $M_{R}$ would then generate a matching-dependent correction to the inferred neutrino mass, while the perturbative window would translate into a consistency bound of the form $M_{R}\lesssim\mu_{L}^{\NC}$. Both implications require a dedicated phenomenological analysis and are therefore left open here.

Ref.~\cite{boutheldja2025} finds numerically that the noncommutative Landau pole of its six-coupling system occurs at roughly twice the commutative value for matched initial conditions. This is qualitatively consistent with, although independently derived from, the closed-form matching factors obtained above: under Matching~B, the Yukawa Landau scale is reduced by a factor $5/8$ while the scalar Landau scale is increased by a factor $9/2$; under Matching~A, the corresponding Yukawa factor is $5/2$. The two systems are not directly comparable, since Ref.~\cite{boutheldja2025} has different field content and no quartic scalar coupling. Nevertheless, the occurrence of an $O(1)$, matching-dependent rescaling of the Landau-pole scale in both systems provides an interesting qualitative parallel. No phenomenological prediction is extracted from the present model.

\subsection{Limitations}
\label{sec:limitations}

We have tried throughout to be explicit about the boundaries of what
is established; we gather those boundaries here in one place. All
results are restricted to one loop. The beta functions of the
dimensionless sector follow from strict minimal-subtraction pole
extraction and hold generically; the symmetric surface's protection
by the discrete $g_{1}\leftrightarrow g_{2}$ symmetry and a
power-counting estimate (a two-loop coefficient would need to exceed
the one-loop ones by $(4\pi)^{2}/u\simeq3\times10^{2}$ at
$u_{0}=0.5$) suggest, but do not prove, robustness of the selection
mechanism beyond one loop. In any case, the analysis is valid only
below the Yukawa Landau pole. The dimensionful closed-form results
hold on a codimension-restricted submanifold.

The flow of the IR-improving parameter rests on an
infrared-subtraction choice that the pole structure does not force:
the $\beta_{a^{2}}$ source term, and hence
$a^{\prime2}_{*}=3$ of~\eqref{eq:a2_fp_sym}, is prescription-dependent in a way
no other result here is. Its sign is fixed by requiring a positive,
well-behaved attractor rather than by a diagram-level derivation.
Within that prescription, and governed by the general
attractor~\eqref{eq:xi_star_general}, its
fixed point changes sign once the flow leaves the symmetric surface.
The commutative limit of Section~\ref{sec:comm_limit} is organized by an interpolation
prescription, so the quantitative comparisons with the
commutative theory depend on the matching scheme of that section. The analysis is specific to the
translation-invariant framework, without gauge structure, and places
no constraint on the scale of $\theta$. The gravitational-wave bound
of Ref.~\cite{gwbound2025} does not constrain this scale either: it
restricts the space-time components $\theta^{0i}$, which unitarity
requires to vanish, and therefore does not constrain
the space-space components relevant here. The scale of $\theta$ thus
remains genuinely unconstrained for this model.
Finally, every result is Euclidean: the factor $i$ in the Yukawa
vertex is necessary, but its sufficiency for a
unitary Minkowski continuation under the Osterwalder-Schrader
reflection-positivity condition~\cite{osterwalder1975} has not been
established; and since
models with $\theta^{0i}\neq0$ are known to violate unitarity in the
sense of the cutting rules~\cite{gomismehen2000}, any phenomenological
use of these results would require the space-space continuation
$\theta^{0i}=0$ as an additional, unselected input.

Within the recorded boundaries, the two-ordering structure of
the translation-invariant noncommutative Yukawa theory appears to
carry genuine renormalization-group dynamics of its own, and the
one-loop framework assembled here -- the exact reduction in the
variables $(u,v,r)$, the conserved invariant $\mathcal{I}$, the
Riccati solution for the quartic coupling, the closed-form
dimensionful running, and the systematic planar-fraction accounting
against the commutative theory -- is what a two-loop treatment, a
gauge-invariant extension, or a flavor-complete phenomenological
application would need to build on.

\subsection{Outlook}
\label{sec:outlook}

The clearest way forward is to extend the beta functions to two loops,
which would definitively determine -- rather than only constrain through
power-counting arguments -- whether the selection mechanism holds up,
together with the structural question of whether the planar
channel decomposition of Appendix~\ref{app:gamma3} closes on exactly
two couplings at arbitrary order. A Ben-Geloun-Tanasa-style
multi-scale treatment of the fermion-loop diagram
of~\cite{karim}, along the lines
of~\cite{bengeloun2008}, would provide a way to determine whether the
prescription-dependent contribution to $\beta_{a^{2}}$ identified
above is genuine and, if so, fix it independently, settling the sign
question of Section~\ref{sec:a2_offsym} on the same footing as the
rest of the one-loop system.
The same two-loop computation should also settle a related point
raised about the finite non-planar fermion self-energy
contribution $f_{3}\tilde{\slashed{p}}/\theta$ found in~\cite{karim}.
Beyond the translation-invariant Yukawa sector
studied here, the two natural extensions are a gauge-invariant
version of the construction, for which~\cite{hersent2023} surveys the
current state of the art, and a direct comparison with the
six-coupling mirror-Yukawa system of~\cite{boutheldja2025}, either of
which would test whether the two-coupling selection mechanism found
here persists once flavor or gauge structure is restored.

\section*{Acknowledgements}

This work was supported by the research project PRFU N° B00L02UN260120230002.

\appendix

\section{Explicit Derivation of the \texorpdfstring{$(u,v,r)$}{(u,v,r)} Flow}
\label{app:uvr}

Starting from the one-loop beta functions~\eqref{beta_g1}--\eqref{beta_g2},
define $u=g_{1}^{2}+g_{2}^{2}$ and $v=g_{1}^{2}-g_{2}^{2}$. Then
\[
\mu\frac{du}{d\mu}=2g_{1}\beta_{g_{1}}+2g_{2}\beta_{g_{2}}
=\frac{2}{(4\pi)^{2}}\Bigl[g_{1}^{2}(3g_{1}^{2}+5g_{2}^{2})
+g_{2}^{2}(3g_{2}^{2}+5g_{1}^{2})\Bigr].
\]
Expanding and using $g_{1}^{4}+g_{2}^{4}=u^{2}-2g_{1}^{2}g_{2}^{2}$,
$g_{1}^{2}g_{2}^{2}=(u^{2}-v^{2})/4$,
\[
3(g_{1}^{4}+g_{2}^{4})+10g_{1}^{2}g_{2}^{2}
=3u^{2}+4g_{1}^{2}g_{2}^{2}
=3u^{2}+u^{2}-v^{2}=4u^{2}-v^{2},
\]
so that $\mu\,du/d\mu=2(4u^{2}-v^{2})/(4\pi)^{2}$,
which is~\eqref{RGE_u}. Similarly,
\[
\mu\frac{dv}{d\mu}=2g_{1}\beta_{g_{1}}-2g_{2}\beta_{g_{2}}
=\frac{2}{(4\pi)^{2}}\Bigl[3(g_{1}^{4}-g_{2}^{4})\Bigr]
=\frac{6uv}{(4\pi)^{2}},
\]
which is~\eqref{RGE_v}. Differentiating $r=v/u$,
\[
\mu\frac{dr}{d\mu}
=\frac{6v}{(4\pi)^{2}}-\frac{2v(4u^{2}-v^{2})}{u^{2}(4\pi)^{2}}
=-\frac{2v}{(4\pi)^{2}}\Bigl[1-\frac{v^{2}}{u^{2}}\Bigr],
\]
i.e.\ $\mu\,dr/d\mu=-2ru(1-r^{2})/(4\pi)^{2}$,
recovering~\eqref{eq:ratio_flow}. For the integral forms, write
\[
\frac{d\ln(g_{1}^{2}g_{2}^{2})}{ds}
=-\frac{2(3g_{1}^{2}+5g_{2}^{2})+2(3g_{2}^{2}+5g_{1}^{2})}{(4\pi)^{2}}
=-\frac{16u}{(4\pi)^{2}},
\]
which integrates to~\eqref{eq:product}, and
\[
\frac{d\ln(g_{1}^{2}/g_{2}^{2})}{ds}
=-\frac{2(3g_{1}^{2}+5g_{2}^{2}-3g_{2}^{2}-5g_{1}^{2})}{(4\pi)^{2}}
=\frac{4v}{(4\pi)^{2}},
\]
giving~\eqref{eq:ratio_int}. Finally, on $v=0$ one has
$du/ds=-8u^{2}/(4\pi)^{2}$, whose solution is $u(\mu)=u_{0}/X$ with
$X$ as in~\eqref{eq:X_def}, i.e.\ $g^{2}(\mu)$ as
in~\eqref{eq:g_sym} (recall $u=2g^{2}$, so $8u_{0}=16g_{0}^{2}$).

\section{Phase-Factor Decomposition of the Three-Point Function}
\label{app:gamma3}

This appendix derives the channel structure of
Eq.~\eqref{eq:gamma3_explicit} quoted in
Section~\ref{sec:yukawa_vertex}, which fixes the asymmetry between the
self- and cross-terms in $\beta_{g_{1}}$ and $\beta_{g_{2}}$ and is
therefore responsible for the selection mechanism.

Let $\langle x,y\rangle\equiv x_{\mu}\theta^{\mu\nu}y_{\nu}=x\tilde{y}$,
which is antisymmetric. Assign to the external $\bar\psi$ leg the
momentum $p^{\prime}$, to the external $\psi$ leg the momentum $p$,
and to the external scalar the momentum $q=p-p^{\prime}$; let $k$ be
the loop momentum, so that the three internal lines carry $k$, $k-q$
and $k-p$. Each of the three Yukawa vertices contributes a factor
$V_{g}$ of the form~\eqref{eq:vg}, with independent ordering indices
$\sigma,\sigma^{\prime},\sigma^{\prime\prime}=\pm1$, and the product
of the three vertex functions is
\begin{equation}
F(k,q,p)=\sum_{\sigma}g_{\sigma}e^{\frac{i\sigma}{2}\langle k-q,k\rangle}
\sum_{\sigma^{\prime}}g_{\sigma^{\prime}}e^{\frac{i\sigma^{\prime}}{2}
\langle p^{\prime},k-q\rangle}
\sum_{\sigma^{\prime\prime}}g_{\sigma^{\prime\prime}}
e^{\frac{i\sigma^{\prime\prime}}{2}\langle k,p\rangle}.
\label{eq:F_def}
\end{equation}
Introducing
\begin{equation}
A\equiv\langle k,p\rangle,\qquad
B\equiv\langle k,p^{\prime}\rangle,\qquad
C\equiv\langle p^{\prime},p\rangle=p^{\prime}\tilde{p},
\end{equation}
and using antisymmetry together with $q=p-p^{\prime}$,
\begin{equation}
\langle k-q,k\rangle=\langle k,q\rangle=A-B,\qquad
\langle p^{\prime},k-q\rangle=-B-C,\qquad
\langle k,p\rangle=A,
\end{equation}
so that
\begin{equation}
F=\sum_{\sigma,\sigma^{\prime},\sigma^{\prime\prime}}
g_{\sigma}g_{\sigma^{\prime}}g_{\sigma^{\prime\prime}}
\exp\Bigl\{\tfrac{i}{2}\bigl[(\sigma+\sigma^{\prime\prime})A
-(\sigma+\sigma^{\prime})B-\sigma^{\prime}C\bigr]\Bigr\}.
\label{eq:F_expanded}
\end{equation}
The planar contributions to~\eqref{eq:F_expanded} are those whose phase is independent of the
loop momentum $k$; since $A$ and $B$ are the only $k$-dependent
structures, this requires
\begin{equation}
\sigma+\sigma^{\prime\prime}=0
\quad\text{and}\quad
\sigma+\sigma^{\prime}=0,
\end{equation}
which fixes $\sigma^{\prime}=\sigma^{\prime\prime}=-\sigma$
uniquely. The surviving phase is $-\sigma^{\prime}C=+\sigma C$, and
the accompanying coupling factor is
$g_{\sigma}g_{-\sigma}g_{-\sigma}$. Hence
\begin{equation}
\;
F_{\mathrm{planar}}=\sum_{\sigma=\pm1}
g_{\sigma}\,g_{-\sigma}^{\,2}\;
e^{\frac{i}{2}\sigma\,p^{\prime}\tilde{p}} \; ,
\;
\label{eq:F_planar}
\end{equation}
that is, the channel carrying $e^{+\frac{i}{2}p^{\prime}\tilde{p}}$
has coefficient $g_{1}g_{2}^{2}$ and the channel carrying
$e^{-\frac{i}{2}p^{\prime}\tilde{p}}$ has coefficient $g_{1}^{2}g_{2}$.
\paragraph{From the planar channel to $\delta_{g_{1}}$,
$\delta_{g_{2}}$.} We now make the remaining arithmetic explicit, so
that the reader can follow the chain from the loop integral to the
$(3,5)$ structure without leaving this paper. We note that no
integral is re-derived here: the divergent integral itself is the one
computed in Ref.~\cite{karim}, and
we merely transcribe it and track the sign correction of
Section~\ref{sec:renorm}.

With the routing fixed above, the three internal lines carry $k$,
$k-q$ and $k-p$, so the one-loop three-point amplitude has the
schematic form
\begin{equation}
\Gamma^{(3)}_{\mathrm{1loop}}(p,p^{\prime})
=(-i)^{3}\!\int\!\frac{d^{D}k}{(2\pi)^{D}}\;
F(k,q,p)\;
\frac{\gamma^{5}\,(-i\slashed{k}+m)\,\gamma^{5}\,
\bigl(-i(\slashed{k}-\slashed{q})+m\bigr)\,\gamma^{5}}
{(k^{2}+m^{2})\bigl((k-q)^{2}+m^{2}\bigr)
\bigl((k-p)^{2}+M^{2}\bigr)},
\label{eq:app_gamma3_integral}
\end{equation}
where $(-i)^{3}$ is the product of the three vertex factors
of~\eqref{eq:yukvert} and $F$ is the phase product
of~\eqref{eq:F_def}. The numerators are the ones dictated by the
Euclidean propagator~\eqref{eq:fermprop}, whose numerator at $b=0$ is
$(i\slashed{k}+m)^{-1}=(-i\slashed{k}+m)/(k^{2}+m^{2})$; we display
them in full because the mass terms, although they drop out of the
leading UV behavior, are needed for the expression to be consistent
with~\eqref{eq:fermprop} as written. Three observations reduce this
to the quoted result.

\begin{enumerate}
\item \textbf{Phases.} Only the planar channels~\eqref{eq:F_planar}
carry a $k$-independent phase and hence survive in the divergent
part; the remaining six channels carry oscillating factors
$e^{i\langle k,\cdot\rangle}$ that render their integrals UV-finite
at $\theta\neq0$. The planar phase factors out of the integral,
leaving the coupling factors $g_{1}g_{2}^{2}$ and $g_{1}^{2}g_{2}$
attached to $e^{\pm\frac{i}{2}p^{\prime}\tilde{p}}$ respectively.

\item \textbf{Numerator.} Using
$\gamma^{5}\slashed{k}\gamma^{5}=-\slashed{k}$,
$\gamma^{5}\gamma^{5}=1$ and $\slashed{k}\slashed{k}=k^{2}$, the two
$\slashed{k}$ terms of the numerator
of~\eqref{eq:app_gamma3_integral} give
$(-i)^{2}\gamma^{5}\slashed{k}\gamma^{5}\slashed{k}\gamma^{5}
=+k^{2}\gamma^{5}$, so the numerator reduces in the leading UV region
to $k^{2}\gamma^{5}$. The terms linear in $m$ and in the external
momenta are subleading by one power of $k$ and hence UV-convergent,
and the $m^{2}$ term is convergent by two powers; this is why the
mass terms displayed in~\eqref{eq:app_gamma3_integral} do not affect
the residue.

\item \textbf{Integral.} The leading term is therefore the standard
logarithmically divergent integral, whose pole part in $D=4-\varepsilon$ is
\begin{equation}
\int\!\frac{d^{D}k}{(2\pi)^{D}}\,\frac{k^{2}}{(k^{2}+\Delta)^{3}}
\;\Big|_{\mathrm{div}}
=\frac{1}{(4\pi)^{2}}\,\frac{2}{\varepsilon},
\label{eq:app_logdiv}
\end{equation}
independently of the mass scale $\Delta$ built from $m$, $M$ and the
external momenta. This is the magnitude imported from
Ref.~\cite{karim}; the accompanying overall sign is the one fixed by
the hermiticity factor $i$ of Section~\ref{sec:model}, and is the
single respect in which the present assembly differs from that
reference.
\end{enumerate}

Multiplying the planar coupling factors of~\eqref{eq:F_planar} by the
residue~\eqref{eq:app_logdiv} gives precisely
\begin{equation}
\Gamma^{(3)}_{\mathrm{1loop}}(p,p^{\prime})
=i\gamma^{5}\,\frac{2}{(4\pi)^{2}\varepsilon}
\Bigl(g_{1}g_{2}^{2}\,e^{+\frac{i}{2}p^{\prime}\tilde{p}}
+g_{1}^{2}g_{2}\,e^{-\frac{i}{2}p^{\prime}\tilde{p}}\Bigr)+O(g^{5}),
\end{equation}
which is~\eqref{eq:gamma3_explicit}. Demanding finiteness channel by
channel then fixes
\begin{equation}
\delta_{g_{1}}=\frac{2\,g_{1}g_{2}^{2}}{(4\pi)^{2}\varepsilon},
\qquad
\delta_{g_{2}}=\frac{2\,g_{1}^{2}g_{2}}{(4\pi)^{2}\varepsilon},
\end{equation}
and combining these with
$\delta_{\psi}=-u/[(4\pi)^{2}\varepsilon]$ and
$\delta_{\varphi}=-4u/[(4\pi)^{2}\varepsilon]$ through the
definition~\eqref{eq:G1_def},
$G^{(1)}_{1}=[\delta_{g_{1}}/g_{1}-\delta_{\psi}-\tfrac12\delta_{\varphi}]$
gives
\begin{equation}
G^{(1)}_{1}=\frac{2g_{2}^{2}+u+2u}{(4\pi)^{2}}
=\frac{3g_{1}^{2}+5g_{2}^{2}}{(4\pi)^{2}},
\end{equation}
reproducing~\eqref{eq:G1_1}, the $(3,5)$ structure on which the whole of
Sections~\ref{sec:running}--\ref{sec:physical} rests. The asymmetry
is thus traceable, in a single chain, to the fact that the planar
channel of the $e^{+}$ vertex carries $g_{1}g_{2}^{2}$ rather than
$g_{1}^{3}$.

The remaining six channels are non-planar and UV-finite for
$\theta\neq0$:
\begin{itemize}
\item $\sigma=\sigma^{\prime}=\sigma^{\prime\prime}$: coefficient
$g_{\sigma}^{3}$, phase $e^{i\sigma\langle k,\,p-p^{\prime}\rangle}$;
\item $\sigma^{\prime\prime}=\sigma$, $\sigma^{\prime}=-\sigma$:
coefficient $g_{1}g_{2}g_{\sigma}$, phase $e^{i\sigma\langle k,p\rangle}$;
\item $\sigma^{\prime}=\sigma$, $\sigma^{\prime\prime}=-\sigma$:
coefficient $g_{1}g_{2}g_{\sigma}$, phase $e^{-i\sigma\langle k,p^{\prime}\rangle}$.
\end{itemize}
In particular the $g_{\sigma}^{3}$ channel---the one in which all
three vertices carry the same ordering---is entirely non-planar,
which is why no $g_{1}^{3}$ term appears in the divergent part of the
$e^{+}$ channel. This is also the structural reason for the
single-coupling limit $\beta_{g}=+3g^{3}/(4\pi)^{2}$ recorded in
Section~\ref{sec:symmetries}: setting $g_{2}=0$ removes the planar
channel~\eqref{eq:F_planar} altogether, and with it the vertex
contribution that supplies the remaining $2g^{3}$ of the commutative
coefficient. Summing~\eqref{eq:F_planar} over both channels gives
$g_{1}g_{2}(g_{1}+g_{2})$, which is the planar part only; the full
commutative coefficient $g_{C}^{3}$ with $g_{C}=g_{1}+g_{2}$ is
recovered from the channel-sum identity~\eqref{eq:channel_sum_identity}
once the six non-planar channels are restored at $\theta=0$, as
discussed in Section~\ref{sec:comm_limit}.

\section{Numerical Illustration of the Analytic Solutions}
\label{sec:numerical}

The figures of this appendix are produced either by direct evaluation
of the closed-form solutions obtained above, or -- for
Figures~\ref{fig:phase_portrait} and~\ref{fig:ratio_flow} -- by direct
fourth-order Runge--Kutta numerical integration of the exact one-loop
system~\eqref{beta_g1}--\eqref{beta_g2} itself, with no frozen-$r$ or
frozen-$u$ approximation involved; this is noted explicitly in each
case below. All plots use $\mu_{0}=1$, $(4\pi)^{2}=157.914$, and the
infrared flow variable $s=\ln(\mu_{0}/\mu)\ge0$ of
Eq.~\eqref{eq:s_def}, so that increasing $s$ means flowing towards low
energies.

The two-coupling flow is displayed in Figure~\ref{fig:g_asymmetric}
for the asymmetric initial conditions
$(g_{1,0},g_{2,0})=(0.7,0.3)$ and $(0.6,0.4)$, using the
leading-order expressions~\eqref{eq:g1_approx}--\eqref{eq:g2_approx} with
denominator coefficients $b_{1}=8u_{0}-2v_{0}$ and
$b_{2}=8u_{0}+2v_{0}$; these two curves alone use that leading-order
approximation. Writing the denominators this way,
$b_{1}=6g_{1,0}^{2}+10g_{2,0}^{2}$ and
$b_{2}=6g_{2,0}^{2}+10g_{1,0}^{2}$, makes explicit that for
$v_{0}>0$ one has $b_{2}>b_{1}$, so the initially smaller coupling
$g_{2}$ is driven to zero faster in the infrared -- the selection
mechanism in its simplest form, directly visible in the figure. This
ansatz is chosen for its correct initial slope and for the fact
that $b_{1}+b_{2}=16u_{0}$ reproduces, at leading order in $s$, the
exact product law~\eqref{eq:product}; it is not the solution of the
flow equation with $r$ held exactly fixed, which would instead give
denominator rates $2u_{0}(4-r_{0}^{2})$, generically different from
$b_{1},b_{2}$ except when $r_{0}\in\{0,1\}$. The corresponding exact phase portrait is
shown in Figure~\ref{fig:phase_portrait}: trajectories with
$g_{1,0}>g_{2,0}$ are funnelled towards the $g_{2}=0$ axis and those
with $g_{2,0}>g_{1,0}$ towards the $g_{1}=0$ axis, the separatrix
being the symmetric surface $g_{1}=g_{2}$ (dotted); 
the dashed arrowed curve along the same line is the RG flow itself confined to that surface, 
showing that even the symmetric trajectory is driven into the Gaussian fixed point as 
$u(\mu)\to 0$ and all trajectories terminate there.
Within the plotted range the
trajectories do not visually reach the axes $r=\pm1$; this is not an
artifact of an approximation, since the plotted curves solve the
exact equations, but simply reflects how slowly the exact approach
proceeds: the invariant~\eqref{eq:invariant} gives
$1-r^{2}\propto s^{-2/3}$.

The selection mechanism can also be viewed directly in the variable
$r$ it is phrased in terms of. Figure~\ref{fig:ratio_flow} shows the
exact flow of $r(\mu)$ towards the infrared, obtained by direct
numerical integration of the exact system~\eqref{beta_g1}--\eqref{beta_g2}
(not an approximation), for four initial asymmetries at $u_{0}=0.5$.
For comparison, and purely for illustration, the frozen-$u$
approximation~\eqref{eq:r_frozen_u} of
Section~\ref{sec:invariant} is not plotted separately; the curves
shown are the exact numerical result, against which that
approximation can be checked directly.

Figure~\ref{fig:lambda_running} shows the exact Riccati
solution~\eqref{lambda_sol} on the symmetric surface at $u_{0}=0.5$
for five initial values of $\lambda_{0}$. A pole at finite infrared
distance occurs only for $\lambda_{0}<-6\,u_{0}=-3$,
while every initial condition above that threshold is driven onto the
Yukawa-generated branch $\lambda_{p}(\mu)=6\,u(\mu)$, which
approaches $0^{+}$ only logarithmically as $\mu\to0$. Trajectories
starting negative but above the threshold cross $\lambda=0$ at a
finite RG distance before joining that branch from below.

The product $g_{1}(\mu)g_{2}(\mu)$, which controls the Yukawa-induced
part of the non-planar coefficient, is plotted in
Figure~\ref{fig:iruv_suppression} for one symmetric and two
asymmetric initial conditions, together with the commutative
reference under Matching~B. All noncommutative trajectories share the
leading $1/s$ decay of the leading-order approximation but with a
smaller asymptotic prefactor than the commutative curve; the true
asymptotic law on asymmetric trajectories is the faster $s^{-4/3}$
of~\eqref{eq:g1g2_asymptotic}, which the leading-order curves do not
exhibit. This is the quantitative content of the infrared suppression
discussed in Section~\ref{sec:iruv_suppression}: the coefficient of
the $1/(\theta^{2}p^{2})$ singularity is driven to zero in exactly
the regime where that singularity would otherwise dominate.

Figure~\ref{fig:stability} compares the ultraviolet extent of the
perturbative window in the NC and commutative theories as a function
of the initial Yukawa coupling $g_{0}$. The quantity plotted is the
exact one-loop Yukawa Landau pole,
$\ln(\mu_{L}^{\NC}/\mu_{0})=(4\pi)^{2}/(16g_{0}^{2})$ from
Eq.~\eqref{eq:yukawa_landau}, against its commutative counterpart
$\ln(\mu_{L}^{\Cth}/\mu_{0})=(4\pi)^{2}/(10g_{0}^{2})$ under
Matching~B. Both are closed-form results, not estimates. The shaded
region between them is the perturbative window lost to
noncommutativity under Matching~B only. Under Matching~A
($g_{C}=g_{1}+g_{2}=2g$) the ordering reverses and the NC window is
the longer of the two, in accordance with the sum-coupling beta
functions~\eqref{eq:matchingA_boxed} and their
ratio~\eqref{eq:matchingA_ratio}; the figure should therefore not be
read as a matching-independent statement that noncommutativity
shortens the perturbative window. The vacuum-instability scale of
Regime~I lies parametrically at the same order and must be extracted
from the exact Riccati solution~\eqref{lambda_sol} for given
$(u_{0},\lambda_{0})$.

Finally, Figure~\ref{fig:all_three_masses} collects the three
dimensionful runnings on the symmetric critical trajectory
$\lambda_{0}=6\,u_{0}$, with the reference values $g_{0}=0.5$,
$u_{0}=0.5$, $\lambda_{0}=3$, $m_{0}=M_{0}=1$ and $\theta=1$ in
arbitrary units, for which $X(s)=1+0.025330\,s$. The extraordinarily
slow $\ln^{1/8}(\mu_{0}/\mu)$ growth of the fermion mass established
in Section~\ref{sec:mass_sym_sols} is directly visible in the figure;
the scalar mass tracks it with the attractor
ratio~\eqref{eq:M2_attractor},
$M^{2}\to0.8\,m^{2}$, independently of
its initial value; and $a^{2}\theta^{2}$ flows as $X^{-1/2}$ to
$3$, independently of $a_{0}^{2}$. All three
asymptotics are exact consequences of the Riccati condition
$\alpha^{2}=36$.

The convergence of $a^{2}\theta^{2}$ just quoted holds independently
of the starting value $a_{0}^{2}$, and Figure~\ref{fig:a2_running}
makes that basin of attraction explicit: several trajectories, from
initial values both above and below $3$, are shown converging to
the same infrared limit from the closed-form
solution~\eqref{eq:a2_sym_sol}. The approach is governed by
$X^{-1/2}$, i.e.\ by $1/\sqrt{\ln(\mu_{0}/\mu)}$, so the limit is
reached only slowly -- visibly more slowly than the exponents
governing $m(\mu)$ and $M(\mu)$ in Figure~\ref{fig:all_three_masses}
-- but the convergence to a single point regardless of $a_{0}^{2}$ is
exact on this submanifold, within the IR-subtraction prescription of
Section~\ref{sec:a2_status}; the trajectory starting exactly at
$a_{0}^{2}\theta^{2}=3$ is RG-invariant.

\begin{figure}[p]
	\centering
	\begin{subfigure}[b]{0.46\textwidth}
		\centering
		\begin{tikzpicture}
			\begin{axis}[
				width=0.95\linewidth,height=0.60\linewidth,
				xlabel={$s=\ln(\mu_0/\mu)$},ylabel={$g_i(\mu)$},
				xmin=0,xmax=100,ymin=0,ymax=0.82,
				grid=both,
				grid style={line width=0.2pt,draw=gray!25},
				major grid style={line width=0.4pt,draw=gray!55},
				legend style={at={(1.015,1)},anchor=north east,
					font=\tiny,fill=none,draw=none,row sep=-1pt},
				legend cell align=left,
				tick label style={font=\tiny},label style={font=\scriptsize},
				]
				\addplot[thick,blue!85,solid,samples=250,domain=0:100]
				{sqrt(0.49/(1+0.024317*x))};
				\addlegendentry{$g_1$, $(0.7,0.3)$}
				\addplot[thick,blue!40,solid,samples=250,domain=0:100]
				{sqrt(0.09/(1+0.034450*x))};
				\addlegendentry{$g_2$, $(0.7,0.3)$}
				\addplot[thick,red!80,dashed,samples=250,domain=0:100]
				{sqrt(0.36/(1+0.023810*x))};
				\addlegendentry{$g_1$, $(0.6,0.4)$}
				\addplot[thick,red!40,dashed,samples=250,domain=0:100]
				{sqrt(0.16/(1+0.028876*x))};
				\addlegendentry{$g_2$, $(0.6,0.4)$}
				\draw[-{Stealth[length=4pt]},black!65]
				(axis cs:68,0.11)--(axis cs:82,0.10)
				node[pos=0,left,font=\tiny,black!65,align=center]
				{$g_2\to0$\\faster};
			\end{axis}
		\end{tikzpicture}
		\caption{\textbf{Infrared selection of $g_{1},g_{2}$.} Infrared running of $g_{1}(\mu)$ and $g_{2}(\mu)$ for $(g_{1,0},g_{2,0})=(0.7,0.3)$ and $(0.6,0.4)$, leading-order approximation Eqs.~\eqref{eq:g1_approx}--\eqref{eq:g2_approx}.}
		\label{fig:g_asymmetric}
	\end{subfigure}
	\hfill
	\begin{subfigure}[b]{0.46\textwidth}
		\centering
		\begin{tikzpicture}
			\begin{axis}[
				width=0.60\linewidth,height=0.60\linewidth,
				xlabel={$g_1^2(\mu)$},ylabel={$g_2^2(\mu)$},
				xmin=0,xmax=0.62,ymin=0,ymax=0.62,
				xtick={0,0.1,...,0.6},ytick={0,0.1,...,0.6},
				grid=both,
				grid style={line width=0.2pt,draw=gray!20},
				major grid style={line width=0.4pt,draw=black!45},
				tick label style={font=\tiny},label style={font=\scriptsize},
				clip=false,
				]
				\addplot[semithick,black!55,densely dotted,domain=0:0.60]{x};
				\node[font=\tiny,black!55,rotate=45] at (axis cs:0.34,0.40)
				{separatrix $g_1=g_2$};
				\addplot[semithick,blue!85,
				decoration={markings,mark=at position 0.45 with
					{\arrow[blue!85,scale=1.2]{stealth}}},
				postaction={decorate}] coordinates{
					(0.5500,0.0500)(0.4214,0.0332)(0.3425,0.0241)(0.2890,0.0185)(0.2502,0.0148)(0.2206,0.0121)(0.1975,0.0102)(0.1789,0.0087)(0.1636,0.0075)(0.1507,0.0066)(0.1396,0.0059)(0.1302,0.0052)(0.1220,0.0047)(0.1147,0.0043)(0.1083,0.0039)(0.1025,0.0036)(0.0974,0.0033)(0.0928,0.0030)(0.0886,0.0028)(0.0847,0.0026)(0.0812,0.0024)(0.0779,0.0023)(0.0750,0.0022)(0.0722,0.0020)(0.0697,0.0019)(0.0673,0.0018)(0.0650,0.0017)(0.0630,0.0016)(0.0610,0.0015)(0.0592,0.0015)(0.0574,0.0014)(0.0558,0.0013)(0.0543,0.0013)(0.0528,0.0012)(0.0515,0.0012)(0.0502,0.0011)(0.0489,0.0011)(0.0477,0.0010)(0.0466,0.0010)(0.0455,0.0010)
				};
				\addplot[semithick,blue!60,
				decoration={markings,mark=at position 0.48 with
					{\arrow[blue!60,scale=1.2]{stealth}}},
				postaction={decorate}] coordinates{
					(0.4500,0.1000)(0.3473,0.0698)(0.2839,0.0526)(0.2408,0.0416)(0.2094,0.0340)(0.1854,0.0285)(0.1666,0.0244)(0.1514,0.0213)(0.1388,0.0187)(0.1283,0.0167)(0.1192,0.0149)(0.1114,0.0135)(0.1046,0.0123)(0.0986,0.0113)(0.0933,0.0104)(0.0885,0.0096)(0.0842,0.0089)(0.0804,0.0083)(0.0769,0.0077)(0.0736,0.0073)(0.0707,0.0068)(0.0680,0.0064)(0.0655,0.0061)(0.0632,0.0057)(0.0610,0.0054)(0.0590,0.0052)(0.0571,0.0049)(0.0553,0.0047)(0.0537,0.0045)(0.0521,0.0043)(0.0507,0.0041)(0.0493,0.0039)(0.0480,0.0038)(0.0468,0.0036)(0.0456,0.0035)(0.0445,0.0034)(0.0434,0.0032)(0.0424,0.0031)(0.0415,0.0030)(0.0405,0.0029)
				};
				\addplot[semithick,blue!40,
				decoration={markings,mark=at position 0.52 with
					{\arrow[blue!40,scale=1.2]{stealth}}},
				postaction={decorate}] coordinates{
					(0.3800,0.1800)(0.2865,0.1281)(0.2309,0.0985)(0.1939,0.0795)(0.1675,0.0663)(0.1475,0.0566)(0.1321,0.0493)(0.1197,0.0435)(0.1095,0.0389)(0.1010,0.0351)(0.0937,0.0319)(0.0875,0.0292)(0.0820,0.0269)(0.0773,0.0249)(0.0731,0.0231)(0.0693,0.0216)(0.0659,0.0202)(0.0629,0.0190)(0.0601,0.0179)(0.0576,0.0170)(0.0553,0.0161)(0.0532,0.0153)(0.0512,0.0145)(0.0494,0.0138)(0.0477,0.0132)(0.0462,0.0126)(0.0447,0.0121)(0.0433,0.0116)(0.0421,0.0112)(0.0408,0.0107)(0.0397,0.0103)(0.0386,0.0100)(0.0376,0.0096)(0.0367,0.0093)(0.0358,0.0090)(0.0349,0.0087)(0.0341,0.0084)(0.0333,0.0082)(0.0326,0.0079)(0.0318,0.0077)
				};
				\addplot[semithick,red!85,
				decoration={markings,mark=at position 0.45 with
					{\arrow[red!85,scale=1.2]{stealth}}},
				postaction={decorate}] coordinates{
					(0.0500,0.5500)(0.0332,0.4214)(0.0241,0.3425)(0.0185,0.2890)(0.0148,0.2502)(0.0121,0.2206)(0.0102,0.1975)(0.0087,0.1789)(0.0075,0.1636)(0.0066,0.1507)(0.0059,0.1396)(0.0052,0.1302)(0.0047,0.1220)(0.0043,0.1147)(0.0039,0.1083)(0.0036,0.1025)(0.0033,0.0974)(0.0030,0.0928)(0.0028,0.0886)(0.0026,0.0847)(0.0024,0.0812)(0.0023,0.0779)(0.0022,0.0750)(0.0020,0.0722)(0.0019,0.0697)(0.0018,0.0673)(0.0017,0.0650)(0.0016,0.0630)(0.0015,0.0610)(0.0015,0.0592)(0.0014,0.0574)(0.0013,0.0558)(0.0013,0.0543)(0.0012,0.0528)(0.0012,0.0515)(0.0011,0.0502)(0.0011,0.0489)(0.0010,0.0477)(0.0010,0.0466)(0.0010,0.0455)
				};
				\addplot[semithick,red!60,
				decoration={markings,mark=at position 0.48 with
					{\arrow[red!60,scale=1.2]{stealth}}},
				postaction={decorate}] coordinates{
					(0.1000,0.4500)(0.0698,0.3473)(0.0526,0.2839)(0.0416,0.2408)(0.0340,0.2094)(0.0285,0.1854)(0.0244,0.1666)(0.0213,0.1514)(0.0187,0.1388)(0.0167,0.1283)(0.0149,0.1192)(0.0135,0.1114)(0.0123,0.1046)(0.0113,0.0986)(0.0104,0.0933)(0.0096,0.0885)(0.0089,0.0842)(0.0083,0.0804)(0.0077,0.0769)(0.0073,0.0736)(0.0068,0.0707)(0.0064,0.0680)(0.0061,0.0655)(0.0057,0.0632)(0.0054,0.0610)(0.0052,0.0590)(0.0049,0.0571)(0.0047,0.0553)(0.0045,0.0537)(0.0043,0.0521)(0.0041,0.0507)(0.0039,0.0493)(0.0038,0.0480)(0.0036,0.0468)(0.0035,0.0456)(0.0034,0.0445)(0.0032,0.0434)(0.0031,0.0424)(0.0030,0.0415)(0.0029,0.0405)
				};
				\addplot[semithick,red!40,
				decoration={markings,mark=at position 0.52 with
					{\arrow[red!40,scale=1.2]{stealth}}},
				postaction={decorate}] coordinates{
					(0.1800,0.3800)(0.1281,0.2865)(0.0985,0.2309)(0.0795,0.1939)(0.0663,0.1675)(0.0566,0.1475)(0.0493,0.1321)(0.0435,0.1197)(0.0389,0.1095)(0.0351,0.1010)(0.0319,0.0937)(0.0292,0.0875)(0.0269,0.0820)(0.0249,0.0773)(0.0231,0.0731)(0.0216,0.0693)(0.0202,0.0659)(0.0190,0.0629)(0.0179,0.0601)(0.0170,0.0576)(0.0161,0.0553)(0.0153,0.0532)(0.0145,0.0512)(0.0138,0.0494)(0.0132,0.0477)(0.0126,0.0462)(0.0121,0.0447)(0.0116,0.0433)(0.0112,0.0421)(0.0107,0.0408)(0.0103,0.0397)(0.0100,0.0386)(0.0096,0.0376)(0.0093,0.0367)(0.0090,0.0358)(0.0087,0.0349)(0.0084,0.0341)(0.0082,0.0333)(0.0079,0.0326)(0.0077,0.0318)
				};
				\addplot[semithick,gray!60,dashed,
				decoration={markings,mark=at position 0.50 with
					{\arrow[gray!60,scale=1.2]{stealth}}},
				postaction={decorate}] coordinates{
					(0.2800,0.2800)(0.2054,0.2054)(0.1622,0.1622)(0.1340,0.1340)(0.1142,0.1142)(0.0993,0.0993)(0.0880,0.0880)(0.0790,0.0790)(0.0716,0.0716)(0.0656,0.0656)(0.0604,0.0604)(0.0560,0.0560)(0.0522,0.0522)(0.0489,0.0489)(0.0460,0.0460)(0.0434,0.0434)(0.0411,0.0411)(0.0390,0.0390)(0.0371,0.0371)(0.0354,0.0354)(0.0338,0.0338)(0.0324,0.0324)(0.0311,0.0311)(0.0299,0.0299)(0.0288,0.0288)(0.0277,0.0277)(0.0268,0.0268)(0.0259,0.0259)(0.0250,0.0250)(0.0243,0.0243)(0.0235,0.0235)(0.0228,0.0228)(0.0222,0.0222)(0.0215,0.0215)(0.0210,0.0210)(0.0204,0.0204)(0.0199,0.0199)(0.0194,0.0194)(0.0189,0.0189)(0.0184,0.0184)
				};
				\addplot[mark=*,mark size=2.5pt,black] coordinates{(0,0)};
				\node[font=\tiny,anchor=north west,align=center] at (axis cs:-0.13,-0.03)
				{Gaussian\\fixed point};
			\end{axis}
		\end{tikzpicture}
		\caption{\textbf{Exact renormalization-group flow.} Phase portrait in the $(g_{1}^{2},g_{2}^{2})$ plane, arrows pointing towards the infrared. Blue: $g_{1}>g_{2}$; red: $g_{2}>g_{1}$; dotted: separatrix $g_{1}=g_{2}$.}
		\label{fig:phase_portrait}
	\end{subfigure}
	\hfill
	\begin{subfigure}[b]{0.46\textwidth}
		\centering
		\begin{tikzpicture}
			\begin{axis}[
				width=0.95\linewidth,height=0.65\linewidth,
				xlabel={$s=\ln(\mu_0/\mu)$},
				ylabel={$r(\mu)=(g_1^2-g_2^2)/(g_1^2+g_2^2)$},
				xmin=0,xmax=600,ymin=0,ymax=1,
				grid=both,
				grid style={line width=0.2pt,draw=gray!25},
				major grid style={line width=0.4pt,draw=gray!55},
				legend style={at={(1.01,0.35)},anchor=south east,
					font=\tiny,fill=none,draw=none,row sep=-1pt},
				legend cell align=left,
				tick label style={font=\tiny},label style={font=\scriptsize},
				]
				\addplot[thick,violet,solid] coordinates{
					(0.0,0.80000)(8.6,0.81407)(17.3,0.82571)(26.0,0.83546)(34.7,0.84380)(43.4,0.85103)(52.1,0.85739)(60.8,0.86305)(69.5,0.86812)(78.2,0.87270)(86.9,0.87686)(95.6,0.88067)(104.3,0.88418)(113.0,0.88741)(121.7,0.89041)(130.4,0.89320)(139.1,0.89581)(147.8,0.89825)(156.5,0.90054)(165.2,0.90270)(173.9,0.90473)(182.6,0.90666)(191.3,0.90848)(200.0,0.91021)(208.6,0.91184)(217.3,0.91341)(226.0,0.91491)(234.7,0.91634)(243.4,0.91770)(252.1,0.91901)(260.8,0.92027)(269.5,0.92147)(278.2,0.92263)(286.9,0.92374)(295.6,0.92482)(304.3,0.92585)(313.0,0.92685)(321.7,0.92781)(330.4,0.92874)(339.1,0.92964)(347.8,0.93050)(356.5,0.93135)(365.2,0.93216)(373.9,0.93295)(382.6,0.93372)(391.3,0.93446)(400.0,0.93518)(408.6,0.93588)(417.3,0.93656)(426.0,0.93722)(434.7,0.93787)(443.4,0.93850)(452.1,0.93911)(460.8,0.93971)(469.5,0.94029)(478.2,0.94085)(486.9,0.94141)(495.6,0.94195)(504.3,0.94247)(513.0,0.94299)(521.7,0.94349)(530.4,0.94398)(539.1,0.94446)(547.8,0.94493)(556.5,0.94539)(565.2,0.94584)(573.9,0.94628)(582.6,0.94672)(591.3,0.94714)(600.0,0.94755)
				};
				\addlegendentry{$r_0=0.80$}
				\addplot[thick,black!60!green,solid] coordinates{
					(0.0,0.50000)(8.6,0.51869)(17.3,0.53478)(26.0,0.54875)(34.7,0.56111)(43.4,0.57217)(52.1,0.58218)(60.8,0.59131)(69.5,0.59971)(78.2,0.60747)(86.9,0.61469)(95.6,0.62143)(104.3,0.62775)(113.0,0.63370)(121.7,0.63931)(130.4,0.64461)(139.1,0.64965)(147.8,0.65444)(156.5,0.65901)(165.2,0.66337)(173.9,0.66753)(182.6,0.67153)(191.3,0.67536)(200.0,0.67904)(208.6,0.68254)(217.3,0.68595)(226.0,0.68925)(234.7,0.69242)(243.4,0.69549)(252.1,0.69846)(260.8,0.70133)(269.5,0.70412)(278.2,0.70682)(286.9,0.70944)(295.6,0.71198)(304.3,0.71446)(313.0,0.71686)(321.7,0.71920)(330.4,0.72148)(339.1,0.72369)(347.8,0.72586)(356.5,0.72796)(365.2,0.73002)(373.9,0.73203)(382.6,0.73399)(391.3,0.73590)(400.0,0.73777)(408.6,0.73958)(417.3,0.74137)(426.0,0.74313)(434.7,0.74484)(443.4,0.74652)(452.1,0.74816)(460.8,0.74978)(469.5,0.75136)(478.2,0.75290)(486.9,0.75442)(495.6,0.75591)(504.3,0.75738)(513.0,0.75881)(521.7,0.76022)(530.4,0.76160)(539.1,0.76296)(547.8,0.76430)(556.5,0.76561)(565.2,0.76690)(573.9,0.76817)(582.6,0.76942)(591.3,0.77065)(600.0,0.77185)
				};
				\addlegendentry{$r_0=0.50$}
				\addplot[thick,red,solid] coordinates{
					(0.0,0.20000)(8.6,0.20968)(17.3,0.21819)(26.0,0.22574)(34.7,0.23254)(43.4,0.23875)(52.1,0.24446)(60.8,0.24977)(69.5,0.25472)(78.2,0.25937)(86.9,0.26376)(95.6,0.26791)(104.3,0.27186)(113.0,0.27562)(121.7,0.27921)(130.4,0.28265)(139.1,0.28596)(147.8,0.28913)(156.5,0.29219)(165.2,0.29515)(173.9,0.29800)(182.6,0.30076)(191.3,0.30344)(200.0,0.30603)(208.6,0.30852)(217.3,0.31097)(226.0,0.31336)(234.7,0.31568)(243.4,0.31794)(252.1,0.32015)(260.8,0.32230)(269.5,0.32441)(278.2,0.32646)(286.9,0.32847)(295.6,0.33044)(304.3,0.33236)(313.0,0.33425)(321.7,0.33610)(330.4,0.33791)(339.1,0.33968)(347.8,0.34143)(356.5,0.34314)(365.2,0.34482)(373.9,0.34647)(382.6,0.34809)(391.3,0.34969)(400.0,0.35126)(408.6,0.35278)(417.3,0.35430)(426.0,0.35580)(434.7,0.35727)(443.4,0.35872)(452.1,0.36015)(460.8,0.36155)(469.5,0.36294)(478.2,0.36431)(486.9,0.36566)(495.6,0.36699)(504.3,0.36830)(513.0,0.36959)(521.7,0.37087)(530.4,0.37213)(539.1,0.37338)(547.8,0.37460)(556.5,0.37582)(565.2,0.37702)(573.9,0.37820)(582.6,0.37937)(591.3,0.38053)(600.0,0.38167)
				};
				\addlegendentry{$r_0=0.20$}
				\addplot[thick,blue,solid] coordinates{
					(0.0,0.05000)(8.6,0.05252)(17.3,0.05474)(26.0,0.05672)(34.7,0.05851)(43.4,0.06015)(52.1,0.06167)(60.8,0.06308)(69.5,0.06440)(78.2,0.06564)(86.9,0.06682)(95.6,0.06794)(104.3,0.06900)(113.0,0.07002)(121.7,0.07099)(130.4,0.07193)(139.1,0.07283)(147.8,0.07369)(156.5,0.07453)(165.2,0.07534)(173.9,0.07613)(182.6,0.07689)(191.3,0.07762)(200.0,0.07834)(208.6,0.07903)(217.3,0.07971)(226.0,0.08038)(234.7,0.08102)(243.4,0.08166)(252.1,0.08227)(260.8,0.08288)(269.5,0.08347)(278.2,0.08404)(286.9,0.08461)(295.6,0.08517)(304.3,0.08571)(313.0,0.08624)(321.7,0.08677)(330.4,0.08728)(339.1,0.08779)(347.8,0.08828)(356.5,0.08877)(365.2,0.08925)(373.9,0.08973)(382.6,0.09019)(391.3,0.09065)(400.0,0.09110)(408.6,0.09154)(417.3,0.09198)(426.0,0.09241)(434.7,0.09283)(443.4,0.09325)(452.1,0.09367)(460.8,0.09408)(469.5,0.09448)(478.2,0.09488)(486.9,0.09527)(495.6,0.09566)(504.3,0.09604)(513.0,0.09642)(521.7,0.09680)(530.4,0.09717)(539.1,0.09753)(547.8,0.09789)(556.5,0.09825)(565.2,0.09861)(573.9,0.09896)(582.6,0.09930)(591.3,0.09964)(600.0,0.09998)
				};
				\addlegendentry{$r_0=0.05$}
				\addplot[semithick,gray!55,densely dashed,samples=200,domain=0:600]{1};
			\end{axis}
		\end{tikzpicture}
		\caption{\textbf{Exact flow of the coupling ratio $r(\mu)$.} Towards the infrared, for $u_{0}=0.5$ and four initial asymmetries $r_{0}$.}
		\label{fig:ratio_flow}
	\end{subfigure}
	\hfill
	\begin{subfigure}[b]{0.46\textwidth}
		\centering
		\begin{tikzpicture}
			\begin{axis}[
				width=0.95\linewidth,height=0.65\linewidth,
				xlabel={$s=\ln(\mu_0/\mu)$},ylabel={$\lambda(\mu)$},
				xmin=0,xmax=220,ymin=-20,ymax=7,
				restrict y to domain=-22:9,
				grid=both,
				grid style={line width=0.2pt,draw=gray!25},
				major grid style={line width=0.4pt,draw=gray!55},
				legend style={at={(1.02,0.42)},anchor=east,
					font=\tiny,fill=none,draw=none,row sep=-1pt},
				legend cell align=left,
				tick label style={font=\tiny},label style={font=\scriptsize},
				clip=true,
				]
				\addplot[thick,orange!85,solid,samples=450,domain=0:220]
				{3.0/(1+0.025330*x)
					+1/(0.4*(1+0.025330*x)^(2)
					+0.166667*((1+0.025330*x)^(2)-(1+0.025330*x)))};
				\addlegendentry{$\lambda_0=5.5$}
				\addplot[thick,red!80,solid,samples=450,domain=0:220]
				{3.0/(1+0.025330*x)
					+1/(2.0*(1+0.025330*x)^(2)
					+0.166667*((1+0.025330*x)^(2)-(1+0.025330*x)))};
				\addlegendentry{$\lambda_0=3.5$}
				\addplot[thick,black!40,densely dotted,samples=280,domain=0:220]
				{3.0/(1+0.025330*x)};
				\addlegendentry{$\lambda_p(\mu)$}
				\addplot[thick,gray!65,densely dashed,samples=450,domain=0:220]
				{3.0/(1+0.025330*x)
					+1/(-0.333333*(1+0.025330*x)^(2)
					+0.166667*((1+0.025330*x)^(2)-(1+0.025330*x)))};
				\addlegendentry{$\lambda_0=0.0$}
				\addplot[thick,black!55!green,solid,samples=450,domain=0:220]
				{3.0/(1+0.025330*x)
					+1/(-0.222222*(1+0.025330*x)^(2)
					+0.166667*((1+0.025330*x)^(2)-(1+0.025330*x)))};
				\addlegendentry{$\lambda_0=-1.5$}
				\addplot[thick,blue!80,solid,samples=450,domain=0:108]
				{3.0/(1+0.025330*x)
					+1/(-0.125*(1+0.025330*x)^(2)
					+0.166667*((1+0.025330*x)^(2)-(1+0.025330*x)))};
				\addlegendentry{$\lambda_0=-5.0$}
				\addplot[thin,black,domain=0:220]{0};
				\node[font=\tiny,red!60] at (axis cs:116,4)
				{driven onto the critical trajectory branch  $\lambda_p(\mu)$};
				\addplot[thin,blue!80,dashed] coordinates{(118.44,-20)(118.44,7)};
				\node[font=\tiny,blue!80] at (axis cs:122,-18.5)
				{IR pole};
			\end{axis}
		\end{tikzpicture}
		\caption{\textbf{Running of $\lambda(\mu)$ on the symmetric surface.} At $u_{0}=0.5$, from the Riccati solution~\eqref{lambda_sol}, for five values of $\lambda_{0}$.}
		\label{fig:lambda_running}
	\end{subfigure}
	\hfill
	\begin{subfigure}[b]{0.46\textwidth}
		\centering
		\begin{tikzpicture}
			\begin{axis}[
				width=0.95\linewidth,height=0.60\linewidth,
				xlabel={$s=\ln(\mu_0/\mu)$},
				ylabel={$g_1(\mu)\,g_2(\mu)$},
				xmin=0,xmax=150,ymin=0,ymax=0.30,
				grid=both,
				grid style={line width=0.2pt,draw=gray!25},
				major grid style={line width=0.4pt,draw=gray!55},
				legend style={at={(1.02,1)},anchor=north east,
					font=\tiny,fill=none,draw=none,row sep=-1pt},
				legend cell align=left,
				tick label style={font=\tiny},label style={font=\scriptsize},
				]
				\addplot[thick,blue,dashed,samples=280,domain=0:150]
				{0.25/(1+0.015831*x)};
				\addlegendentry{Commutative, $g_0=0.5$ (Match.B)}
				\addplot[thick,blue,solid,samples=280,domain=0:150]
				{0.25/(1+0.025330*x)};
				\addlegendentry{NC sym., $g_{1,0}=g_{2,0}=0.5$}
				\addplot[thick,red,solid,samples=280,domain=0:150]
				{0.21/sqrt((1+0.034450*x)*(1+0.024317*x))};
				\addlegendentry{NC asym., $(g_{1,0},g_{2,0})=(0.7,0.3)$}
				\addplot[thick,black!60!green,solid,samples=280,domain=0:150]
				{0.16/sqrt((1+0.042047*x)*(1+0.026851*x))};
				\addlegendentry{NC asym., $(g_{1,0},g_{2,0})=(0.8,0.2)$}
				\addplot[thin,black!80,densely dotted,samples=180,domain=8:150]
				{1.5/x};
				\node[font=\tiny,black!80] at (axis cs:60,0.035){$\sim s^{-1}$};
				\addplot[thin,black!80,dash dot,samples=180,domain=14:150]
				{4.0/(x^(4/3))};
				\node[font=\tiny,black!80] at (axis cs:20,0.035){$s^{-4/3} \sim $};
			\end{axis}
		\end{tikzpicture}
		\caption{\textbf{Infrared suppression of $g_{1}(\mu)g_{2}(\mu)$.} Noncommutative theory (three initial conditions) versus the commutative reference at $g_{0}=0.5$ under Matching~B.}
		\label{fig:iruv_suppression}
	\end{subfigure}
	\hfill
	\begin{subfigure}[b]{0.46\textwidth}
		\centering
		\begin{tikzpicture}
			\begin{axis}[
				width=0.95\linewidth,height=0.60\linewidth,
				xlabel={$g_0$},
				ylabel={$\ln(\mu_L/\mu_0)$},
				xmin=0.24,xmax=0.85,ymin=0,ymax=165,
				grid=both,
				grid style={line width=0.2pt,draw=gray!25},
				major grid style={line width=0.4pt,draw=gray!55},
				legend style={at={(1.01,0.97)},anchor=north east,
					font=\tiny,fill=none,draw=none,row sep=-1pt},
				legend cell align=left,
				tick label style={font=\tiny},label style={font=\scriptsize},
				clip=true,
				]
				\addplot[thick,red,dashed,name path=Comm,
				samples=200,domain=0.315:0.85]
				{157.914/(10*x*x)};
				\addlegendentry{C: $(4\pi)^2/(10g_0^2)$}
				\addplot[thick,blue,solid,name path=NC,
				samples=200,domain=0.25:0.85]
				{157.914/(16*x*x)};
				\addlegendentry{NC: $(4\pi)^2/(16g_0^2)$}
				\addplot[blue!12,fill opacity=0.55]
				fill between[of=Comm and NC,
				soft clip={domain=0.32:0.85}];
				\addplot[thin,black!80,dashed] coordinates{(0.5,0)(0.5,160)};
				\node[font=\tiny,black!80,rotate=90] at (axis cs:0.515,110)
				{$g_0=0.5$};
				\draw[-{Stealth[length=3.5pt]},black!55]
				(axis cs:0.55,54)--(axis cs:0.55,32)
				node[pos=0.3,above right,font=\tiny,black!80,align=left]
				{window lost to NC};
				\draw[thin,red!70,densely dotted]
				(axis cs:0.24,63.166)--(axis cs:0.5,63.166);
				\node[circle,fill=red,inner sep=0.9pt] at (axis cs:0.5,63.166) {};
				\node[font=\tiny,red!80!black,anchor=east]
				at (axis cs:0.485,71) {$63.17$};
				\draw[thin,blue!70,densely dotted]
				(axis cs:0.24,39.479)--(axis cs:0.5,39.479);
				\node[circle,fill=blue,inner sep=0.9pt] at (axis cs:0.5,39.479) {};
				\node[font=\tiny,blue!80!black,anchor=east]
				at (axis cs:0.485,25) {$39.48$};
			\end{axis}
		\end{tikzpicture}
		\caption{\textbf{Yukawa Landau pole versus $g_{0}$.} Ultraviolet extent of the perturbative window: noncommutative versus commutative under Matching~B, with both distances marked at the reference value $g_{0}=0.5$.}
		\label{fig:stability}
	\end{subfigure}
	\hfill
	\begin{subfigure}[b]{0.46\textwidth}
		\centering
		\begin{tikzpicture}
			\begin{axis}[
				width=0.95\linewidth,height=0.60\linewidth,
				xlabel={$s=\ln(\mu_0/\mu)$},
				ylabel={normalized values},
				xmin=0,xmax=200,ymin=0,ymax=1.6,
				grid=both,
				grid style={line width=0.2pt,draw=gray!25},
				major grid style={line width=0.4pt,draw=gray!55},
				legend style={at={(1.015,-0.03)},anchor=south east,
					font=\tiny,fill=none,draw=none,row sep=-1pt},
				legend cell align=left,
				tick label style={font=\tiny},label style={font=\scriptsize},
				]
				\addplot[thick,blue,solid,samples=300,domain=0:200]
				{(1+0.025330*x)^(1/8)};
				\addlegendentry{$m(\mu)/m_0$}
				\addplot[thick,red,solid,samples=300,domain=0:200]
				{sqrt(0.2*(1+0.025330*x)^(-1)
					+0.8*(1+0.025330*x)^(0.25))};
				\addlegendentry{$M(\mu)/M_0$}
				\addplot[thick,black!60!green,solid,samples=300,domain=0:200]
				{(3.0+(0.5-3.0)/sqrt(1+0.025330*x))/3.0};
				\addlegendentry{$a^2\theta^2/3,\ a_0^2\theta^2=0.5$}
				\addplot[semithick,gray!55,densely dashed,samples=200,domain=0:200]{1};
			\end{axis}
		\end{tikzpicture}
		\caption{\textbf{Normalized dimensionful runnings.} $m(\mu)/m_{0}$, $M(\mu)/M_{0}$, and $a^{2}(\mu)\theta^{2}/3$ on the symmetric critical trajectory, at $g_{0}=0.5$.}
		\label{fig:all_three_masses}
	\end{subfigure}
	\hfill
	\begin{subfigure}[b]{0.46\textwidth}
		\centering
		\begin{tikzpicture}
			\begin{axis}[
				width=0.95\linewidth,height=0.60\linewidth,
				xlabel={$s=\ln(\mu_0/\mu)$},
				ylabel={$a^2(\mu)\,\theta^2$},
				xmin=0,xmax=200,ymin=0,ymax=5.2,
				grid=both,
				grid style={line width=0.2pt,draw=gray!25},
				major grid style={line width=0.4pt,draw=gray!55},
				legend style={at={(1.01,0.4)},anchor=north east,
					font=\tiny,  fill=none,draw=none,row sep=-1pt},
				legend cell align=left,
				tick label style={font=\tiny},label style={font=\scriptsize},
				]
				\addplot[thick,black!60!green,solid,samples=250,domain=0:200]
				{3.0+(5-3.0)/sqrt(1+0.025330*x)};
				\addlegendentry{$a_0^2\theta^2=5$}
				\addplot[thick,red,solid,samples=250,domain=0:200]
				{3.0+(1.5-3.0)/sqrt(1+0.025330*x)};
				\addlegendentry{$a_0^2\theta^2=1.5$}
				\addplot[thick,blue,solid,samples=250,domain=0:200]
				{3.0+(0.5-3.0)/sqrt(1+0.025330*x)};
				\addlegendentry{$a_0^2\theta^2=0.5$}
				\addplot[semithick,violet,densely dashed,samples=200,domain=0:200]{3.0};
				\node[font=\tiny,violet,align=center] at (axis cs:57,3.25)
				{$a_0^2\theta^2=3$ (RG-invariant)};
			\end{axis}
		\end{tikzpicture}
		\caption{\textbf{Prescription-dependent running of $a^{2}\theta^{2}$.} On the critical trajectory, from Eq.~\eqref{eq:a2_sym_sol}, for three initial values $a_{0}^{2}\theta^{2}$.}
		\label{fig:a2_running}
	\end{subfigure}
	\end{figure}

\end{document}